\documentclass[a4paper,11pt]{article}

\usepackage[utf8]{inputenc}
\usepackage[british]{babel}
\usepackage{csquotes}

\usepackage{manfnt}
\usepackage{braket}

\newcommand{\COMMENTOOK}[1]{}

\usepackage{graphicx}
\usepackage{subcaption} 
\usepackage{color} 

\usepackage{authblk}
\usepackage{hyperref}
\usepackage{amsmath}
\usepackage{amssymb}
\usepackage{amsfonts}
\usepackage{enumitem}
\numberwithin{equation}{section}

\newcommand{\myscale}[1]{\scalebox{0.8}{#1}}
\newcommand{\myscaleM}[1]{\scalebox{0.8}{#1}}
\newcommand{\myscaleR}[1]{\scalebox{1.0}{#1}}
\newcommand{\myscaleT}[1]{\scalebox{0.8}{#1}}

\usepackage{ytableau}

\newcommand\Ytr[3]{
  {\begin{ytableau}
      #1 & #2 & #3
   \end{ytableau}
  }
  }

\newcommand\Yto[3]{
  {\begin{ytableau}
      #1 & #2 \\
      #3
   \end{ytableau}
  }
  }

\newcommand\Yt[2]{
  {\begin{ytableau}
      #1 & #2 
   \end{ytableau}
  }
  }
\newcommand\Ytfrom[3]{
  {\begin{ytableau}
      #1 & #2 
   \end{ytableau}
  }_{(#3)}
  }

\newcommand\Yo[1]{
  {\begin{ytableau}
      #1 
   \end{ytableau}
  }
  }

\newcommand\Yoo[2]{
  {\begin{ytableau}
      #1 \\ #2 
   \end{ytableau}
  }
  }

\newcommand\Yooo[3]{
  {\begin{ytableau}
      #1 \\ #2 \\ #3
   \end{ytableau}
  }
  }

\newcommand\YsymmOV[1]{
\overbrace{ \smallY{\ydiagram{2}} \dots  \smallY{\ydiagram{1}} }
^{ #1 }
}

\newcommand\YmixOVUN[3]{
  {\begin{ytableau}
       \scriptstyle 2_1 & \none[\dots] & \none[\dots] & \scriptstyle 2_{#1}
       & \scriptstyle 1_1 & \none[\dots] & \scriptstyle 1_{#2}
       \\
       \scriptstyle 1_1 & \none[\dots] & \scriptstyle 1_{#3}
   \end{ytableau}
  }
}

\newcommand\smallY[1]{
\ytableausetup{boxsize=0.5em}
#1
\ytableausetup{boxsize=1em}
}

\usepackage{cleveref}

\usepackage{pdflscape} 
\usepackage{afterpage} 
\usepackage{geometry}
\usepackage[
backend=biber,
style=numeric-comp, 
sorting=none 
]{biblatex}
\newcommand{\nzm}{{n.z.m.} }
\newcommand{\wrt}{{w.r.t.} }

\newcommand{\lc}{{lightcone} }

\newcommand{\oh}{ \frac{1}{2} }
\newcommand{\ap}{ \alpha' }

\newcommand{\sap}{ \sqrt{\alpha'} }
\newcommand{\sdap}{ \sqrt{2\alpha'} }

\newcommand{\cN}{{\cal N} }

\newcommand{\cS}{{\cal S} }

\newcommand{\cV}{{\cal V} }

\newcommand{\sN}[1]{ { [#1] } }  

\newcommand{\bce}{{(b\cdot e)}}

\newcommand{\GL}{``GL''}
\newcommand{\ualpha}{ {\underline \alpha}}

\newcommand{\uA}{ {\underline A}}

\newcommand{\uk}{ {\underline k}}

\title{
The bosonic string spectrum and the explicit states up to level $10$
from the \lc and the chaotic behavior of certain string
amplitudes
}

\author[1]{Igor Pesando}

\affil[1]
{%
  Dipartimento di Fisica, Universit\`{a} di Torino \authorcr
  and I.N.F.N. -- Sezione di Torino \authorcr
  Via P.\ Giuria 1, I-10125 Torino, Italy 
}

\hypersetup
{%
  pdfauthor={
  Igor Pesando},
  pdfsubject={String Theory},
  pdfkeywords={string theory, physical states},
}

\begin{document}

\maketitle

\begin{abstract}
  We compute the irreps and their multiplicities of bosonic string
  spectrum up to level $10$ and we give explicitly the on shell top
  level \lc states which make the irreps.
  For the irreps up to three indexes and all the totally antisymmetric ones
  we give the general recipe and
  the full irreps.
  It turns out that \lc is quite efficient in building these low
  indexes irreps once the top level states are known.

  For scalars and vectors we compute the multiplicity up to level $22$
  and $19$ respectively.
The first scalar at odd level appears at level $11$.

For the bosonic string in non critical dimensions we argue that at level $N$
there are always states transforming as tensors with $s\ge \oh N$ indices.

Only in critical dimensions there are states with $s\le \oh N$.

Looking at the explicit coefficients of the combinations needed to
make the irreps from the \lc states we trace the origin of the chaotic
behavior of certain cubic amplitudes considered in literature to the extremely
precise and sensitive mixtures of states.
For example the vectors at level $N=19$ are a linear
combinations of states and when the coefficients are normalized to be
integer some of them have more than 1200 figures.

\end{abstract}

\section{Introduction}
String theory is probably the best candidate for quantum gravity and,
as such, it should be able to tell something about both spacelike and
timelike singularities in General Relativity.
Until recent years most of the research activity 
has been devoted to massless states.
A very likely reason is that massive states are unstable (see for
example \cite{Iengo:2006gm}).

More recently the attention has turned also to massive states.
There are many reasons for that, some of these are the followings.
\begin{itemize}
  \item
They are responsible for the amplitude divergences in
some temporal orbifolds \cite{Arduino:2020axy,Hikida:2005ec} and the
non existence of the effective theory \cite{Pesando:2022amk}.

\item
They may be identified with with some Black Holes microstates
when we take into account gravitational self-interaction
\cite{Horowitz:1997jc,Damour:1999aw}
in order to try to match the non-free massive string entropy
with that of a Black Holes.

\item
They are involved in the chaotic behavior of a class of amplitudes
both in the bosonic and NSR string computed using the DDF formalism
(see, for example,
\cite{Hindmarsh:2010if,Skliros:2011si,Skliros:2016fqs,Bianchi:2019ywd,Aldi:2019osr,Gross:2021gsj,Rosenhaus:2021xhm,Firrotta:2022cku,Bianchi:2022mhs,Firrotta:2023wem,Bianchi:2023uby,Firrotta:2024qel}).
See also the recent reformulation of DDF and Brower operators
\cite{Biswas:2024unn} which gives more compact expressions for
amplitudes \cite{BiswasMarottaPesando}.

\item
Finally they have been used to try to build theories with higher spin
massless particles in flat space (see \cite{Rahman:2015pzl} for a review).

\end{itemize}

Therefore, a better understanding of the massive string spectrum is
required.

Some work in that direction has already been done
\cite{DelGiudice:1970dr,Manes:1988gz,Hanany:2010da,Markou:2023ffh}
in covariant formalism
but it is mostly limited to the description of the spectrum and up to $N=6$.

In this paper we would like to go a step further and give a
description of the bosonic string spectrum up to level $N=10$ but,
most importantly, an explicit construction in the rest frame
of the states in \lc formalism,
at least for all the irreps with at most three indexes or
totally antisymmetric.
For all the other irreps we give the explicit states with at most one
index in direction $1$ (when the \lc is in directions $0$ and $1$) and
all the others transeverse.  
For the scalars and vectors we count them up to level $22$ and $19$
respectively.

Physical string states can be described either in the \lc formalism or
in the covariant formalism.  The \lc formalism yields the full
physical spectrum, but this comes at the expense of losing explicit
Lorentz covariance.  On the other hand, the covariant formalism
requires one to select the physical states using Virasoro conditions.
Therefore in \lc formalism we need tackling the issue of
reconstructing the covariant states and this is performed in this
paper with the help of a CAS, maxima.
From the results of this paper it turns out that \lc is more efficient
in building low spin states wherereas the covariant approach is more
efficient in building higher spin states.

The paper is organized as follows.
In section \ref{sec:main} we give the main result for the spectrum,
i.e for the bosonic string up to level $10$ we give the list of all
irreps and their multiplicity as long as their dimensions and the
dimensions of the vector space where the associated symmetric group
is represented.
We give also the scalar up to level $ss$ and vector spectrum up to level $19$.

In section \ref{sec:general_approach} we explain the general ideas on
how to tackle brute force the spectrum problem.
We have not tried any optimization, such as considering which states
may or may not contibute to a goven irrep but done all in almost the most
straighforward way.
We notice that with the help of Brower states the analysis could be
performed off shell.
Using a general approach we argue that in every dimensions at level
$N$ there are states with $s\ge \oh N$ indices.
The argument is very simple since we are dealing with a linear algebra
problem: simply the counting of linear constraints and independent variables.
The constraints are the equations needed to find a \lc $\GL(D-2)$
massive state\footnote{
\label{foot:GL_vs_O}
We write $\GL(*)$ and not $SO(*)$ because we are not imposing the
tracelessness condition or duality conditions on the free indexes.
We use this notation despite we have only $so(D-1)$ generators
and the states involve $O(D-2)$ scalar
products in order to stress that we do not impose any condition on the trace.
}with $s$ indices which is a \lc $\GL(D-1)$ state and not
image of a massive state with more indices under the boost $M^{- i}$.
The independent variables are all the possible \lc $\GL(D-2)$ states
with $s$ indices.
We then state the dimensions of the vector spaces of the true \lc
$\GL(D-1)$ states with $s$ indices in the critical dimension $D=26$
in eq. \ref{eq:D26_true_tensors_vector_spaces_dimensions}.
Looking at the numbers we notice some regularities.
Most of them fail when going to higher levels but one resists.
This relation is present for any dimensions
and if it true means that knowing the number of scalars at all
levels $N$ allows to compute the dimensions of the vector spaces where the
symmetric groups act for all $N$ and $s$.
This is not the same of knowing the $SO(D-1)$ irreps but puts strong
constraints.

It may be associated with the idea of raising trajectories stressed in \cite{Markou:2023ffh}.

The previous vector spaces of states with $s$ indices which are true
\lc $\GL(D-1)$ states with $s$ indices are acted upon the symmetric
group $S_s$ and split into \lc $\GL(D-1)$ irreps.
The algorithm used to perform this task is explained in section
\ref{sec:details_on_irreps}.
In the same section we note a recurring pattern of increasingly big
numbers (and prime numbers) in the linear combinations needed to
build \lc $\GL(D-1)$ states whose irrep has few indices $s\le \oh N$.
The intuitive idea is that to find these states
requires making a number of constraint combinations of the order of
independent variables which grow exponentially, thus transforming
small numbers of the order of the independent variables into numbers
with thousands of figures at level $N\sim 20$.

Still in section \ref{sec:details_on_irreps} we consider the problem
of computing the $SO(D-1)$ massive irreps from $\GL(D-1)$ irreps.
While the problem is well defined the general solution possible,
it depends heavily on the irrep and the explicit $\GL(D-1)$ states.
\ytableausetup{boxsize=0.5em}
We limit therefore the analysis to the lowest spin irreps
$\bullet$,
$\ydiagram{1}$, 
$\ydiagram{2}$, $\ydiagram{1,1}$,
$\ydiagram{3}$, $\ydiagram{2,1}$ and
$\ydiagram{1,1,1}$ (and all higher spin antisymmetric ones)
but for all possible levels $N$.
There is no unique way of choosing a basis for most of the previous
irreps and we discuss some of them.

Finally in section \ref{sec:chaos_from_big_numbers} we discuss how the
presence of enormous numbers (which seem to grow more than
exponentially with the level) in the $\GL(D-1)$ states with a small $s$
irrep is the cause of chaos in some three point massive string amplitudes.

In appendix \ref{app:Lorentz_algebra}
we discuss some constraints and relations amog the matrices describing
the Lorentz boost.

In appendix \ref{app:irrep_dims} we give the dimensions of the
$SO(25)$ and $S_s$ irreps
needed for checking the correctness of the table \ref{tab:1}

In appendixes \ref{app:level3} and \ref{app:level4} we give the full
results for the level $N=3$ and $N=4$.
All the other levels are in separated TeX files since they are very big.

In appendix \ref{app:scalars_up_to_10} we give however the explicit
form of the scalars up to level $10$ and how the coefficients
factorize over integers. Already for these low levels the prime
numbers involved are very big.

\section{Main result: the irreps up to level $N=10$}
\label{sec:main}

We can summarize the irreps up to $N=10$ in the table \ref{tab:1}
while in appendix \ref{app:level3} and \ref{app:level4}
we have given the explicit states for the different irreps for the
levels $N=3$ and $N=4$.
All the other levels have a separate TeX file since they are quite long.
The summary for the number of scalars and vectors up to $N=22$ is
\begin{equation}
  \begin{array}
    {c||
      c| c| c| c| c| c|
      c| c| c| c|
      c| c| c| c|
      c| c| c| c|
      c| c| c| c| c|} 
    s \backslash N
    & 0  & 1  & 2  & 3 & 4 & 5
    & 6  & 7  & 8  & 9
    & 10 & 11 & 12 & 13
    & 14 & 15 & 16 & 17
    & 18 & 19 & 20 & 21 & 22
    \\
    \hline 
    0
    & 1  & 0  & 0  & 0 & 1 & 0
    & 1  & 0  & 2  & 0 
    & 3  & 1  & 6  & 2
    & 9  & 6  & 16 & 11 
    & 27 & 22 & 46 & 42 & 76
    \\
    \hline 
    1
    & 0  & 1  & 0  & 0 & 1 & 0
    & 1  & 2  & 2  & 4 
    & 4  & 7  & 8  & 14
    & 16 & 25 & 31 & 47 
    & 58 & 85 &107?&153?&195?
  \end{array}
  ,
  \label{eq:scalars_and_vectors_up_to_19}
\end{equation}
where the last three vectors have been guessed from the rule that
the number of vectors at level $N+1$ is equal to the sum of scalars
and vectors at level $N$.
The explicit states may be easily extracted from the associated lisp
data file but are probably useless at the moment since their
amplitudes cannot be computed in reality.

\newgeometry{a4paper,landscape,left=0.1in,right=0.5in,top=1in,bottom=0.5in,nohead}
%
\ytableausetup{boxsize=0.3em}
\scalebox{0.5}{
%
\parbox{0.9\textheight}{ 
%
\begin{equation*}
%
\begin{array}{ 
c || 
c | 
c | 
c | 
c | 
c | 
c | 
c | 
c | 
c | 
c | 
c | 
c | 
} 
\\ 
s  &  
N= 0  &  
N= 1  &  
N= 2  &  
N= 3  &  
N= 4  &  
N= 5  &  
N= 6  &  
N= 7  &  
N= 8  &  
N= 9  &  
N= 10 
\\ \hline
0  & 
\bullet &
&
&
&
\bullet &
&
\bullet &
&
2\,\bullet&
&
3\,\bullet
\\ 
SO(25)  & 
1 &
&
&
&
1 &
&
1 &
&
2*1&
&
3*1
\\ 
S_0  & 
1 &
&
&
&
1 &
&
1 &
&
2*1 &
&
3*1
\\ \hline 
%
1  & 
&
\ydiagram{1}&
&
&
&
\ydiagram{1}&
\ydiagram{1}&
2\,\ydiagram{1}&
2\,\ydiagram{1}&
4\,\ydiagram{1}&
4\,\ydiagram{1}
\\ 
%
SO(25)  & 
&
25&
&
&
&
25&
25&
2* 25&
2* 25&
4* 25&
4* 25
\\ 
%
S_1  & 
&
1&
&
&
&
1&
1&
2* 1&
2* 1&
4* 1&
4* 1
\\ \hline 
2  & 
&
&
\ydiagram{2}&
&
\ydiagram{2}
&
&
2\,\ydiagram{2}
&
\ydiagram{2}
&
4\,\ydiagram{2}
&
3\,\ydiagram{2}
&
8\,\ydiagram{2}
%
\\ 
 & 
&
&
&
\ydiagram{1,1}&
&
\ydiagram{1,1}&
&
\oplus 2\,\ydiagram{1,1}&
\oplus\ydiagram{1,1}&
\oplus4\,\ydiagram{1,1}&
\oplus3\,\ydiagram{1,1}
\\ 
SO(25) & 
&
&
324
&
300&
324 
&
300 
&
2* 324 
&
324 
+2* 300 
&
4* 324 
+300 
&
3* 324 
+ 4* 300 
&
8*324 
+ 3*300 
\\ 
S_2 & 
&
&
1 
&
1 
&
1 
&
1 
&
2* 1 
&
1 
+ 2* 1 
&
4*1 
+ 1 
&
3*1 
+4*1 
&
8* 1 
+3* 1 
\\ \hline 
%
3  & 
&
&
&
\ydiagram{3}
&
&
\ydiagram{3}
&
\ydiagram{3}
&
2\,\ydiagram{3}
&
2\,\ydiagram{3}
&
5\,\ydiagram{3}
&
5\,\ydiagram{3}
%
\\ 
%
  & 
&
&
&
&
\ydiagram{2,1}&
\oplus\ydiagram{2,1}&
\oplus\ydiagram{2,1}
&
\oplus2\,\ydiagram{2,1}&
\oplus3\,\ydiagram{2,1}
&
\oplus4\,\ydiagram{2,1}
&
\oplus7\,\ydiagram{2,1}
%
\\ 
%
  & 
&
&
&
&
&
&
\oplus\ydiagram{1,1,1}&
&
\oplus\ydiagram{1,1,1}&
\oplus\ydiagram{1,1,1}
&
\oplus2\, \ydiagram{1,1,1}
\\ 
%
SO(25)  & 
&
&
&
2900 
&
5175 
&
2900 + 5175 
&
2900 + 5175 
+ 2300 
&
2* 2900 + 2* 5175 
&
2* 2900 + 3* 5175 
+ 2300 
&
5* 2900 + 4* 5175 
+ 2300 
&
5* 2900 + 7* 5175 
+ 2* 2300 
\\ 
%
S_3  & 
&
&
&
1
&
2
&
1 +  2 
&
1 + 2 
+1 
&
2* 1 + 2* 2 
&
2* 1 + 3* 2 
+1 
&
5* 1 + 4* 2 
+1 
&
5* 1 
+ 7* 2 
+2* 1 
\\ \hline 
4  & 
&
&
&
&
\ydiagram{4}
&
&
\ydiagram{4}
&
\ydiagram{4}
  &
3\,\ydiagram{4}
  &
2\,\ydiagram{4}
&
6\,\ydiagram{4}
%
\\ 
  & 
&
&
&
&
&
\ydiagram{4,1}
&
\oplus\ydiagram{3,1}\oplus\ydiagram{2,2}
&
\oplus2\,\ydiagram{3,1}
  &
\oplus2\,\ydiagram{3,1}\oplus2\,\ydiagram{2,2}
  &
\oplus5\,\ydiagram{3,1}\oplus\ydiagram{2,2}
&
\oplus6\,\ydiagram{3,1}\oplus4\,\ydiagram{2,2}
%
\\ 
  & 
&
&
&
&
&
&
&
  \oplus\ydiagram{2,1,1}
  &
  \oplus\ydiagram{2,1,1}
  &
  \oplus2\,\ydiagram{2,1,1}
  &
\oplus2\,\ydiagram{2,1,1}
%
\\ 
  & 
&
&
&
&
&
&
&
  &
  &
  &
\oplus\ydiagram{1,1,1,1}
\\ 
SO(25)  & 
&
&
&
&
20150 
&
52026 
&
\begin{array}{l} 20150 + 52026 \\ + 32175 \end{array}
&
\begin{array}{l} 20150 + 2* 52026 \\ + 44550 \end{array}
  &
\begin{array} {l } 3* 20150 + 2* 52026 \\ + 2* 32175 + 44550 \end{array}
  &
\begin{array}{l}2* 20150 + 5* 52026 \\ + 32175 + 2* 44550 \end{array}
  &
\begin{array} {l} 6* 20150 + 6* 52026 \\ + 4* 32175 + 2* 44550 + 12650 \end{array}
%
\\ 
S_4  & 
&
&
&
&
1 
&
3 
&
1 + 3 + 2 
&
1 + 2* 3 
+ 3 
  &
3* 1 + 2* 3 + 2* 2 
+ 3
  &
2* 1 + 5* 3 +2 
+ 2* 3 
  &
6* 1
+ 6* 3 + 4* 2
+2* 3 
+ 1
\\ \hline 
5  & 
&
&
&
&
&
\ydiagram{5}
&
&
\ydiagram{5}
&
\ydiagram{5}
&
3\,\ydiagram{5}
&
3\,\ydiagram{5}
%
\\ 
  & 
&
&
&
&
&
&
\ydiagram{4,1}
&
\oplus\ydiagram{4,1}\oplus\ydiagram{3,2}
&
\oplus2\,\ydiagram{4,1}\oplus\ydiagram{3,2}
&
\oplus3\,\ydiagram{4,1}\oplus2\,\ydiagram{3,2}
&
\oplus5\,\ydiagram{4,1}\oplus3\,\ydiagram{3,2}
%
\\ 
  & 
&
&
&
&
&
&
&
&
\oplus\ydiagram{3,1,1}
&
\oplus\ydiagram{3,1,1}\oplus\ydiagram{2,2,1}
&
\oplus3\,\ydiagram{3,1,1}\oplus\ydiagram{2,2,1}
\\ 
SO(25)  & 
&
&
&
&
&
115830 
&
385020 
&
\begin{array}{l} 115830 + 385020 \\ + 430650  \end{array}
&
\begin{array}{l} 115830 + 2* 385020 \\ + 430650 + 476905 \end{array}
&
\begin{array}{l} 3* 115830 + 3* 385020 \\ + 2* 430650 + 476905
\\ + 368550 \end{array}
&
\begin{array}{l} 3* 115830 + 5* 385020 \\ + 3* 430650 + 3* 476905
\\ + 368550 \end{array}
%
\\ 
S_5  & 
&
&
&
&
&
1 
&
4 
&
1 + 4 + 5
&
1 + 2* 4 + 5 + 6
&
3* 1 + 3* 4 + 2* 5 + 6 + 5
&
3* 1 + 5* 4 + 3* 5 + 3* 6 + 5
%
\\ \hline 
%
6  & 
&
&
&
&
&
&
\ydiagram{6}
&
&
\ydiagram{6}
&
\ydiagram{6}
&
3\,\ydiagram{6}
%
\\ 
%
  & 
&
&
&
&
&
&
&
\ydiagram{5,1}
&
\oplus\ydiagram{5,1}\oplus\ydiagram{4,2}
&
\oplus2\ydiagram{5,1}\oplus\ydiagram{4,2}
\oplus\ydiagram{3,3}
&
\oplus3\,\ydiagram{5,1}\oplus3\,\ydiagram{4,2}
%
\\ 
%
  & 
&
&
&
&
&
&
&
&
&
\oplus\ydiagram{4,1,1}
&
\oplus\ydiagram{4,1,1}\oplus\ydiagram{3,2,1}
\\ 
%
SO(25)  & 
&
&
&
&
&
&
573300 
&
2302300 
&
573300 + 2302300 + 3580500 
&
\begin{array}{c} 573300 + 2* 2302300 + 3580500
  \\ 1848924 + 3670524 \end{array}
&
\begin{array}{c} 3* 573300 + 3* 2302300 + 3* 3580500
  \\ 3670524 + 5252625 \end{array}
%
\\ 
%
S_6  & 
&
&
&
&
&
&
1 
&
5 
&
1 + 5 + 9
&
1 + 2* 5 + 9 + 5 + 10
&
3* 1 + 3* 5 + 3* 9 + 10 + 16
%
\\ \hline 
%
7  & 
&
&
&
&
&
&
&
\ydiagram{7}
&
&
\ydiagram{7}
&
\ydiagram{7}
%
\\ 
%
  & 
&
&
&
&
&
&
&
&
\ydiagram{7,1}
&
\oplus\ydiagram{6,1}\oplus \ydiagram{5,2}
&
\oplus2\,\ydiagram{6,1}\oplus\ydiagram{5,2}\oplus\ydiagram{4,3}
%
\\ 
%
  & 
&
&
&
&
&
&
&
&
&
&
\oplus\ydiagram{5,1,1}
\\ 
%
SO(25)  & 
&
&
&
&
&
&
&
2510820 
&
11705850 
&
\begin{array}{c} 2510820 + 11705850 \\+  22808500 \end{array}
&
\begin{array}{c} 2510820 + 2* 11705850
  \\+  22808500 + 20470230
\\ + 22542300 \end{array}
%
\\ 
%
S_7  & 
&
&
&
&
&
&
&
1 
&
6 
&
1 + 6 + 14
&
1 + 2* 6 + 14 + 14 + 15
%
\\ \hline
%
8  & 
&
&
&
&
&
&
&
&
\ydiagram{8}
&
&
\ydiagram{8}
%
\\ 
%
  & 
&
&
&
&
&
&
&
&
&
\ydiagram{7,1}
&
\oplus\ydiagram{7,1}\oplus\ydiagram{6,2}
\\ 
%
SO(25)  & 
&
&
&
&
&
&
&
&
9924525 
&
52272675 
&
\begin{array}{c} 9924525 + 52272675 \\ + 120656250 \end{array}
%
\\ 
%
S_8  & 
&
&
&
&
&
&
&
&
1 
&
7 
&
1 + 7 + 20
%
\\ \hline
%
9  & 
&
&
&
&
&
&
&
&
&
\ydiagram{9}
&
%
\\ 
%
  & 
&
&
&
&
&
&
&
&
&
&
\ydiagram{9,1}
\\ 
%
  & 
&
&
&
&
&
&
&
&
&
35937525 
&
209664780 
\\ 
%
S_9  & 
&
&
&
&
&
&
&
&
&
1 
&
8
%
\\ \hline
%
10  & 
&
&
&
&
&
&
&
&
&
&
\ydiagram{10}
\\ 
%
SO(25)  & 
&
&
&
&
&
&
&
&
&
&
120609840 
\\ 
%
S_{10}  & 
&
&
&
&
&
&
&
&
&
&
1 
\\ \hline
\end{array}
\end{equation*}
} 
} 
%
\begin{table}
  \caption{In the following table $s$ is the number of indices,
    $SO(25)$ refers to the irreps dimensions and similarly for the
    symmetric group $S_s$ irreps.}
\label{tab:1}
\end{table}
\restoregeometry

\section{On the massive spectrum: constraints from the \lc}
\label{sec:general_approach}

We would like to show that there are no massive scalars and
vectors, actually tensors with  roughly $s\le\oh N$ indices
at level $N$ in the spectrum in {\sl non critical } dimension.
On the contrary they are present in critical dimension as the previous
table \ref{tab:1} shows.
In particular  scalars are present for all even levels for $s \ge 4$
and odd levels for $s\ge 11$ in critical dimensions
and vectors for level $s\ge 5$.

We give a simple counting argument for presence  of massive tensors with $s$
indices at level $N$ in {\sl non critical } dimension.
The upshot is that at level $N$ there are always tensors with
$s\sim \oh N, \dots N$ indices.

For critical dimension we must rely on the explicit computation since
the states which transform as tensors with $s<\oh N$ indices are very
precise and sensitive mixtures.
One example for all: there is a scalar at level $N=22$ where some
coefficients have more than $2000$ digits when we normalize it to have
integer coefficients.

\subsection{Constraints from \lc on the spectrum: an overview}

We now describe the approach to get constraints from \lc in the on shell case.
We start giving an overview for massive scalars and then we proceed to
massive tensors.

The idea is very simple.

A Poincar\'e  group massive scalar in the rest frame is a $SO(D-1)$
massive scalar.
Then  a $SO(D-1)$ massive scalar is obviously a $SO(D-2)$
massive scalar, i.e. a scalar \wrt the transverse coordinates.

If we consider a $SO(D-2)$ massive scalar and infinitesimally boost it
using $M^{- i}$ we can get a state which is a vector.
This happens for example because the component $v^1$ is a scalar \wrt
$SO(D-2)$.
Similarly the tensor component $t^{1 1\dots 1}$ appears to be a
$SO(D-2)$ scalar.

Technically this happens because the \lc expression for $M^{- i}$
contains cubic terms, in particular there are terms with two creators
and one annihilator.

If the original would be scalar acquires an index under a boost then
this $SO(D-2)$ massive scalar is not a $SO(D-1)$ massive scalar,
i.e. a Poincar\'e group massive scalar in the rest frame rather it is
a piece of $SO(D-1)$ massive vector or tensor.

We can therefore find the true $SO(D-1)$ massive scalars by
considering the most general linear combination of $SO(D-2)$ massive
scalars and requiring that the boost by $M^{- i}$ does not yield a
vector, i.e. that $M^{- i}$ annihilates the state.

Explicitly at level $N$ we can start from a basis of $SO(D-2)$ scalar
states at level $N$ 
\begin{align}
  T_{N,  0} =
  \left\{
    \prod_{a=1}^k\,
    ( \vec \ualpha_{-n_{2 a} } \cdot \vec \ualpha_{-n_{2 a-1} } )
    |~~
    n_A\ge n_{A-1}
    ,~~~~
    \sum_{A=1}^{2 k} n_A = N
    \right\}
    ,
\end{align}
where $\ualpha^i_n$ ($i=2, \dots D-1 = 25$) are the \lc bosonic oscillators.
We then consider the generic linear combination, i.e
the generic $SO(D-2)$ massive scalar
\begin{equation}
  \sum_{ \{ n_A\}\, }
    c_{ \{ n_A \} }\,
    \delta_{\sum_A n_A, N}\,
    \prod_{a=1}^k\,
    ( \vec \ualpha_{-n_{2 a} } \cdot \vec \ualpha_{-n_{2 a-1} } )
    |\uk\rangle
,
\end{equation}
in rest frame i.e. with $\uk^i=0$ and infinitesimally boost it.

We act with $M^{- i}$ and require that the image vector is zero.
This gives a generically overdetermined set of homogeneous equations
which only in some special dimensions has solutions.
This happens in the critical dimension $D=26$ for $N\ge 4$.

In a similar way a massive vector for the Poincar\'e group is
a massive vector for $SO(D-1)$ in rest frame
because of the transversality condition.
As done for the massive scalar 
we can write a basis of $SO(D-2)$ vectors at level $N$ as
\begin{equation}
  T_{N, 1}
  =
  \left\{
  \prod_{a=1}^k
( \vec \ualpha_{-n_{2 a} } \cdot \vec \ualpha_{-n_{2 a-1} } )
\ualpha_{-n}^i
|\uk\rangle
|~~
n_A \ge n_{A-1}
,~~~~
\sum_{A=1}^{2 k} n_A +n = N
\right\}
,
\end{equation}
consider the generic linear combination
and determine the possible massive vectors
by the requirement that the infinitesimal boost of this linear combination
does not contain a two index tensor.

The same approach can be pursued with all tensors, however there are
cases which may, and not must,
be treated differently since they do not involve any
product $( \vec \ualpha_{-n } \cdot \vec \ualpha_{-m } )$.
They are associated with Young tableaux which first appear at a
certain level \cite{Markou:2023ffh}.

For example the $\ydiagram{2,1}$ appears first at level $N=4$ as the
projection of the state
\begin{equation}
  \ualpha^i_{-1}\,\ualpha^j_{-1}\,\ualpha^k_{-2}\,|\uk\rangle
  ~~\rightarrow~~
    \left(
  \ualpha^i_{-1}\,\ualpha^j_{-1}\,\ualpha^k_{-2}
  \right)_{\ydiagram{2,1}}\,
  |\uk\rangle
  ,
\end{equation}
\ytableausetup{boxsize=1em}
to the Young tableau
$\begin{ytableau} i & j \\ k \end{ytableau}$.

The minimal level of a Young diagram is easily determined
because we want the lowest $N$
and this implies that we fill the first and longest line with
$\ualpha_{-1}$,
the second line with $\ualpha_{-2}$,
the third line with $\ualpha_{-3}$ and so on.

It should then be easy to see that its infinitesimal boost
does not involve any tensor with an index more.
Then from this state we should easily track a Regge trajectory of the
form \cite{Markou:2023ffh}
\begin{equation}
  \left(
  \ualpha^i_{-1}\,\ualpha^j_{-1}\,\ualpha^k_{-2}
  \right)_{\ydiagram{2,1}}\,
  \prod_{a=1}^{N-4}\,
  \ualpha^{l_a}_{-1}
  \,|\uk\rangle
  .
\end{equation}

\subsection{Constraints from \lc on the spectrum: details}

We now describe in more details the approach in the on shell case.
\begin{enumerate}
\item
  We choose the simplest frame allowed by DDF construction, i.e.
  \begin{equation}
    k^i= k^1= 0,
    ~~~
    k^+= k^- \ne 0
  ,
\end{equation}
i.e. the rest frame.  
\item
  In the rest frame
  a  massive scalar for the Poincar\'e group is a $SO(D-1)$ scalar.
  
  Similarly in the rest frame because of the transversality condition
  a massive vector for the Poincar\'e group is a $SO(D-1)$ vector
  and so on for all the other tensors.

\item
  A $SO(D-1)$ scalar is also a scalar \wrt the transverse $SO(D-2)$.
  
  Generically a $SO(D-1)$ tensor with $s$ indices decomposes
  as  $SO(D-1)$ tensors with $s_1$ indices with $s\ge s_1 \ge 0$,
  i.e. $T^{I_1 \dots I_s} \rightarrow 
  T^{i_1 \dots i_s}
  \oplus T^{1 i_2 \dots i_s}
  \oplus \dots
  \oplus T^{i_1 \dots i_{s-1} 1}
  \dots
  \oplus
  T^{1 1 \dots 1}$
  with $1\le I \le D-1$ and $2 \le i \le D-1$
  (where some components of the $SO(D-2)$ tensors
  may be zero because of some symmetry in the
  original tensor).

Notice however that we are not taking about $SO(D-2)$ irreps since we
are not considering the trace.
We are actually considering $\GL(D-2)$ irreps.
In the following we will write $\GL(D-2)$ in order to stress this
point despite the states involve $O(D-2)$ scalar
products and we act with $M^{i -}$ only.

\item
  We consider and count the basis elements for
  transverse $SO(D-2)$ scalars at level $N$.
  Then we build the most general linear combination.

  Similarly for a (reducible) $SO(D-2)$ tensor with $s$ indices. 

\item
  Now the key tool is to consider the action of $M^{1 i}$.
  This is not $M^{- i}$ but the action of $M^{+ i}$ is only non zero on the
  zero modes
  therefore we can use $M^{- i}|_{n.z.m.}$ restricted to non zero modes.
  In the following we omit the specification $|_{n.z.m.}$.

  Notice that the generators $M^{- i}$ commute with each other so the
  conditions we get are from them are formally different
  but give exactly the same
  constraints since the conditions for different $i$ are in one to one
  correspondence.
  
  The key observation is that the action of $M^{- i}$ on a state with
  $s$ $SO(D-2)$ indices
  yields generically a state which is the sum of
  a state with $s+1$ $SO(D-2)$ indices
  and
  a state with $s-1$ $SO(D-2)$ indices.
  In the following we call the states with $s-1$ indices descendants.

  %
  In particular
  if the original state was a $SO(D-1)$ massive scalar
  then all vectors which are
  created by a boost $M^{- i}$, i.e. the part of the variation
  with an extra index must be zero.
  A generic generic $SO(D-2)$ scalar at level $N$
  is transformed into a $SO(D-2)$ vector at level $N$ by a \lc boost.

  We require that the boost at most decreases the indices by one,
  i.e we require that the $SO(D-2)$ tensor with $s+1$ indices is zero.

\item
  Consider the conditions for the $SO(D-1)$ scalars.
  Under the previous hypothesis,
  the number of basis elements for $SO(D-2)$ vectors at level $N$
  is the number of
  homogeneous linear equations in the coefficients of the generic
  $SO(D-2)$ scalar at level $N$.
  In order to have a solution in all dimensions
  we must require that the dimension of
  the vector space of the $SO(D-2)$ scalars at level $N$ is strictly
  greater   the dimension of
  the vector space of the $SO(D-2)$ vectors at level $N$.

  Similarly for tensors with $s$ indices.

  In critical dimensions solutions appear even when there are none in
  generic dimensions.

\item
  Finally when we find a combination of $SO(D-2)$ basis tensors with
  $s$ indices at level $N$ which transforms as a $SO(D-1)$ tensor
  we can compute its descendants, i.e. the images under the {\sl part}
  of the boost with less $SO(D-2)$ indices.
  Said differently we can start from
  $T^{i_1 \dots i_s}$ and we use a sequence of boosts to compute
  $T^{1 i_2 \dots i_s}, \dots T^{i_1 \dots i_{s-1} 1}$
  down to $  T^{1 1 \dots 1}$.
  This prrocedure is not so immediate to implement when two or more
  $i_k$s are equal because of traceless condition.
  We have implemented it in details for some simple irreps in section \ref{sec:details_on_irreps}.
  For the other irreps we have done it for the first boost only.

  While intuitively obvious and expected for the consistency
  it is not immediate to show that the
  descendants of a tensor with $s$ indices cannot be raised above
  a tensor with $s$ indices.
  It can however be verified explicitly
   
\end{enumerate}

The previous algorithm can be extended off shell using DDF and Brower
operators, i.e with the inclusion of
the contributions from $\tilde \uA^-(E)$s which give raise to null states
on shell.
For example the on shell scalar basis is
$T_{N=3,s=0}=\{ \uA^k_{-2}\, \uA^k_{-1}\} $
while the off shell is
$T_{N=3,s=0}=\{
\uA^k_{-2}\, \uA^k_{-1},\,
\uA^k_{-1}\, \uA^k_{-1}\, \tilde \uA^{-}_{-1},\,
\tilde \uA^{-}_{-3}
\} $.

\subsection{Details on the $M^{- i}$ action}

The most important step involves the action of $i \ualpha^+_0\, M^{-i}$
on \nzm and
in the rest frame on states with mass $M$.
In this case we can use\footnote{
See appendix \ref{app:Lorentz_algebra} for more details, but in order
to get the proper normalization the main point is that
$M^{+ i}|_{\nzm}=0$ on \lc.
Moreover the signs of $\delta^{i }$ action are
$  \delta^{i } |j 1 \rangle =
(-\sdap  M) \left( \delta_{i j}\, |1 1 \rangle - | j i \rangle \right)$
.
}
\begin{equation}
  \delta^{i }(*)
  =
  [ i \ualpha^+_0\, M^{- i}|_{\nzm},\, *]
  =
  [ i \sdap M\, M^{i 1}|_{\nzm},\, *]
  =
  [ : \sum_{n\ne 0} \sum_{m\ne 0}
    \frac{1}{ 2 n}\ualpha^k_{-n-m} \ualpha^k_{m} \ualpha^i_{n}\, : ,\,
    *]
 . 
\end{equation}

At the same time given a level $N$ and $s$ indexes we have basis elements
$e^{\sN {N, s, a}}_{i_1\dots i_s} \in T_{N, s}$
where $a=1,\dots dim\, T_{N, s}$ labels the element in $T_{N, s}$
which is the set of basis elements.
These basis elements span the vector space $V_{N, s}=span\, T_{N, s}$.
See eq.s \eqref{eq:scalar_basis}, \eqref{eq:vector_basis}
 and  \eqref{eq:spin2_basis} for explicit examples.

It is important to stress that these elements are $GL(D-2)$ tensors
with $s$ indexes
but may be linear superposition of $GL(D-1)$ tensors with a number of
indexes bigger or equal to $s$.
The reason is simple: there may be some index $1$ which is hidden.

For this reason we denote
$| i_1\dots i_s \gg$ the \lc states which have a proper transformation
under $\GL(D-2)$
and
$| I_1\dots I_s \gg$ the \lc states which have a proper transformation
under $\GL(D-1)$.

We are interested in the action of $\delta^i$ on these basis elements
and to compare with the known action of $SO(D-1)$ generators.

The action of $SO(D-1)$ generators on a true $GL(D-1)$ tensor $T_{I_1 \dots I_s}$
with $s$ indexes ($I,J,\dots=1,\dots D-1$, $i,j,\dots=2,\dots D-1$) is
\begin{align}
i M^{L M} T_{I_1\dots I_s}
=&
\sum_{p=1}^s
\delta_{M, I_p}\,
T_{I_1\dots I_{p-1} L I_{p+1} \dots I_s}
-
\sum_{p=1}^s
\delta_{L, I_p}\,
T_{I_1\dots I_{p-1} M I_{p+1} \dots I_s}
,
\end{align}
so in particular
\begin{align}
i M^{m 1} T_{I_1\dots I_s}
=&
\sum_{p=1}^s
\delta_{m, I_p}\,
T_{I_1\dots I_{p-1} 1 I_{p+1} \dots I_s}
 -
 \sum_{p=1}^s
\delta_{1, I_p}\,
 T_{I_1\dots I_{p-1} m I_{p+1} \dots I_s}
\label{eq:Mml_action_on_true_element}
\end{align}

To proceed in the analysis of the action of $\delta^{i }$
we split $\delta^{i }$ according to the number of creators and
annihilators as
\begin{equation}
  \delta^{i }
  =
  \delta^{(--)(+) i }
  +
  \delta^{(-+)(-)i }
  +
  \delta^{(++)(-)i }
  +
  \delta^{(-+)(+)i }
  ,
\end{equation}
where f.x. $  \delta^{(-+)(+)i }$ means that there is one creator and
one annihilator in $\tilde \ualpha^-$ and that $\ualpha^i$ is an annihilator.

To compute the action on a state we use Wick theorem and we compute
the contraction of  all possible couples of annihilators and creators.

We notice that when there are two creators
the action on a state could be  computed considering
one $\ualpha^j_{-n}$ at a time.
This is not possible when there are two annihilators.
Moreover the states we are going to consider have two different
building blocks for which we use the short hand notations
\begin{equation}
  n^j \leftrightarrow \ualpha_{-n}^j,
  ~~~~
  (m, n) \leftrightarrow
  \vec\ualpha_{-m} \cdot \vec\ualpha_{-n} =\ualpha_{-m}^j\ualpha_{-n}^j
  ,
\end{equation}
so we can write
\begin{equation}
\prod_{a=1}^s \alpha^{j_a}_{-n_a}
\dots
\prod_{c} \vec\ualpha_{-m_{2 c-1}} \cdot \vec\ualpha_{-m_{2 c}}
| \uk^0= M, \vec \uk=0 \rangle
=
\prod_{a=1}^s n_a^{j_a}\, \prod_{c} (m_{2c-1}, m_{2c} )
,
\end{equation}
with $\ap M^2 = \sum_a n_a + \sum_c (m_{2c-1}+ m_{2c}) -1 $. 

Therefore the action of $M^{-i}$ is better discussed using these
building blocks.
In view of Wick's theorem we have the following actions
when only one annihilator is present in $\delta^{ i }$
\begin{align}
  \delta^{(--)(+) i }& n_1^{j_1}\dots \underset{\uparrow}{n_k^{j_k}} \dots n_s^{j_s}
  \, (m_1,\,m_2)\, \dots (m_{2l-1},\,m_{2l}) \dots (m_{2c-1},\,m_{2c})
  ,
  \nonumber\\
  \delta^{(--)(+) i }&
   n_1^{j_1} 
   \dots               
  \, (m_1,\,m_2)\, \dots
  (\underset{\uparrow}{m_{2l-1}},\,m_{2l}) \dots (m_{2c-1},\,m_{2c})
\nonumber\\
        &+
  \delta^{(--)(+) i }
   n_1^{j_1} 
   \dots               
  \, (m_1,\,m_2)\, \dots
  (m_{2l-1},\, \underset{\uparrow}{m_{2l}}) \dots (m_{2c-1},\,m_{2c})
  ,
  \nonumber\\
\end{align}
and similarly for $  \delta^{(-+)(-) i }$.
The $\uparrow$ means that the pointed creator is annihilated by
the annihilator in $\delta^{ i }$.

In the case of two annihilators there are more cases.
For example for $\delta^{(-+)(+) i }$ we have
\begin{align}
  \delta^{(-+)(+) i }&
  n_1^{j_1}\dots \underset{\uparrow}{n_k^{j_k}}
  \dots \underset{\uparrow}{n_n^{j_n}}\dots
  \, (m_1,\,m_2)\, \dots (m_{2c-1},\,m_{2c})
  ,
  \nonumber\\
  \delta^{(-+)(+) i }& n_1^{j_1}\dots \underset{\uparrow}{n_k^{j_k}}
  \dots
   (\underset{\uparrow}{m_{2l-1}},\,m_{2l}) \dots (m_{2c-1},\,m_{2c})
 \nonumber\\
&+      
  \delta^{(-+)(+) i }
  n_1^{j_1}\dots \underset{\uparrow}{n_k^{j_k}}\dots
  (m_{2l-1},\, \underset{\uparrow}{m_{2l}}) \dots (m_{2c-1},\,m_{2c})
  ,
  \nonumber\\
  \delta^{(--)(+) i } &
   n_1^{j_1} 
   \dots               
  \, (m_1,\,m_2)\, \dots
  (\underset{\uparrow}{m_{2l-1}},\, \underset{\uparrow}{m_{2l}}) \dots (m_{2c-1},\,m_{2c})
  ,
\end{align}     
and similarly for $\delta^{(++)(-) i }$.

The possible actions on the building blocks which increase the number
of indices are 
\begin{align}
\delta^{i \uparrow}\,  n_1^j
=&
-n_1\, \sum_{l=1}^{n_1-1} \frac{1}{l}\, l^i\, (n_1-l)^j
,
\end{align}
and
\begin{align}
\delta^{i \uparrow}\, (n_1,\, n_2)
=&
-n_1 \sum_{l=1}^{n_1 -1} \frac{1}{l} (n_1-l,\, n_2)\, l^i
-n_2 \sum_{l=1}^{n_2 -1} \frac{1}{l} (n_1,\, n_2-l)\, l^i
\nonumber\\
%
&
-\oh \sum_{l=1}^{n_1 -1} (n_1-l,\, l)\, n_2^i
-\oh \sum_{l=1}^{n_2 -1} (l,\, n_2-l)\, n_1^i
\nonumber\\
%
&
+\left( - \frac{n_1 n_2}{n_1+n_2} d + n_1 +n_2\right)
\, (n_1+n_2)^i
%
%
,
\end{align}
and
\begin{align}
\delta^{i \uparrow}\, (n_1,\, n_2)\, (m_1,\, m_2)
=&
- n_1
\left(
\frac{m_1}{n_1+m_1} (m_2,\, n_2)\, (n_1+m_1)^i
-(n_1+m_1,\, n_2)\, m_2^i
\right)
\nonumber\\
&
- n_1
 \left(
+\frac{m_2}{n_1+m_2} (m_1,\, n_2)\, (n_1+m_2)^i
-(n_1+m_2,\, n_2)\, m_1^i
\right)
\nonumber\\
&
- n_2
\left(
\frac{m_1}{n_2+m_1} (m_2,\, n_1)\, (n_2+m_1)^i
-(n_2+m_1,\, n_1)\, m_2^i
\right)
\nonumber\\
&
-n_2
\left(
+\frac{m_2}{n_2+m_2} (m_1,\, n_1)\, (n_2+m_2)^i
-(n_2+m_2,\, n_1)\, m_1^i
\right)
\nonumber\\
%
&
+
m_1\, (n_1+m_1,\, m_2)\, n_2^i
+
m_2\, (m_1,\, n_1+m_2)\, n_2^i
\nonumber\\
&
+m_1\, (n_2+m_1,\, m_2)\, n_1^i
+m_2\, (m_1,\, n_2+m_2)\, n_1^i
\end{align}

\begin{align}
\delta^{i \uparrow}\,  n_1^j\, (m_1,\, m_2)
=&
- n_1
\left(
\frac{m_1}{n_1+m_1}\, (n_1+m_1)^i\, m_2^j
-m_2^i\, (n_1+m_1)^j
\right)
\nonumber\\
&\phantom{-n_1}
-n_1
\left(
+\frac{m_2}{n_1+m_2}\, (n_1+m_2)^i\, m_1^j
-m_1^i\, (n_1+m_2)^j
\right)
.
\end{align}
Notice that these actions are ``anomalous'' from the $\GL(D-2)$
perspective since there is an increase of number of indexes.
These actions can on the contrary be explained from the $\GL(D-1)$
point of view as the presence of an hidden ``1'' index.
For example we have the variation
$\delta^{i} |{j 1}\rangle
= (- \sdap M)
( \delta_{i j} |{1 1}\rangle - |{j i }\rangle)$
of a 2 index $\GL(D-1)$ state which appears as
$\delta^{i } |{j }\gg =
(- \sdap M)( \delta_{i j} | \emptyset \gg - |{j i}\gg )$
from the $\GL(D-2)$ point of view.

The possible actions on the building blocks which decrease the number
of indices are

\begin{align}
\delta^{i \downarrow}\,  n_1^j\, 
=&
 \delta^{i j} \sum_{l=1,n_1} \oh (n_1-l,\, l)
,
\nonumber\\
\delta_R^{i \downarrow}\,  n_1^{j_1}\, n_2^{j_2}\, 
=&
+ n_2 (n_1+n_2)^{j_2}\, \delta^{j_1 i}
+ n_1 (n_1+n_2)^{j_1}\, \delta^{j_2 i}
,
\nonumber\\
\delta_A^{i \downarrow}\,  n_1^{j_1}\, n_2^{j_2}\, 
=&
- \frac{n_1 n_2}{n_1+n_2} (n_1+n_2)^i\, \delta^{j_1 j_2}
,
\nonumber\\
\delta^{i \downarrow}\,  n_1^{j_1}\, (m_1,\, m_2) 
=&
  + m_1 (n_1+m_1,\, m_2)\, \delta^{i j_1}
  + m_2 (n_1+m_2,\, m_1)\, \delta^{i j_1} 
,
\end{align}
where the action $\delta_A^{i \downarrow}\,  n_1^{j_1}\, n_2^{j_2}$ is
again ``anomalous'' since it is associated with a rotation
in the $1 i$ plane which acts on a
``hidden" $I=1$ indexes.

Before discussing the meaning of the previous statement we define
symbolically the almost true $\GL(D-1)$ tensor states  as
\begin{equation}
\delta^{i \uparrow} |i_1\dots i_s\gg = 0
,
\label{eq:true_spin_s_states}
\end{equation}
or more precisely for a state at level $N$ with $s$ indexes
as the linear combination of the basis elements
for which
\begin{align}
\sum_a b_{\sN{N\,s\,a}}\,
\delta^{i \uparrow} e^{\sN {N, s, a}}_{i_1 \dots i_s}
&=0
.
\end{align}
The reason of the almost true will become clear shortly.

Now  we compare the variation of 
a state with two equal $\GL(D-2)$ indexes $| i i \gg$
with that of
a state with two different $\GL(D-2)$ indexes
$| i j \gg$  ($i\ne j$)
and we suppose that both are almost true 2 index states as
defined in \eqref{eq:true_spin_s_states}.
Then because of $\delta_A^{i \downarrow}$ we realize that
$| i j \gg$  transforms under $SO(D-1)$ rotations
as a $\GL(D-1)$  state $| i j \rangle$
while
$| i i \gg$  transforms under $SO(D-1)$ rotations
as a superposition of $\GL(D-1)$  states like
$| i i \rangle   + \sum_j |i i j j\rangle + \sum_{j, l} |i i j j l l\rangle
+ \dots$.
This happens because
$\GL(D-1)$ states like $\sum_j |i i j j\rangle$
behave as 
a state $|i i\rangle$ under a $SO(D-2)$ rotation.
Notice that states like $| i i 1 \rangle $ or $| i i 1 1 \rangle $
which also behave as $|i i\rangle$  are absent in the
superposition because $|i i \gg$ is an almost
true 2 index state.
As it will become clear when constructing the irreps
the need of finding a state $| i i\rangle$ will enforce the
tracelessness conditions.

We can now state the actions of the different pieces of $\delta^i$ on
the basis elements.
The action of an increasing $\delta^{l \uparrow}$ operator is
defined as
\begin{align}
\delta^{l \uparrow}\, 
e^{\sN {N, s, a} }_{i_1\dots i_s}
=
U^{\sN {N, s}}_{a b}\,
e^{\sN {N, s+1, b} }_{i_1\dots i_s\, l}
.
\end{align}
The action of  decreasing $\delta^{m \downarrow}$ operator is more
complex and defined as
\begin{align}
\delta^{m \downarrow}\, 
e^{\sN {N, s, a} }_{i_1\dots i_s}
=&
\delta_{m, i_1}
D^{\sN {N, s, 1}}_{a b}\,
e^{\sN {N, s-1, b} }_{i_2\dots i_s}
+
\delta_{m, i_2}
D^{\sN {N, s, 2}}_{a b}\,
e^{\sN {N, s-1, b} }_{i_1\, i_3\dots i_s}
+
\dots
\nonumber\\
&
+
\delta_{m, i_s}
D^{\sN {N, s, s}}_{a b}\,
e^{\sN {N, s-1, b} }_{i_1\, i_2\dots i_{s-1}}
\nonumber\\
=&
\sum_{p=1}^{s}
\delta_{m, i_p}
D^{\sN {N, s, p}}_{a b}\,
e^{\sN {N, s-1, b} }_{i_1\dots i_{p-1}\,i_{p+1} \dots i_{s}}
.
\end{align}
The action of  decreasing $\delta_A^{m \downarrow}$ operator is even more
complex and defined as
\begin{align}
\delta_A^{m \downarrow}\, 
e^{\sN {N, s, a} }_{i_1\dots i_s}
=&
\delta_{i_1, i_2}
A^{\sN {N, s, 1 2}}_{a b}\,
e^{\sN {N, s-1, b} }_{m\,i_3\dots i_s}
+
\delta_{i_1, i_3}
A^{\sN {N, s,1 3}}_{a b}\,
e^{\sN {N, s-1, b} }_{m\, i_2\dots i_s}
+
\dots
\nonumber\\
&
+
\delta_{i_p, i_q}
A^{\sN {N, s, p q}}_{a b}\,
e^{\sN {N, s-1, b} }_{m i_1\dots i_{p-1}\,i_{p+1} \dots i_{q-1}\,i_{q+1}\dots i_{s}}
+
\dots
\nonumber\\
=&
\sum_{p=1}^{s-1} \sum_{q=p+1}^s\,
\delta_{i_p, i_q}
A^{\sN {N, s, p q}}_{a b}\,
e^{\sN {N, s-1, b} }_{m i_1\dots i_{p-1}\,i_{p+1} \dots i_{q-1}\,i_{q+1}\dots i_{s}}
.
\end{align}
Not all $D$s and $A$s matrices are independent.
Actually only
$D^{\sN {s,1}}_{a b}$ and
$A^{\sN {s,1 2}}_{a b}$ are independent as shown in
appendix \ref{app:Lorentz_algebra}.
They are the only ones reported in the
supplementary material.

Because of this the almost true tensors can be computed as
\begin{align}
\sum_a \hat b_{\sN{N\,s\,a}} \, U^{\sN{N\,s}}_{a b}
=
0
.
\label{eq:almost_true_tensor}
\end{align}
Here and in the following we use $\hat b$ for the almost true tensors
not projected using a Young symmetrizer while we reserve $b$ the
almost true tensors projected with the appropriate Young symmetrizer.
The Young symmetrizer depends in the context.

For the true states at level $N$  we have (all $i$s different)
\begin{equation}
\delta^{i_1 \uparrow} \delta^{i_1 \downarrow}|i_1\dots i_s\gg =
- 2(N-1) |i_1\dots i_s\gg
,
\end{equation}
because the $\delta^{i}$ normalization includes a $i \alpha^+_0 =
i \sap M$ in the rest frame as discussed in
appendix \ref{app:Lorentz_algebra}.
Obviously for all the other states we need to consider
\begin{equation}
\left(
\delta^{i_1 \uparrow} \delta^{i_1 \downarrow} +
\delta^{i_1 \downarrow} \delta^{i_1 \uparrow}
\right)
|i_1\dots i_s\gg
,
\end{equation}
as discussed in appendix \ref{app:Lorentz_algebra}.

However the 

\subsection{On the absence of massive scalars in non critical dimension}

We start by counting the independent $SO(D-2)$ (or $\GL(D-2)$ that is
the same) scalars at different levels
\begin{equation}
  \begin{array}{ c || c | c || c}
    N & \mbox{basic} & \mbox{composite} & dim\,T_{N,\, 0}
    \\
    \hline
    2
    & (1,1)
    &
    & 1
    \\
    \hline
    3
    & (1,2)
    &
    & 1
    \\
    \hline
    4
    & (1,3),\, (2,2)
    & (1,1)^2
    & 3
    \\
    \hline
    5
    & (1,4),\, (2,3)
    & (1,1)(1,2)
    & 3
    \\
    \hline
    6
    & (1,5),\, (2,4),\, (3,3)
    & (1,1)(1,3),\, (1,1)(2,2),\, (1,2)^2,\, (1,1)^3
    & 7
    \\
    \hline
  \end{array}
  \label{eq:scalar_basis}
\end{equation}

We denote the basis at level $N$ as $T_{N,0}\equiv S_N$
and the vector spaces it generates as
$V_{N, 0}=span\, T_{N,0}$.

We notice that the basic couples at level $N=2k$ and $N=2k+1$ are $k$.

Then the generating function for the scalars is
\begin{align}
  {\cal T}^{\sN 0}
  =&
  \cS(x)
  =
  \prod_{k=1}^\infty
  \left[ \frac{1}{1-x^{2k}} \frac{1}{1-x^{2k+1}} \right]^k
  \nonumber\\
  =&
  1+x^2+x^3+3\,x^4+3\,x^5+7\,x^6+8\,x^7+16\,x^8+20\,x^9+35\,x^{10}+46
  \,x^{11}+77\,x^{12}
  \nonumber\\
  &+102\,x^{13}+161\,x^{14}+220\,x^{15}+334\,x^{16}+
  457\,x^{17}+678\,x^{18}+930\,x^{19}
  \nonumber\\
  &+1351\,x^{20}+1855\,x^{21}
 +\cdots
.
\end{align}

We can now proceed to list the basis of the $SO(D-2)$ (or again
$\GL(D-2)$ that is the same) vectors
\begin{equation}
  \begin{array}{ c || c || c}
    N & \mbox{vector}  & dim\, T_{N,\, 1}
    \\
    \hline
    2
    & 2^i \emptyset,\, 1^i \{S_1\}
    & 1+0=1
    \\
    \hline
    3
    & 3^i \emptyset,\, 2^i \{S_1\},\, 1^i \{S_2\}
      & 1 + 1 + 0 =2
   \\
    \hline
    4
    & 4^i \emptyset,\, 3^i \{S_1\},\, 2^i \{S_2\},\, 1^i \{S_3\}
    & 1 + 0 + 1 +1 =3
        \\
    \hline
  \end{array}
  ,
    \label{eq:vector_basis}
\end{equation}
where $\{S_2\}$ means any scalar at level $N=2$
and
$3^i$ means $\ualpha^i_{-3}$ as explained above.

Similarly we denote the basis at level $N$ as $T_{N,1} \equiv V_N$
and the vector spaces it generates as $V_{N, 1}=span\, T_{N,1}$.

The generating function of the basic vectors $1^i, 2^i, \dots$ is
\begin{equation}
  \cV_0
  = \frac{x}{1-x}
  = x+ x^2+ x^3+ \dots
  .
\end{equation}
Then the generating function for the vectors
is
\begin{align}
  {\cal T}^{\sN 1}(x)
  =&
  \cV(x)
  =
  \cV_0(x) \cS(x)
  =
  \frac{x}{1-x}
    \prod_{k=1}^\infty
  \left[ \frac{1}{1-x^{2k}} \frac{1}{1-x^{2k+1}} \right]^k
  \nonumber\\
  =&
  x+x^2+2\,x^3+3\,x^4+6\,x^5+9\,x^6+16\,x^7+24\,x^8+40\,x^9
  +60\,x^{10}
  \nonumber\\
  &
  +95\,x^{11}+141\,x^{12}+218\,x^{13}+320\,x^{14}+481\,x^{15}+701\,x
  ^{16}+1035\,x^{17}
  \nonumber\\
  &+1492\,x^{18}+2170\,x^{19}+3100\,x^{20}+4451\,x^{
    21}+\cdots
  .
\end{align}

We expect that the image of a linear combination of
$SO(D-2)$ scalars at level $N$ under a boost $M^{i -}$
be a generic combination of $SO(D-2)$ vectors at level $N$.
Therefore we can compute the naively expected number of scalars by
considering how many constraints we have \wrt to how many free
coefficients we have.
The number of homogeneous equations exceeding
the possible coefficients for the scalars is then 
\begin{align}
  \Delta^{\sN 0}(x)
    =&
  \cV(x) -\cS(x)
  =
  (\cV_0(x) -1)* \cS(x)
  \nonumber\\
  =&
  -1+x
  +0 x^2
  +x^3
  +0 x^4
  +3\,x^5+2\,x^6+8\,x^7+8\,x^8+20\,x^9+25\,x^{10}
  \nonumber\\
  &+49\,x^{11}+64\,x^{12}+116\,x^{13}+159\,x^{14}+261\,x^{15}+367\,x^{16}
  \nonumber\\
  &+578\,x^{
   17}+814\,x^{18}+1240\,x^{19}+1749\,x^{20}+2596\,x^{21}+\cdots
 .
\end{align}
Therefore for $N\ge 1$ there are always more equations than coefficients and 
we expect no massive scalars in the bosonic open string spectrum
if there are no hidden symmetries or at special dimensions.
In facts even in absence of any hidden symmetry when the number of
constraints is equal to the number of coefficients
and some coefficients depend on
the dimension (this happens when the a basic object collapses to
number times $\ualpha$ under a boost)
we can find a possible solution by choosing the dimension so that the
system has a solution.

Let us see what asserted in an explicit non trivial case, i.e. the
$N=4$ level scalar which exists only in some special dimensions ($d=D-2$):
\begin{align}
  &\delta^{i \uparrow} (c_3 (1,3) +\, c_2  (2,2)+\, c_1 (1,1)^2)
  %
  \nonumber\\
  =& 
  1^i\, (1,2)
  \left(-2 c_3  - 4 c_2 + 8 c_1\right)
  +
  2^i\, (1,1)
  \left(  - \frac{3}{2} c_3 + c_2 + c_1(-d+2) \right)
\nonumber\\
&+
  4^i\,
  \left( \left( -\frac{3}{4}d+4 \right) c_3 + (-d+4) c_2 \right)
  ,
\end{align}
then the associated matrix is 
\begin{equation}
  U^{\sN {N=4, s=0} }
  =
  \left(
  \begin{array}{c | c c c}
           & 1^i\, (1,2) & 2^i\, (1,1) & 4^i
    \\
    \hline
   (1,1)^2 & 8           & -d+2       & 0
    \\
   (2,2)   & -4          & 1          & -d+4
    \\
   (1,3)   & -2          &-\frac{3}{2}& -\frac{3}{4} d +4
  \end{array}
  \right)
  ,
\end{equation}
whose determinant is $(d-24)(d-4)$ and so for $d=24$ there is a scalar. 
It has left eigenvector $(c_1, c_2, c_3)=(1,7,-10)$.

Notice once again that we have considered the constraints from
only one possible $i$ since any $i$ gives exactly the same set of equations.

\subsection{On the absence of massive vector irreps in non critical dimensions}

We can proceed as done for the massive scalars.

Again constraints arise from the action of the boost which leads to an addition
of an index $i$,
i.e starting from a generic combination of states with $1$ index the
action of interest is
the tensor multiplication of vector $\GL(D-2)$ irrep for a $\GL(D-2)$
vector irrep associate with $i$.
We do not write $SO(D-2)$ since we are not imposing the traceless
property in any way.
This means that for finding the constraints on the vector irrep we
need to consider all 2-index tensors.
These are not 2 indices irreps and  may be decomposable into irreps.

Let us now list the basis for $\GL(D-2)$ $2$ indices tensors 
\begin{equation}
  \begin{array}{ c || c || c}
    N & \mbox{basic tensor}  & T_{(0)\, N, 2}
    \\
    \hline
    2
    & 1^i 1^j 
    & 1 
    \\
    \hline
    3
    & 1^i 2^j,\, 2^i 1^j 
    & 2 
    \\
    \hline
    4
    & 3^i 1^j,\, 2^i 2^j,\, 1^i 3^j 
    & 3 
   \\
    \hline
    5
    & 4^i 1^j,\, 3^i 2^j,\, 2^i 3^j,\, 1^i 4^j 
    & 4 
    \\
    \hline
  \end{array}
  ,
    \label{eq:spin2_basis}
\end{equation}
where the number of basic tensor is counted keeping in mind that the
action of $M^{- i}$ on $\ualpha^j$  gives the structures like $1^i 2^j$
but the equations are for the coefficients and these equations are 
independent on $i$ and $j$!
Using the previous notation $ (-1)^i \equiv \ualpha_{+1}^i$ we have e.g.
for the action of the infinitesimal boost $M^{- i}$ on the level $N=3$
generic $SO(D-2)$ vector
\begin{align}
  &\delta^{- i} (c_2 3^j +\, c_1 1^j (1,1))
%
  \nonumber\\
  =&
  ( 2 c_2 + \frac{-(d-2) }{2} c_1) 2^i 1^j
  + (- 3 c_2 + 2 c_1) 1^i 2^j
 ,
\end{align}
then the associated matrix is 
\begin{equation}
  U^{\sN{N=3, s=1}}
  =
  \left(
  \begin{array}{c | c c}
            & 2^i\, 1^j & 1^i\, 2^j
    \\
    \hline
   (1,1) 1^j & -\oh d+1         & 2
    \\
    3^j & -\frac{3}{2} & -3
   \end{array}
  \right)
  ,
\end{equation}
whose determinant is $\frac{3}{2}d$ and so for $d=0$ there is a vector.

The generating function of the basic 2-index tensors $1^i 1^j, 2^i
1^j, \dots$ is
\begin{equation}
  {\cal T}_0^{\sN 2}(x)
  = 
  \cV_0^2
  = 
  (\frac{x}{1-x})^2
  =
  ( x^2+2\,x^3+3\,x^4+4\,x^5+5\,x^6+ \dots )
  .
\end{equation}
Then the generating function for the 2-index tensors
is
\begin{align}
  {\cal T}^{\sN 2}(x)
  =&
  {\cal T}_0^{\sN 2}(x) \cS(x)
  =
  \left( \frac{x}{1-x} \right)^2
  \prod_{k=1}^\infty
  \left[ \frac{1}{1-x^{2k}} \frac{1}{1-x^{2k+1}} \right]^k
  \nonumber\\
  =&
  x^2+2\,x^3+4\,x^4+7\,x^5+13\,x^6+22\,x^7+38\,x^8+62\,x^9+102\,x^{10
  }
  \nonumber\\
  &
  +162\,x^{11}+257\,x^{12}+398\,x^{13}+616\,x^{14}+936\,x^{15}+1417\,
  x^{16}
  \nonumber\\
  &+2118\,x^{17}+3153\,x^{18}+4645\,x^{19}+6815\,x^{20}+9915\,x^{
    21}+14366\,x^{22}+\cdots
  .
\end{align}
Finally we can compute the number of homogeneous equations exceeding
the possible coefficients for the vectors as
\begin{align}
    \Delta^{\sN 1}
    =&
  {\cal T}^{\sN 2}(x) -\cV(x)
  =
  \cV_0(x) (\cV_0(x) -1) \cS(x)
  \nonumber\\
  =&
  -x
  +0 x^2 +0 x^3
  + x^4+x^5+4\,x^6+6\,x^7+14\,x^8+22\,x^9+42\,x^{10}
  \nonumber\\
  &+67\,x^{11}+116\,x^{12}+180\,x^{13}+296\,x^{14}+455\,x^{15}+716\,x^{16}
  \nonumber\\
  &+1083\,x^{17}+1661\,x^{18}+2475\,x^{19}+3715\,x^{20}+5464\,x^{21}+\cdots
 ,
\end{align}
hence we expect no massive vectors for $N\ge4$.
Wit the possible exception for $N=2, 3$ in some special dimensions.


\subsection{Constraints on the number of tensors with $s\ge 2$ indices}

Let us consider some examples of higher tensors.
We start with the simplest case $N=3$ $s=2$:
%
%
\begin{align}
  &\delta^{- k} (c_1 1^i 2^j +\, c_2 1^j 2^i)
  %
  %
=
  -2 (c_1 + c_2) 1^i 1^j 1^k
  ,
\end{align}
from which we see that there is only one constraint and always a
solution $(c_1, c_2)=(1, -1)$.

Next we consider the next simplest example $N=4$ $s=2$:
\begin{align}
  &\delta^{i \uparrow} ( c_4 2^{j_2}\, 2^{j_1} + c_3 1^{j_2}\, 3^{j_1}  + c_2 1^{j_1}\, 3^{j_2} + c_1
  (1,1) 1^{j_1} 1^{j_2})
  \nonumber\\
  =&
  2^{j_1}\, 1^{i}\, 1^{j_2}\,
  \left(-2c_1+3c_3+2c_4 \right)
  +
  2^{j_2}\, 1^{j_1}\, 1^{i}\,
  \left( -2c_1+3c_2+2c_4\right)
  +
    2^i\, 1^{j_1}\, 1^{j_2}\,
  \left( -\oh d c_1 - \frac{3}{2} c_2 - \frac{3}{2} c_3 \right)
,
\end{align}
then the associated matrix is 
\begin{equation}
  U^{\sN{N=3, s=2}}
  =
  \left(
  \begin{array}{c | c c c}
            &   2^{j_1}\,1^{j_2}\,1^{i}  & 1^{j_1}\,2^{j_2}\,1^{i} &
                1^{j_1}\,1^{j_2}\,2^{i}
    \\
    \hline
    (1,1)\, 1^{j_1}\, 1^{j_2}
        & -\oh d  & 2  & 2
    \\
    1^{j_1}\, 3^{j_2}
    & -\frac{3}{2} & 0 &-3
    \\
    3^{j_1}\, 1^{j_2}
    & -\frac{3}{2} & -3 & 0
    \\
    2^{j_1}\, 2^{j_2}
    & 0 & -2 &-2
   \end{array}
  \right)
  ,
\end{equation}
which has one zero eigenvalue
$(c_1, c_2, c_3, c_4)=(1, -\frac{d}{6}, -\frac{d}{6}, \frac{d}{4}+1)$.

All the previous examples fall into the table for the basic 3 index tensors
\begin{equation}
  \begin{array}{ c || c || c}
    N & \mbox{basic tensor}  & T_{(0)\, N, 3}
    \\
    \hline
    3
    & 1^i 1^j 1^k 
    & 1 
    \\
    \hline
    4
    & 1^i 1^j 2^k,\, 1^i 2^j 1^k,\, 2^i 1^j 1^k 
    & 3 
    \\
    \hline
    5
    & 1^i 1^j 3^k,\, 1^i 3^j 1^k,\, 3^i 1^j 1^k,\,
    1^i 2^j 2^k,\,  2^i 1^j 2^k,\, 2^i 2^j 1^k\,
    & 6
   \\
    \hline
    6
    & 1^i 1^j 4^k,\, 1^i 4^j 1^k,\, 4^i 1^j 1^j,\,
    1^i 2^j 3^k,\, 2^i 1^j 3^k,\, 2^i 3^j 1^k,\,
    1^i 3^j 2^k,\, 3^i 1^j 2^k  ,\, 3^i 2^j 1^k      
    & 10
    \\
    \hline
  \end{array}
  ,
\end{equation}
whose generating function is
\begin{align}
  {\cal T}_0^{\sN s}(x)
  =&
  \cV_0^s(x)
  .
\end{align}
We get then generating function for the s index tensors
is
\begin{align}
  {\cal T}^{\sN s}(x)
  =&
  {\cal T}_0^{\sN s}(x) \cS(x)
  =
  \left( \frac{x}{1-x} \right)^s
  \prod_{k=1}^\infty
  \left[ \frac{1}{1-x^{2k}} \frac{1}{1-x^{2k+1}} \right]^k
  %
  .
  \label{eq:gen fun dim T N s}
\end{align}
Therefore generalizing naively the previous result we get that
the number of possible tensors with $s$ indices is encoded into
the negative numbers of
\begin{align}
  \Delta^{\sN s}
  =
  (\cV_0^{s+1} -\cV_0^s) \cS
  =
  \cV_0^s
  (\cV_0 - 1) \cS
  =
  \cV_0^s \, \Delta^{\sN 0}
.
  \label{eq:gen fun dim Delta N s}
\end{align}  

For example for $s=2$ we get
\begin{align}
  (\cV_0^{s+1} -\cV_0^s)|_{s=3} \cS
  =&
  -x^2-x^3-x^4+x^6+5\,x^7+11\,x^8+25\,x^9+47\,x^{10}
  \nonumber\\
  &+89\,x^{11}+156\,
  x^{12}+272\,x^{13}+452\,x^{14}+748\,x^{15}
  \nonumber\\
  &
  +1203\,x^{16}+1919\,x^{17}
  +3002\,x^{18}+4663\,x^{19}+7138\,x^{20}+10853\,x^{21}
  +\cdots
  ,
\end{align}
so for generic dimension we expect $1$ 2-index tensor at levels
$N=2,3,4$ only
since he coefficients are more than the constraints only in these cases.

Similarly for $s=3$ we get
\begin{align}
  (\cV_0^{s+1} -\cV_0^s)|_{s=4} \cS
  =&
  -x^3-2\,x^4-3\,x^5-3\,x^6-2\,x^7+3\,x^8+14\,x^9
+ \dots
  ,
\end{align}
so we expect at least 
$1$ 3-index tensor at level $N=3,4,5,6,7$. 

Let us explain in more details the meaning of the previous numbers.
We see that for $3\le N \le 7$ the vector space of solutions has
different dimension.

These vector spaces are where the symmetric group $S_s$ with $s=3$
acts and can and must be split into $S_3$ irreps.
These irreps correspond to $\GL(D-2)$ irreducible tensors with $s=3$ indices.

We know that  $S_3$ irreps and their dimensions are
\begin{align}
  \ydiagram{3} \rightarrow dim=1,
  ~~~~
  \ydiagram{1,1,1} \rightarrow dim=1,
  ~~~~
  \ydiagram{2,1} \rightarrow dim=2,
.
\end{align}
Using this knowledge we see that at level $N=3$ we have either 
$ \ydiagram{3}$ or $\ydiagram{1,1,1}$.
Looking to the possible tensors we see that actually we have
$\ydiagram{3}$, i.e. a state on the leading Regge trajectory.

At level $N=4$ we could in principle have
two irreps chosen among either $ \ydiagram{3}$ or $\ydiagram{1,1,1}$
or simply  $\ydiagram{2,1}$.
From the knowledge of the explicit tensors we know that we have
a subleading Regge state in the $\GL(D-2)$  irrep $\ydiagram{2,1}$.

For higher level things get more complex and the easiest thing is to
proceed brute force. 

\subsection{Summary of the naive approach up to $N=24$}

We can now easily get an idea of which tensors are present in the
generic dimension by simply examining the generating functions.
The experimental result is that for generic dimension at level $N$ we
have $\GL(D-2)$ physical states with $\oh N \le s \le N$  indices.
Since we are considering the states with the highest number of indices
these can be identified with $SO(D-1)$ states in the rest frame as
discussed below.

The generating functions for the basis of $\GL(D-2)$ tensors with
$0\le s\le 24$ indices ${\cal T}^{[s]}$ are given by

\begin{align}
{\cal T}^{[0]}=&1+x^2+x^3+3\,x^4+3\,x^5+7\,x^6+8\,x^7+16\,x^8+20\,x^9+35\,x^{10}
\nonumber\\
&+46\,x^{11}+77\,x^{12}+102\,x^{13}+161\,x^{14}+220\,x^{15}+334\,x^{16}+457\,x^{17}
\nonumber\\
&+678\,x^{18}+930\,x^{19}+1351\,x^{20}+1855\,x^{21}+2647\,x^{22}+3629\,x^{23}+5117\,x^{24}+\cdots
.
\end{align}

\begin{align}
{\cal
  T}^{[1]}=&x+x^2+2\,x^3+3\,x^4+6\,x^5+9\,x^6+16\,x^7+24\,x^8+40\,x^9
+60\,x^{10}
\nonumber\\
&+95\,x^{11}+141\,x^{12}+218\,x^{13}+320\,x^{14}+481\,x^{15}+701\,x^{16}
+1035\,x^{17}
\nonumber\\
&
+1492\,x^{18}+2170\,x^{19}+3100\,x^{20}+4451\,x^{21}+6306\,x^{22}+8953\,x^{23}+12582\,x^{24}+\cdots 
\end{align}

\begin{align}
  {\cal T}^{[2]}=&
  x^2+2\,x^3+4\,x^4+7\,x^5+13\,x^6+22\,x^7+38\,x^8+62\,x^9+102\,x^{10}
  \nonumber\\
  &+162\,x^{11}+257\,x^{12}+398\,x^{13}+616\,x^{14}+936\,x^{15}+1417\,x^{16}
  +2118\,x^{17}
  \nonumber\\
  &+3153\,x^{18}+4645\,x^{19}+6815\,x^{20}+9915\,x^{21}+14366\,x^{22}+20672\,x^{23}+29625\,x^{24}+\cdots 
\end{align}

\begin{align}
  {\cal T}^{[3]}=&
  x^3+3\,x^4+7\,x^5+14\,x^6+27\,x^7+49\,x^8+87\,x^9+149\,x^{10}
  \nonumber\\
  &
  +251\,x^{11}+413\,x^{12}+670\,x^{13}+1068\,x^{14}+1684\,x^{15}+2620\,x^{16}+4037\,x^{17}
  \nonumber\\
  &
  +6155\,x^{18}+9308\,x^{19}+13953\,x^{20}+20768\,x^{21}+30683\,x^{22}+45049\,x^{23}+65721\,x^{24}+\cdots 
\end{align}

\begin{align}
  {\cal T}^{[4]}=&
  x^4+4\,x^5+11\,x^6+25\,x^7+52\,x^8+101\,x^9+188\,x^{10}
  \nonumber\\
  &+337\,x^{11}+588\,x^{12}+1001\,x^{13}+1671\,x^{14}+2739\,x^{15}+4423\,x^{16}+7043\,x^{17}
  \nonumber\\
  &
  +11080\,x^{18}+17235\,x^{19}+26543\,x^{20}+40496\,x^{21}+61264\,x^{22}+91947\,x^{23}+136996\,x^{24}+\cdots 
\end{align}

\begin{align}
  {\cal T}^{[5]}=&
  x^5+5\,x^6+16\,x^7+41\,x^8+93\,x^9+194\,x^{10}
  \nonumber\\
  &
  +382\,x^{11}+719\,x^{12}+1307\,x^{13}+2308\,x^{14}+3979\,x^{15}+6718\,x^{16}+11141\,x^{17}
  \nonumber\\
  &
  +18184\,x^{18}+29264\,x^{19}+46499\,x^{20}+73042\,x^{21}+113538\,x^{22}+174802\,x^{23}+266749\,x^{24}+\cdots 
\end{align}

\begin{align}
  {\cal T}^{[6]}=&
  x^6+6\,x^7+22\,x^8+63\,x^9+156\,x^{10}
  \nonumber\\
  &
  +350\,x^{11}+732\,x^{12}+1451\,x^{13}+2758\,x^{14}+5066\,x^{15}+9045\,x^{16}+15763\,x^{17}
  \nonumber\\
  &
  +26904\,x^{18}+45088\,x^{19}+74352\,x^{20}+120851\,x^{21}+193893\,x^{22}+307431\,x^{23}+482233\,x^{24}+\cdots 
\end{align}

\begin{align}
  {\cal T}^{[7]}=&
  x^7+7\,x^8+29\,x^9+92\,x^{10}
  \nonumber\\
  &
  +248\,x^{11}+598\,x^{12}+1330\,x^{13}+2781\,x^{14}+5539\,x^{15}+10605\,x^{16}+19650\,x^{17}
  \nonumber\\&
  +35413\,x^{18}+62317\,x^{19}+107405\,x^{20}+181757\,x^{21}+302608\,x^{22}+496501\,x^{23}+803932\,x^{24}+\cdots 
\end{align}

\begin{align}
  {\cal T}^{[8]}=&
  x^8+8\,x^9+37\,x^{10}
  \nonumber\\
  &
  +129\,x^{11}+377\,x^{12}+975\,x^{13}+2305\,x^{14}+5086\,x^{15}+10625\,x^{16}+21230\,x^{17}
  \nonumber\\
  &
  +40880\,x^{18}+76293\,x^{19}+138610\,x^{20}+246015\,x^{21}+427772\,x^{22}+730380\,x^{23}+1226881\,x^{24}+\cdots 
\end{align}


\begin{align}
{\cal T}^{[9]}=&x^9+9\,x^{10}\nonumber\\&+46\,x^{11}+175\,x^{12}+552\,x^{13}+1527\,x^{14}+3832\,x^{15}+8918\,x^{16}+19543\,x^{17}\nonumber\\&+40773\,x^{18}+81653\,x^{19}+157946\,x^{20}+296556\,x^{21}+542571\,x^{22}+970343\,x^{23}+1700723\,x^{24}+\cdots 
\end{align}

\begin{align}
{\cal T}^{[10]}=&x^{10}\nonumber\\&+10\,x^{11}+56\,x^{12}+231\,x^{13}+783\,x^{14}+2310\,x^{15}+6142\,x^{16}+15060\,x^{17}\nonumber\\&+34603\,x^{18}+75376\,x^{19}+157029\,x^{20}+314975\,x^{21}+611531\,x^{22}+1154102\,x^{23}+2124445\,x^{24}+\cdots 
\end{align}

\begin{align}
{\cal T}^{[11]}=&x^{11}+11\,x^{12}+67\,x^{13}+298\,x^{14}+1081\,x^{15}+3391\,x^{16}+9533\,x^{17}\nonumber\\&+24593\,x^{18}+59196\,x^{19}+134572\,x^{20}+291601\,x^{21}+606576\,x^{22}+1218107\,x^{23}+2372209\,x^{24}+\cdots 
\end{align}

\begin{align}
{\cal T}^{[12]}=&x^{12}+12\,x^{13}+79\,x^{14}+377\,x^{15}+1458\,x^{16}+4849\,x^{17}\nonumber\\&+14382\,x^{18}+38975\,x^{19}+98171\,x^{20}+232743\,x^{21}+524344\,x^{22}+1130920\,x^{23}+2349027\,x^{24}+\cdots 
\end{align}

\begin{align}
{\cal T}^{[13]}=&x^{13}+13\,x^{14}+92\,x^{15}+469\,x^{16}+1927\,x^{17}\nonumber\\&+6776\,x^{18}+21158\,x^{19}+60133\,x^{20}+158304\,x^{21}+391047\,x^{22}+915391\,x^{23}+2046311\,x^{24}+\cdots 
\end{align}

\begin{align}
{\cal T}^{[14]}=&x^{14}+14\,x^{15}+106\,x^{16}+575\,x^{17}\nonumber\\&+2502\,x^{18}+9278\,x^{19}+30436\,x^{20}+90569\,x^{21}+248873\,x^{22}+639920\,x^{23}+1555311\,x^{24}+\cdots 
\end{align}

\begin{align}
{\cal T}^{[15]}=&x^{15}+15\,x^{16}+121\,x^{17}\nonumber\\&+696\,x^{18}+3198\,x^{19}+12476\,x^{20}+42912\,x^{21}+133481\,x^{22}+382354\,x^{23}+1022274\,x^{24}+\cdots 
\end{align}

\begin{align}
{\cal T}^{[16]}=&x^{16}+16\,x^{17}\nonumber\\&+137\,x^{18}+833\,x^{19}+4031\,x^{20}+16507\,x^{21}+59419\,x^{22}+192900\,x^{23}+575254\,x^{24}+\cdots 
\end{align}

\begin{align}
{\cal T}^{[17]}=&x^{17}\nonumber\\&+17\,x^{18}+154\,x^{19}+987\,x^{20}+5018\,x^{21}+21525\,x^{22}+80944\,x^{23}+273844\,x^{24}+\cdots 
\end{align}

\begin{align}
{\cal T}^{[18]}=&x^{18}+18\,x^{19}+172\,x^{20}+1159\,x^{21}+6177\,x^{22}+27702\,x^{23}+108646\,x^{24}+\cdots 
\end{align}

\begin{align}
{\cal T}^{[19]}=&x^{19}+19\,x^{20}+191\,x^{21}+1350\,x^{22}+7527\,x^{23}+35229\,x^{24}+\cdots 
\end{align}

\begin{align}
{\cal T}^{[20]}=&x^{20}+20\,x^{21}+211\,x^{22}+1561\,x^{23}+9088\,x^{24}+\cdots 
\end{align}

\begin{align}
{\cal T}^{[21]}=&x^{21}+21\,x^{22}+232\,x^{23}+1793\,x^{24}+\cdots 
\end{align}

\begin{align}
{\cal T}^{[22]}=&x^{22}+22\,x^{23}+254\,x^{24}+\cdots 
\end{align}

\begin{align}
{\cal T}^{[23]}=&x^{23}+23\,x^{24}+\cdots 
\end{align}

\begin{align}
{\cal T}^{[24]}=&+x^{24}+\cdots 
\end{align}

The generating functions for the excess of constraints (negative numbers
mean existence of a solution) for a tensor with $s$
indices ${\cal T}^{[s]}$ are given by

\begin{align}
\Delta ^{[0]}=&-1+x+x^3+3\,x^5+2\,x^6+8\,x^7+8\,x^8+20\,x^9+25\,x^{10}\nonumber\\&+49\,x^{11}+64\,x^{12}+116\,x^{13}+159\,x^{14}+261\,x^{15}+367\,x^{16}+578\,x^{17}\nonumber\\&+814\,x^{18}+1240\,x^{19}+1749\,x^{20}+2596\,x^{21}+3659\,x^{22}+5324\,x^{23}+7465\,x^{24}+\cdots 
\end{align}

\begin{align}
\Delta ^{[1]}=&-x+x^4+x^5+4\,x^6+6\,x^7+14\,x^8+22\,x^9+42\,x^{10}\nonumber\\&+67\,x^{11}+116\,x^{12}+180\,x^{13}+296\,x^{14}+455\,x^{15}+716\,x^{16}+1083\,x^{17}\nonumber\\&+1661\,x^{18}+2475\,x^{19}+3715\,x^{20}+5464\,x^{21}+8060\,x^{22}+11719\,x^{23}+17043\,x^{24}+\cdots 
\end{align}

\begin{align}
\Delta ^{[2]}=&-x^2-x^3-x^4+x^6+5\,x^7+11\,x^8+25\,x^9+47\,x^{10}\nonumber\\&+89\,x^{11}+156\,x^{12}+272\,x^{13}+452\,x^{14}+748\,x^{15}+1203\,x^{16}+1919\,x^{17}\nonumber\\&+3002\,x^{18}+4663\,x^{19}+7138\,x^{20}+10853\,x^{21}+16317\,x^{22}+24377\,x^{23}+36096\,x^{24}+\cdots 
\end{align}

\begin{align}
\Delta ^{[3]}=&-x^3-2\,x^4-3\,x^5-3\,x^6-2\,x^7+3\,x^8+14\,x^9+39\,x^{10}\nonumber\\&+86\,x^{11}+175\,x^{12}+331\,x^{13}+603\,x^{14}+1055\,x^{15}+1803\,x^{16}+3006\,x^{17}\nonumber\\&+4925\,x^{18}+7927\,x^{19}+12590\,x^{20}+19728\,x^{21}+30581\,x^{22}+46898\,x^{23}+71275\,x^{24}+\cdots 
\end{align}

\begin{align}
\Delta ^{[4]}=&-x^4-3\,x^5-6\,x^6-9\,x^7-11\,x^8-8\,x^9+6\,x^{10}\nonumber\\&+45\,x^{11}+131\,x^{12}+306\,x^{13}+637\,x^{14}+1240\,x^{15}+2295\,x^{16}+4098\,x^{17}\nonumber\\&+7104\,x^{18}+12029\,x^{19}+19956\,x^{20}+32546\,x^{21}+52274\,x^{22}+82855\,x^{23}+129753\,x^{24}+\cdots 
\end{align}

\begin{align}
\Delta ^{[5]}=&-x^5-4\,x^6-10\,x^7-19\,x^8-30\,x^9-38\,x^{10}\nonumber\\&-32\,x^{11}+13\,x^{12}+144\,x^{13}+450\,x^{14}+1087\,x^{15}+2327\,x^{16}+4622\,x^{17}\nonumber\\&+8720\,x^{18}+15824\,x^{19}+27853\,x^{20}+47809\,x^{21}+80355\,x^{22}+132629\,x^{23}+215484\,x^{24}+\cdots 
\end{align}

\begin{align}
\Delta ^{[6]}=&-x^6-5\,x^7-15\,x^8-34\,x^9-64\,x^{10}\nonumber\\&-102\,x^{11}-134\,x^{12}-121\,x^{13}+23\,x^{14}+473\,x^{15}+1560\,x^{16}+3887\,x^{17}\nonumber\\&+8509\,x^{18}+17229\,x^{19}+33053\,x^{20}+60906\,x^{21}+108715\,x^{22}+189070\,x^{23}+321699\,x^{24}+\cdots 
\end{align}

\begin{align}
\Delta ^{[7]}=&-x^7-6\,x^8-21\,x^9-55\,x^{10}\nonumber\\&-119\,x^{11}-221\,x^{12}-355\,x^{13}-476\,x^{14}-453\,x^{15}+20\,x^{16}+1580\,x^{17}\nonumber\\&+5467\,x^{18}+13976\,x^{19}+31205\,x^{20}+64258\,x^{21}+125164\,x^{22}+233879\,x^{23}+422949\,x^{24}+\cdots 
\end{align}

\begin{align}
\Delta ^{[8]}=&-x^8-7\,x^9-28\,x^{10}\nonumber\\&-83\,x^{11}-202\,x^{12}-423\,x^{13}-778\,x^{14}-1254\,x^{15}-1707\,x^{16}-1687\,x^{17}\nonumber\\&-107\,x^{18}+5360\,x^{19}+19336\,x^{20}+50541\,x^{21}+114799\,x^{22}+239963\,x^{23}+473842\,x^{24}+\cdots 
\end{align}

\begin{align}
\Delta ^{[9]}=&-x^9-8\,x^{10}\nonumber\\&-36\,x^{11}-119\,x^{12}-321\,x^{13}-744\,x^{14}-1522\,x^{15}-2776\,x^{16}-4483\,x^{17}\nonumber\\&-6170\,x^{18}-6277\,x^{19}-917\,x^{20}+18419\,x^{21}+68960\,x^{22}+183759\,x^{23}+423722\,x^{24}+\cdots 
\end{align}

\begin{align}
\Delta ^{[10]}=&-x^{10}\nonumber\\&-9\,x^{11}-45\,x^{12}-164\,x^{13}-485\,x^{14}-1229\,x^{15}-2751\,x^{16}-5527\,x^{17}\nonumber\\&-10010\,x^{18}-16180\,x^{19}-22457\,x^{20}-23374\,x^{21}-4955\,x^{22}+64005\,x^{23}+247764\,x^{24}+\cdots 
\end{align}

\begin{align}
\Delta ^{[11]}=&-x^{11}-10\,x^{12}-55\,x^{13}-219\,x^{14}-704\,x^{15}-1933\,x^{16}-4684\,x^{17}\nonumber\\&-10211\,x^{18}-20221\,x^{19}-36401\,x^{20}-58858\,x^{21}-82232\,x^{22}-87187\,x^{23}-23182\,x^{24}+\cdots 
\end{align}

\begin{align}
\Delta ^{[12]}=&-x^{12}-11\,x^{13}-66\,x^{14}-285\,x^{15}-989\,x^{16}-2922\,x^{17}\nonumber\\&-7606\,x^{18}-17817\,x^{19}-38038\,x^{20}-74439\,x^{21}-133297\,x^{22}-215529\,x^{23}-302716\,x^{24}+\cdots 
\end{align}

\begin{align}
\Delta ^{[13]}=&-x^{13}-12\,x^{14}-78\,x^{15}-363\,x^{16}-1352\,x^{17}\nonumber\\&-4274\,x^{18}-11880\,x^{19}-29697\,x^{20}-67735\,x^{21}-142174\,x^{22}-275471\,x^{23}-491000\,x^{24}+\cdots 
\end{align}

\begin{align}
\Delta ^{[14]}=&-x^{14}-13\,x^{15}-91\,x^{16}-454\,x^{17}\nonumber\\&-1806\,x^{18}-6080\,x^{19}-17960\,x^{20}-47657\,x^{21}-115392\,x^{22}-257566\,x^{23}-533037\,x^{24}+\cdots 
\end{align}

\begin{align}
\Delta ^{[15]}=&-x^{15}-14\,x^{16}-105\,x^{17}\nonumber\\&-559\,x^{18}-2365\,x^{19}-8445\,x^{20}-26405\,x^{21}-74062\,x^{22}-189454\,x^{23}-447020\,x^{24}+\cdots 
\end{align}

\begin{align}
\Delta ^{[16]}=&-x^{16}-15\,x^{17}\nonumber\\&-120\,x^{18}-679\,x^{19}-3044\,x^{20}-11489\,x^{21}-37894\,x^{22}-111956\,x^{23}-301410\,x^{24}+\cdots 
\end{align}

\begin{align}
\Delta ^{[17]}=&-x^{17}\nonumber\\&-16\,x^{18}-136\,x^{19}-815\,x^{20}-3859\,x^{21}-15348\,x^{22}-53242\,x^{23}-165198\,x^{24}+\cdots 
\end{align}

\begin{align}
\Delta ^{[18]}=&-x^{18}-17\,x^{19}-153\,x^{20}-968\,x^{21}-4827\,x^{22}-20175\,x^{23}-73417\,x^{24}+\cdots 
\end{align}

\begin{align}
\Delta ^{[19]}=&-x^{19}-18\,x^{20}-171\,x^{21}-1139\,x^{22}-5966\,x^{23}-26141\,x^{24}+\cdots 
\end{align}

\begin{align}
\Delta ^{[20]}=&-x^{20}-19\,x^{21}-190\,x^{22}-1329\,x^{23}-7295\,x^{24}+\cdots 
\end{align}

\begin{align}
\Delta ^{[21]}=&-x^{21}-20\,x^{22}-210\,x^{23}-1539\,x^{24}+\cdots 
\end{align}

\begin{align}
\Delta ^{[22]}=&-x^{22}-21\,x^{23}-231\,x^{24}+\cdots 
\end{align}

\begin{align}
\Delta ^{[23]}=-x^{23}-22\,x^{24}+\cdots 
\end{align}

\begin{align}
\Delta ^{[24]}=-x^{24}+\cdots 
\end{align}

So we can make the following forecast on the minimal dimension of the
vector spaces where the symmetric groups $S_s$ act according to the
level $N$ to be
\begin{equation} 
\begin{array}{ 
c || 
c | 
c | 
c | 
c | 
c | 
c | 
c | 
c | 
c | 
c | 
c | 
c | 
} 
\\ 
s  &  
N= 0  &  
N= 1  &  
N= 2  &  
N= 3  &  
N= 4  &  
N= 5  &  
N= 6  &  
N= 7  &  
N= 8  &  
N= 9  &  
N= 10 
\\ \hline 
0  &  
1  &  
  &  
  &  
  &  
  &  
  &  
  &  
  &  
  &  
  &  
\\ \hline 
1  &  
  &  
1  &  
  &  
  &  
  &  
  &  
  &  
  &  
  &  
  &  
\\ \hline 
2  &  
  &  
  &  
1  &  
1  &  
1  &  
  &  
  &  
  &  
  &  
  &  
\\ \hline 
3  &  
  &  
  &  
  &  
1  &  
2  &  
3  &  
3  &  
2  &  
  &  
  &  
\\ \hline 
4  &  
  &  
  &  
  &  
  &  
1  &  
3  &  
6  &  
9  &  
11  &  
8  &  
\\ \hline 
5  &  
  &  
  &  
  &  
  &  
  &  
1  &  
4  &  
10  &  
19  &  
30  &  
38
\\ \hline 
6  &  
  &  
  &  
  &  
  &  
  &  
  &  
1  &  
5  &  
15  &  
34  &  
64
\\ \hline 
7  &  
  &  
  &  
  &  
  &  
  &  
  &  
  &  
1  &  
6  &  
21  &  
55
\\ \hline 
8  &  
  &  
  &  
  &  
  &  
  &  
  &  
  &  
  &  
1  &  
7  &  
28
\\ \hline 
9  &  
  &  
  &  
  &  
  &  
  &  
  &  
  &  
  &  
  &  
1  &  
8
\\ \hline 
10  &  
  &  
  &  
  &  
  &  
  &  
  &  
  &  
  &  
  &  
  &  
1
\\ \hline
\end{array}
.
\end{equation}
In the previous table we note the following patterns
\begin{itemize}
  \item
\begin{equation}
  dim V_{N, s}  =   dim V_{N-1, s} + dim V_{N-1, s-1},
  ~~~~N \ge 4
  .
\end{equation}
\item
  \begin{equation}
    dim V_{N, N}=1
    .
  \end{equation}
\item
  \begin{equation}
    dim V_{N, N}-1=N-2
    .
  \end{equation}
\end{itemize}

The first pattern can be written as
\begin{equation}
 \Delta^{\sN s}=  x\, \left( \Delta^{\sN s} + \Delta^{\sN {s-1} } \right)
 ,
\end{equation}
and it is actually consequence of
\eqref{eq:gen fun dim T N s} and
\eqref{eq:gen fun dim Delta N s} 
when written as $(1-x) \Delta^{\sN s} = x \Delta^{\sN {s-1}}$.

The other two are simply that the leading Regge trajectory is the
totally symmetric tensor and the subleading is the ``pistol''.


\subsection{Computing the dimensions of the vector spaces of states
  where the symmetric group is represented}
The previous analysis has been performed for a generic dimension
and must therefore be performed in the critical dimension in a very
explicit way since constraint equations may have solutions only in
critical dimension.

This has been done using the symbolic computation program maxima.
The result of the analysis is the following table where the actual
dimension of the vector spaces where the $S_s$ act are
\begin{equation} 
\begin{array}{ 
c ||
c |
c |
c |
c |
c |
c |
c |
c |
c |
c |
c |
}
\\
s &
N= 0 &
N= 1 &
N= 2 &
N= 3 &
N= 4 &
N= 5 &
N= 6 &
N= 7 &
N= 8 &
N= 9 &
N= 10
\\ \hline
0 &
0 &
1 &
 &
 &
1 &
 &
1 &
 &
2 &
 &
3 
\\ \hline
1 & 
 &  
 & 
1 & 
 &
 &
1 &
1 &
2 &
2 &
4 &
4 
\\ \hline
2 &
 &
 &
1 &
1 &
1 &
1 &
2 &
3 &
5 &
7 &
11 
\\ \hline
3 &
 &
 &
 &
1 &
2 &
3 &
4 &
6 &
9 &
14 &
21 
\\ \hline
4 &
 &
 &
 &
 &
1 &
3 &
6 &
10 &
16 &
25 &
39 
\\ \hline
5 &
 &
 &
 &
 &
 &
1 &
4 &
10 &
20 &
36 &
61 
\\ \hline
6 &
 &
 &
 &
 &
 &
 &
1 &
5 &
15 &
35 &
71 
\\ \hline
7 &
 &
 &
 &
 &
 &
 &
 &
1 &
6 &
21 &
56 
\\ \hline
8 &
 &
 &
 &
 &
 &
 &
 &
 &
1 &
7 &
28 
\\ \hline
9 &
 &
 &
 &
 &
 &
 &
 &
 &
 &
1 &
8 
\\ \hline
10 &
 &
 &
 &
 &
 &
 &
 &
 &
 &
 &
1 
\\ \hline
\end{array}
\label{eq:D26_true_tensors_vector_spaces_dimensions}
\end{equation}

In the previous table we note the following patterns
\begin{itemize}
  \item
\begin{equation}
  dim V_{N, s}  =   dim V_{N-1, s} + dim V_{N-1, s-1},
  ~~~~N \ge 4
  .
\end{equation}

\item
  \begin{equation}
    dim V_{N=2n, 1} = 2^{ n - 2 },
      ~~~~N \ge 6
    .
  \end{equation}
  
\item
  \begin{equation}
     dim V_{N=2n-1, 1} = 2^{ n - 2 },
      ~~~~N \ge 5
    .
    \end{equation}
\end{itemize}
Notice that the first pattern is already present in the naive table
but it is not obvious why it should persist when the critical
dimension is chosen.

The third pattern is actually not true but a consequence of the first two.

Moreover the second pattern breaks down at level $N=16$, so the number
of vectors is the following
\begin{equation}
  \begin{array}{c|| c| c| c| c| c| c| c|} 
    s  & N=6  & N=8  & N=10  & N=12  & N=14  & N=16 & N=18
    \\
    \hline
    1 & 1 & 2 & 4 & 8 & 16 & 32-1=31 & 64-6=58
  \end{array}
  .
  \label{eq:scalars_up_to_22}
\end{equation}

Looking to the table \ref{eq:scalars_and_vectors_up_to_19}
it seems that the number of scalars at even and odd levels follows two
distinct successions.
In facts the full sequence has no match in The On-Line Encyclopedia of Integer Sequences. 
On the other side the number of scalars at even levels
seems to follow the sequence
$A327475$ (excluding $N=4$ and starting from $N=6$)
but breaks down at level $N=22$
\begin{equation}
  \begin{array}{c|| c| c| c| c| c| c| c| c| c| c|} 
    s  & N=4   & N=6  & N=8  & N=10
        & N=12 & N=14 & N=16
        & N=18 & N=20 & N=22
    \\
    \hline
 0   & (1) & 1 & 2 & 3 
    & 6 & 9 & 16
    & 27 & 46 & 77-1
  \end{array}
  .
\end{equation}
Similarly the odd levels sequence starting from $N=11$ up to $N=21$
seems the sequence $A083322$
\begin{equation}
  \begin{array}{c|| c| c| c| c| c| c| c|} 
    s   & N=11 & N=13  & N=15  & N=17
        & N=19 & N=21 & N=23
    \\
    \hline
 0  & 1 & 2 & 6 & 11 
    & 22& 42& 85?
  \end{array}
  ,
\end{equation}
where the number of scalars at level $N=23$ has not been computed and
$85$ is the number predicted by the sequence $A083322$.

What seems to resist is the relation among the dimensions of the
vector spaces where the symmetric group is represented
\begin{equation}
  \begin{array}
    {c||
      c| c| c| c|
      c| c| c| c|
      c| c| c| c|
      c| c|} 
    s \backslash N
    & 6  & 7  & 8  & 9
    & 10 & 11 & 12 & 13
    & 14 & 15 & 16 & 17
    & 18 & 19
    \\
    \hline
    0
    & 1  & 0  & 2  & 0 
    & 3  & 1  & 6  & 2
    & 9  & 6  & 16 & 11 
    & 27 & 22
    \\
    \hline
    1
    & 1  & 2  & 2  & 4 
    & 4  & 7  & 8  & 14
    & 16 & 25 & 31 & 47 
    & 58 & 85
  \end{array}
  .
\end{equation}
In particular this means that knowing the number of scalars at all
levels $N$ allows to compute the dimensions of the vector spaces where the
symmetric groups act for all $N$ and $s$.
This is not the same of knowing the $SO(D-1)$ irreps but puts strong
constraints.

\section{From states to $S_s$  and $SO(D-1)$ irreps:
  details on the algorithm used}
\label{sec:details_on_irreps}

Once we have determined the vector spaces where the symmetric groups
act we have to split them into irreps from which we can also deduce the
$SO(D-1)$  irreps .
How to do this is not obvious.
The mathematical literature offers the classification of the possible
irreps using Young tableaux.
It is also possible to find the explicit construction of these irreps
using Specht modules associated to Young tabloids.

\ytableausetup{boxsize=1em}
Our problem is different since we have tensors and we cannot use the
previous results.
For example the swap
$i \leftrightarrow k \equiv (i,\, k)$ acts on Young tableaux
and tabloids  by giving a sign
but
this symmetry of the Young diagram
$
\begin{ytableau} i & j \\ k  \end{ytableau}
=
- \begin{ytableau} j & i \\ k  \end{ytableau}
$
is transferred to tensor in a not so straightforward way.
In facts
if we swap two indices we act both on the Young symmetrizer and the tensor
\begin{align}
  T_{
    \begin{ytableau} i & j \\ k  \end{ytableau}
  }
  =
  Y_{
    \begin{ytableau} i & j \\ k  \end{ytableau}
  }
  T_{ i j k}
  =
  T_{ i j k} +  T_{ j i k}
  -
  T_{ k j i} +  T_{ j k i}
  =
  -
  Y_{
    \begin{ytableau} k & j \\ i  \end{ytableau}
  }
  T_{ k j i}
  \ne
  -
  Y_{
    \begin{ytableau} k & j \\ i  \end{ytableau}
  }
  T_{ i j k}
  ,
\end{align}
since the swap $i \leftrightarrow k $ acts both on the Young
symmetrizer
$
  Y_{
    \begin{ytableau} i & j \\ k  \end{ytableau}
  }
$
and the tensor $T_{ i j k}$.

The normalization of Young symmetrizer we use is the simplest one,
i.e. unity.
For example
\begin{align}
  Y_{
    \begin{ytableau} i_1 & i_2 & i_3 \\ j_1 & j_2  \end{ytableau}
    }
    =&
    A_{i_1\, j_1}\, A_{i_2\, j_2}\,
    S_{i_1\, i_2\, i_3}\,     S_{j_1\, j_2}
    ,
 \end{align}
where  $A$ is an antisymmetrizer like   
\begin{align}
    A_{i_1 i_2}=&      
    \sum_{\sigma\in S_2} (-1)^{\sigma}\,
    P_{i_1 \rightarrow \sigma(i_1),\, i_2 \rightarrow \sigma(i_2)}
    ,
\end{align}
and $S$ is a symmetrizer like
\begin{align}
    S_{i_1 i_2 i_3}=&      
    \sum_{\sigma\in S_3}
    P_{i_1 \rightarrow \sigma(i_1),\, i_2 \rightarrow \sigma(i_2),\, i_3 \rightarrow \sigma(i_3)}
    ,
\end{align}
where
$P_{i_1 \rightarrow \sigma(i_1),\, i_2 \rightarrow \sigma(i_2),\, i_3 \rightarrow \sigma(i_3)}$
perform the swaps on the indexes. 

The algorithm we have used to build the irreps is the following.
Given the level $N$ and the group $S_s$ we have the basis $T_{N,s}$.
We take one Young diagram for each irrep and we apply its associated
symmetrizer  to the basis vectors.
We then extract the independent vectors.
On these vectors we apply the swaps $(1,\, k)$ with $2\le k \le s$
and build an eventually bigger vector space.
On this new vector space we apply again the same swaps until the
dimension of the vector space becomes stable.

We do this procedure for all possible irreps of the given symmetric
group $S_s$.
After we have computed all the corresponding vector spaces we check
that we have not missed anything by counting the dimensions and
changing basis from the original basis to the new basis associated to
the irreps.

It turns out that we can immediately find the $SO(D-1)$ irreps without
the need of going through the construction of $GL(D-1)$ irreps.

Actually this is forced on us by the structure of the ``anomalous''
$ \delta_A^{l \downarrow}$ actions.

\subsection{The simplest non trivial example $\ydiagram{2}$ irrep:
general considerations}
To understand how this happens we consider the simplest non trivial case,
i.e. the $\ydiagram{2}$ irrep.
In particular the first interesting case of this irrep
appears at level $N=4$ since the $N=2$ is simpler as we explain in the
next section.
However the discussion in this section is valid for all levels $N$.

Instead of considering immediately the states let us start looking at the
polarizations.
In rest frame only the $\epsilon_{I J}$ polarizations survive since
they are transverse and we have
\begin{equation}
\epsilon_{I J} = \epsilon_{J I},
~~~~
\sum_I \epsilon_{I I} = 0,
~~~~
I, J = 1,2,\dots D-1
.
\end{equation}  
The last equation says that only $D-2$ ``diagonal'' polarizations
$\epsilon_{I I}$ are independent.
There is no canonical way of choosing them.

Let us now rewrite the previous equations from the $SO(D-2)$ point of
view in a way to show the independent components
\begin{align}
\epsilon_{\Yt i j}&, 
\nonumber\\
\sum_i \epsilon_{\Yt i i }  + \epsilon_{\Yt 1 1 } &=0
.
\end{align}

Because of our definition of \lc coordinates the $I=1$ spacial
direction is special and the minimal and simplest approach to get
the corresponding $D-2$ ``diagonal'' states
is actually to consider (no sum over $i$)
\begin{equation}
\epsilon_{i i} = -\epsilon_{1 1},~~~~ i=2,\dots D-1,
\end{equation}
but then the explicit expression for the unnormalized corresponding states
$| {\Yt i i}_{\GL(D-1)} \rangle - | {\Yt 1 1}_{\GL(D-1)} \rangle$ 
is obtained only after some rotations, i.e. by applying $\delta^i\sim
M^{i 1}$ a
couple of times.

However in \lc if we want to obtain immediately {\sl most} but not all
of these states
and do not want to ``dig'' into the irrep there is a
more natural way which is however not unique, i.e.
\begin{equation}
\epsilon_{I I} = -\epsilon_{2 2},~~~~ I \ne 2,
\end{equation}
so that most of the corresponding states
$| {\Yt i i}_{\GL(D-1)} \rangle - | {\Yt 2 2}_{\GL(D-1)} \rangle$
($i \ne 2$)
do not involve the $I=1$ index but only the transverse ones.
The non uniqueness is due to the fact that
we could replace $\epsilon_{2 2}$ with any other
$\epsilon_{i i}$.
Nevertheless with this approach we still need to ``dig'' into the irrep
to find the state corresponding to $\epsilon_{1 1} = -\epsilon_{2 2}$.

Notice however that this approach is the most natural one when the number of
symmetric indexes is more than $D-2$.
This happens because some indexes must be equal and the $SO(D-1)$ is
traceless which requires to start with a subtracted state.
If it were $D=3$ then we should start with
$|\Ytr 3 3 2\rangle - \frac{1}{3} |\Ytr 2 2 2\rangle$.

%

%
Let us start building the previous $SO(D-1)$ states from the $SO(D-2)$ states 
with the most straightforward approach.
We consider the states
\begin{equation}
|{\Yt i j}_{SO(D-1)} \rangle
\equiv
|{\Yt i j}_{SO(D-2)} \gg
\equiv
|{\Yt i j}_{\GL(D-2)} \gg
,~~~~
i\ne j
,
\end{equation}
where the condition $i \ne j$ allows to forget about the trace
condition and consider the $\GL(D-2)$ states as $SO(D-2)$ states
as $SO(D-1)$ states.
In the following 
$|*\rangle$ are the $SO(D-1)$ states and
$|*\gg$ are the $SO(D-2)$ states.

On these states we apply a sequence of $\delta$s as
\begin{align}
\begin{matrix}
|{\Yt i j}_{SO(D-1)} \rangle
& 
\equiv
&
|{\Yt i j}_{SO(D-2)} \gg
\\\\
\downarrow \delta^{i} / ({-\sdap M}) & &
\\\\
|{\Yt 1 j}_{SO(D-1)} \rangle
& 
\equiv
&
|{\Yo j}_{[1]\, SO(D-2)} \gg
\\\\
\downarrow \delta^{j} / ({-\sdap M}) & &
\\\\
- 2\, |{\Yt j j}_{ (1) SO(D-1)} \rangle
=
|{\Yt 1 1}_{\GL(D-1)} \rangle - |{\Yt j j}_{\GL(D-1)} \rangle
& 
\equiv
&
|{\emptyset_{[1][1]}}_{\GL(D-2)} \gg
-
|{\Yt j j}_{ \GL(D-2)} \gg
\end{matrix}
,
\label{eq:Y2_SOD-1_lc}
\end{align}     
so that we obtain the full irrep.
In the previous expressions
the $(1)$ in $\Ytfrom  {\*} {\*} 1 $  refers to the specific choice
of basis, i.e. the subtraction of
$|{\Yt 1 1}_{\GL(D-1)} \rangle$.
In a similar way
f.x. $\Ytfrom  {\*} {\*} {[1][1]}$ means that
the state is obtained by first varying the first original index and
then the first index of the state obtained after the first variation.
This further specification may be necessary since there may be in
principle some
differences between $|{\emptyset_{[1][1]}}_{\GL(D-2)} \gg$
and $|{\emptyset_{[2][1]}}_{\GL(D-2)} \gg$
even if it is not the case with the fully symmetric irreps.

In particular the $SO(D-1)$ states
\begin{equation}
|{\Yt i i }_{(1) SO(D-1)} \rangle
\equiv
\oh \left(
|{\Yt i i}_{\GL(D-1)} \rangle - |{\Yt 1 1}_{\GL(D-1)} \rangle
\right)
,
\end{equation}
are as suggested by the above polarization argument obtained by taking
the difference of two ${\GL(D-1)}$ states, i.e. states for which there
are no constraints on the trace.
The normalization factor $\oh$ is discussed below.

We then set
\begin{equation}
|\Ytfrom 1 1 1\rangle
=
-\sum_{i} |\Ytfrom i i 1\rangle
.
\end{equation}
While not obvious in this basis the difference is fundamental in
obtaining a state which transforms as ${\Yt  {\*} {\*}}_{SO(D-1)} $ and does not
contain any contribution from states with more than $s=2$ indexes.

This point becomes obvious when we now discuss the second approach.
This amounts to a change of basis.
The ``diagonal'' states with $i\ne 2$ which can be computed
immediately without applying $\delta^l$ are
\begin{align}
|\Ytfrom i i 2 \rangle
=
|\Ytfrom i i 1 \rangle - |\Ytfrom 2 2 1 \rangle
\equiv
\oh \left(
|{\Yt i i}_{\GL(D-2)} \gg - |{\Yt 2 2}_{\GL(D-2)} \gg
\right),
\label{eq:s2_basis_choice}
\end{align}
where similarly as before
the $2$ in $\Ytfrom  {\*} {\*} 2 $  refers to the specific choice
of basis.

Notice that we have written
$|{\Yt i i}_{\GL(D-2)} \gg$
but neither
$|{\Yt i i}_{\GL(D-1)} \gg$
nor
$|{\Yt i i}_{SO(D-2)} \gg$.
The reason is the following.
Each $|{\Yt i i}_{\GL(D-2)} \gg$ transforms under the ``anomalous''
$ \delta^{l \downarrow}_A \subset i M^{1 l}$ ($l \ne i$)
to give a state $| l \gg_A$.
\COMMENTOOK{$|{\Yt i i}_{\GL(D-2)} \gg$ or $| i i \rangle$}
This state has no place in such a ``rotation'' under $i M^{1 l}$
of a $| i i \rangle$ state (no Young symmetrizer applied a priori)
which reads
$\delta_{l\, i}\, | 1 i \rangle
+ \delta_{l\, i}\, | i 1\rangle $.
From the $\GL(D-1)$ point of view we can get such a result only when
acting on a state like
$\sum_l|l l\rangle
+ \sum_{l,m}|l l m m\rangle
+ \dots$.
Despite the weird appearance this kind of states are the right ones
since they do no transform under $SO(D-2)$ rotation $i M^{l m}$ so that
$|{\Yt i i}_{\GL(D-2)} \gg$ transforms as expected
$\delta_{l i} |{\Yt m i}_{\GL(D-2)} \gg
-\delta_{m i} |{\Yt l i}_{\GL(D-2)} \gg
+ \dots
$
under such rotations.
As noticed 
$|{\Yt i i}_{\GL(D-2)} \gg$ are almost true $s=2$ states and therefore
$i M^{1 l}$ does not increase the number of indexes
and no index $1$ is allowed.

To the previous states we need to add
\begin{align}
~~~
|\Ytfrom 1 1 2\rangle
=
&
\oh \left(
|{\Yt 1 1}_{\GL(D-1)} \rangle
- |{\Yt  2 2}_{\GL(D-1)} \rangle
\right)
\nonumber\\
=
-|\Ytfrom 2 2 1\rangle
\equiv
&
\oh \left(
|{\emptyset}_{[1][1]\, \GL(D-2)} \gg
-
|{\Yt 2 2 }_{\GL(D-2)} \gg
\right)
,
\label{eq:s2_basis_choice_11}
\end{align}
and
\begin{equation}
|\Ytfrom 2 2 2\rangle
=
-\sum_{i\ne 2} |\Ytfrom i i 2\rangle
-|\Ytfrom 1 1 2\rangle
,
\label{eq:s2_basis_choice_22}
\end{equation}
in order to complete the irrep.

Finally let us discuss the normalizations.
All the states have the same normalization since they are obtained
one from the other by acting with unitary operators.
Given so we can take whichever explicit representation to compute the
normalizations.
Suppose we write
$ |\Yt I J \rangle  = a^{\dagger I}\, a^{\dagger J}\, |0\rangle$
with $[a^I,\, a^{\dagger J}] = n\, \delta^{I J}$ 
and $I\ne J$.
The normalization is then
$\langle \Yt I J | \Yt I J \rangle = n^2$.
Now the ``diagonal'' state 
$ |{\Yt i i}_{(1)} \rangle  =
\oh \left( {a^{i \dagger}}^2 - {a^{1 \dagger}}^2 \right) \, |0\rangle$
has the same normalization.
The same state with the same normalization may be obtained by acting
with the unitary operators
$U_{(L M)} = \exp \left( \frac{1}{n} \left( 
a^{M \dagger}\, a^{L} - a^{L \dagger}\, a^{M}
\right)\right)
$.

\subsection{The simplest non trivial example $\ydiagram{2}$ irrep:
explicit construction and examples}

In the previous section we have discussed the general approach,
now we consider the explicit constructions and some explicit examples.

Suppose we have solved eq. \eqref{eq:almost_true_tensor} for the
coefficients $\hat b^{\sN{N, s=2, a}}$ which give the
almost true states at level $N$ with $s=2$ indexes.
In the case of multiple solutions we consider one solution at the time, 
explicitly
\begin{align}
| i j \rangle
=&
| i j \gg
=
\sum_a \hat b^{\sN{N, s=2, a}}\,e^{\sN {N, s=2, a} }_{i j}
,
\end{align}
where we have no restrictions on $i$ and $j$ and therefore the states
are not states of an irrep.
Moreover even the states which are true tensors
among these almost true tensors with $s=2$ indexes
are not states of an irrep also because they may be decomposed into
$\ydiagram{1,1}$ and $\ydiagram{2}$ states.
The first level where both irreps appear is $N=7$.
For $N=4$ we have only the irrep $\ydiagram{2}$.

We then apply the Young symmetrizer $Y_{\ydiagram{2}}$ and get
the almost true states in the irrep $\ydiagram{2}$ at level $N$
with coefficients $b^{\sN{N, s=2, a}}$,
explicitly
\begin{align}
| {\Yt i j} \rangle
=&
| {\Yt i j} \gg
\nonumber\\
=&
\cN_b\,
\sum_a \hat b^{\sN{N, s=2, a}}\,
(e^{\sN {N, s=2, a} }_{i j} + e^{\sN {N, s=2, a} }_{j i})
=
\sum_a b^{\sN{N, s=2, a}}\, e^{\sN {N, s=2, a} }_{i j}
,
\label{eq:Y2_b_has_symmetrizer}
\end{align}
where these states may or may not belong to the $SO(D-1)$ $\ydiagram{2}$ irrep
and therefore we have not explicitly indicated this.
States with $i \ne j$ do since the traceless condition is
automatically satisfied but
states with $i = j$ do not because the traceless condition.
We have then normalized the state with $\cN_b$ so that the set of
$b^{\sN{N, s=2, a}}$ has not a common divisor.

Given these initial steps
the $SO(D-1)$ $\ydiagram{2}$ states in \eqref{eq:Y2_SOD-1_lc}
with $i \ne j$
are then explicitly computed as 
\begin{align}
| {\Yt i j}_{SO(D-1)} \rangle
=&
| {\Yt i j}_{SO(D-2)} \gg
\nonumber\\
=&
\sum_a b^{\sN{N, s=2, a}}\, e^{\sN {N, s=2, a} }_{i j}
,
\nonumber\\
| {\Yt 1 j}_{SO(D-1)} \rangle
=&
| {\Yo j}_{[1] SO(D-2)} \gg
\nonumber\\
=&
\frac{-1}{\sdap M}
\sum_{a\, b} b^{\sN{N, s=2, a}}\, D^{\sN{ N, s=2, 1}}_{a b} e^{\sN {N, s=1, b} }_{i}
\nonumber\\
\equiv&
\frac{-1}{\sdap M}
\sum_{a\, b} b^{\sN{N, s=2\rightarrow1, a}}\, e^{\sN {N, s=1, b} }_{i}
,
\nonumber\\
| {\Ytfrom j j 1}_{SO(D-1)} \rangle
=&
-\oh \left[
|{\Yt 1 1}_{\GL(D-1)} \rangle - |{\Yt j j}_{\GL(D-1)} \rangle
\right]
\nonumber\\
=&-\oh
\left(\frac{-1}{\sdap M}\right)^2
\sum_{a\, b} b^{\sN{N, s=2, a}}\,
\Bigg[
\left( D^{\sN{ N, s=2, 1}} D^{\sN{ N, s=1}} \right) _{a b} e^{\sN {N, s=0, b} }
\nonumber\\
&
\phantom{-\oh\left(\frac{-1}{\sdap M}\right)^2}
+
\left( D^{\sN{ N, s=2, 1}} U^{\sN{ N, s=1}} \right) _{a b} e^{\sN {N,s=2, b}}_{ j j }
\Bigg]
\nonumber\\
\equiv&
\left(\frac{-1}{\sdap M}\right)^2
\left[
\sum_{a} b^{\sN{N, s=2 \rightarrow 0 , a}}\, e^{\sN {N, s=0, a} }
+
\sum_{a} b^{\sN{N, s=2\rightarrow 1 \rightarrow 2 \rightarrow 0 ,a}}\,
e^{\sN {N, s=2, a} }_{j j}
\right]
,
\label{eq:irrep_Y2_explicit}
\end{align}
where we have defined the descendants  of $b^{\sN{N, s=2}}$
to be $b^{\sN{N, s=2 \rightarrow 1}}$ and $b^{\sN{N, s=2 \rightarrow 0}}$.

Let us see this explicitly for $N=4$
for which the solution in matricidal form  of
eq. \eqref{eq:almost_true_tensor} is
\begin{align}
\hat b^{\sN{ N=4,\, s=2}}=
b^{\sN{ N=4,\, s=2}}=
 &
\left( 
\begin{matrix} 
-1
&
-7
&
4
&
4
\end{matrix}
\right) 
,  \end{align}
or with the tensor structures displayed
\begin{align}
\bce^{\sN{N=4,\, s=2}}_{i_1 i_2}
=&
\left( 
\begin{matrix} 
-\left(1 , 1\right)\,{1}^{i_{1}}\,{1}^{i_{2}}
&
-7\,{2}^{i_{1}}\,{2}^{i_{2}}
&
+4\,{1}^{i_{2}}\,{3}^{i_{1}}
&
+4\,{1}^{i_{1}}\,{3}^{i_{2}}
\end{matrix}
\right) 
\nonumber\\
=&
\sum_a b^{\sN{N=4,\, s=2, a}}\, e^{\sN {N=4, s=2, a} }_{i_1 i_2}
.
\end{align}
If the Young symmetrizer $Y_{\ydiagram{1,1}}$ is applied we get zero
since this state is obviously symmetric.
So far we have not constrained $i_1$ and $i_2$.
The previous states are all almost true $s=2$ tensor states.
As discussed in the previous section and above the $i_1 \ne i_2$ are true  $s=2$
tensor states but not the  $i_1 = i_2$ ones.

Applying the procedure described above we get
\begin{align}
| {\Yt i j}_{SO(D-1)}
=&
\bce^{\sN{N=4,\, s=2}}_{i j}
=
\left( 
\begin{matrix} 
-\left(1 , 1\right)\,{1}^{i}\,{1}^{j}
&
-7\,{2}^{i}\,{2}^{j}
&
+4\,{1}^{i}\,{3}^{j}
&
+4\,{1}^{i}\,{3}^{j}
\end{matrix}
\right)
\label{eq:N4_Y2_2}
,
\end{align}
and
\begin{align}
| {\Yt 1 j}_{SO(D-1)}
 =&
 \bce^{\sN{N=4,\, s=2\rightarrow 1}}_j
=
\frac{1}{(- \sdap M)}
\left( 
\begin{matrix} 
2\,{1}^{j}\,\left(2 , 1\right)
&
-\frac{9}{2}\,\left(1 , 1\right)\,{2}^{j}
&
+2\,{4}^{j}
\end{matrix}
\right) 
,
\label{eq:N4_Y2_1}
\end{align}
with $\ap M^2 = N-1 = 3$
and the general expression for the ``diagonal'' states (no sum over $i$)
\begin{align}
|\Ytfrom i i 1\rangle
=&
\frac{1}{(- \sdap M)^2}
\left( 
\begin{matrix} 
\oh \left(1 , 1\right)\,{1}^{i}\,{1}^{i}
&
+\frac{7}{2}\,{2}^{i}\,{2}^{i}
&
-4\,{1}^{i}\,{3}^{i}
&
-\frac{3}{16} (1,1)^2
&
+\frac{1}{4} (2,2)
&
-\frac{1}{4} (3,1)
\end{matrix}
\right) 
.
\end{align}     

All these states have the same norm in critical dimension $D=26$
\begin{equation}
\langle\Yt i j| \Yt i j \rangle
=
\langle\Ytfrom i i 1| \Ytfrom i i 1\rangle
=
\langle\Yt i 1| \Yt i 1\rangle
=
348
,
\end{equation}
and are orthogonal.

Because of the identities discussed in
appendix \ref{app:Lorentz_algebra}
from the states in \eqref{eq:N4_Y2_1} we can go back as
\begin{align}
\delta^{j \uparrow} \bce^{\sN{N=4,\, s=2 \rightarrow 1 }}_{i}
=&
(+ \sdap M) \bce^{\sN{N=4,\, s=2 }}_{i j}
\nonumber\\
=&
\frac{1}{- \sdap M}
\sum_{a\, b} b^{\sN{N=4, s=2, a}}\,
\left( D^{\sN{ N=4, s=2, 2}}\, U^{\sN{ N=4, s=1}}\right)_{a b}
e^{\sN {N=4, s=2, b} }_{i j}
.
\end{align}
Notice however that this state is not generically the one entering the last
equation of eq. \eqref{eq:irrep_Y2_explicit} since here we make use of
$D^{\sN{ N=4, s=2, 2}}$ while there of $D^{\sN{ N=4, s=2, 1}}$.
The two states are connected by a reshuffling of the indexes and
therefore are equal in this specific case due to the symmetric nature
of the tensor.

To see explicitly the general discussion of the previous
section on the necessity of taking the difference of states when two
or more indexes are equal  let us consider the action of
$\delta^{l \downarrow}$ on the states \eqref{eq:N4_Y2_2} with $i =j$.

We will consider the ``diagonal'' states
$|\Ytfrom I I 2\rangle$ since they are more representative of the
states with more than $D-1=25$ indexes since
generically these states have at least two equal indexes and there is
no way of writing a $SO(D-1)$ state without a subtraction.

The action of $\delta^{l \downarrow}$ on states $\bce^{\sN{N=4,\, s=2}}_{i i}$ (no sum)
includes the ``anomalous action'' so for $j \ne i$
\begin{align}
\delta_A^{j \downarrow} \bce^{\sN{N=4,\, s=2}}_{i i}
=&
(- \sdap M) a^{\sN{N=4,\, s=2 \rightarrow 1}}_{j}
\nonumber\\
=&
\sum_{a\, b} b^{\sN{N=4,\, s=2, a}}\,
A^{\sN{ N=4,\, s=2, 1 2}}_{a b} e^{\sN {N=4, s=1, b} }_{i}
,
\end{align}
and for $j = i$
\begin{align}
\delta^{i \downarrow} \bce^{\sN{N=4,\, s=2}}_{i i}
=&
2 (- \sdap M) b^{\sN{N=4~ s=2 \rightarrow 1}}_{i}
+
(- \sdap M) a^{\sN{N=4~ s=2 \rightarrow 1}}_{i}
\nonumber\\
=&
2 \sum_{a\, b} b^{\sN{N=4, s=2, a}}\, D^{\sN{ N=4, s=2, 1}}_{a b}
e^{\sN {N=4, s=1, b} }_{i}
\nonumber\\
&+
\sum_{a\, b} b^{\sN{N=4, s=2, a}}\,
A^{\sN{ N=4, s=2, 1 2}}_{a b} e^{\sN {N=4, s=1, b} }_{i}
,
\end{align}
where $\ap M^2 = N-1 = 3$
and
$a^{\sN{N=4~ s=2 \rightarrow 1}}_{j}$ is actually a $SO(D-1)$
tensor with at least $3$ indexes since increasing the number of
indexes with $U^{\sN{ N=4,s=1}}$ does not yield a almost true tensor
with $s=2$ indexes, i.e. applying
$U^{\sN{ N=4,s=1}} U^{\sN{N=4,s=2}}$ does not give zero.
This does not happen for the $N=2$ case since there are no states with
more indexes and it is the reason why we started looking to the $N=4$ case.

In order to get a state which does not have this higher $s$ components
we are forced to consider some combinations which cancel the
``anomalous'' higher $s$ component, f.x.
($i \ne 2 $)
\begin{align}
\oh&\left(
\bce^{\sN{N=4,\, s=2}}_{i i} - \bce^{\sN{N=4,\, s=2}}_{2 2}\right)
=
\nonumber\\
 =&
\oh
\left( 
\begin{matrix} 
-\left(1 , 1\right)\,{1}^{i}\,{1}^{i}
&
-7\,{2}^{i}\,{2}^{i}
&
+4\,{1}^{i}\,{3}^{i}
&
+4\,{1}^{i}\,{3}^{i}
\end{matrix}
\right)
\nonumber\\
&
-\oh
\left( 
\begin{matrix} 
-\left(1 , 1\right)\,{1}^{2}\,{1}^{2}
&
-7\,{2}^{2}\,{2}^{2}
&
+4\,{1}^{2}\,{3}^{2}
&
+4\,{1}^{2}\,{3}^{2}
\end{matrix}
\right)
.
\end{align}
This is the kind of state we discussed above in eq. \eqref{eq:s2_basis_choice}
and satisfies ($i, j\ne 2$)
\begin{align}
\delta^{j \downarrow} ( \bce^{\sN{N=4,\, s=2}}_{i i} - \bce^{\sN{N=4,\, s=2}}_{2 2} )
=
2\, \delta_{i j} (- \sdap M) \bce^{\sN{N=4,\, s=2 \rightarrow 1}}_{i},~~~~
\nonumber\\
\delta^{j=2 \downarrow} ( \bce^{\sN{N=4,\, s=2}}_{i i} - \bce^{\sN{N=4,\, s=2}}_{2 2} )
=
2\, ( \sdap M) \bce^{\sN{N=4,\, s=2 \rightarrow 1}}_{2},~~~~
\end{align}
so that they involve only the $b^{\sN{N=4~ s=2 \rightarrow 1}}_{i}$ vectors.

Up to this point we have found the following $SO(D-1)$ states
\begin{align}
|\Yt i j \rangle = \bce^{\sN{N=4,\, s=2}}_{i j},~~~~
|\Yt i 1 \rangle = \bce^{\sN{N=4~,\, s=2\rightarrow 1}}_{i},~~~~
|\Ytfrom i i 2\rangle = \oh ( \bce^{\sN{N=4,\, s=2}}_{i i} - \bce^{\sN{N=4,\, s=2}}_{22} )
,
\end{align}
where the $\oh$ in $|\Ytfrom i i 2 \rangle$ is due to the normalization
but we are still missing the 
$|\Ytfrom 1 1 2\rangle$ state.
This is easily obtained with a further ``rotation'' as 
\begin{align}
\delta^{i=2} \bce^{\sN{N=4,\, s=2\rightarrow 1}}_{i=2 }
=&
(- \sdap M) \bce^{\sN{N=4,\, s=2 \rightarrow 0}}_{\emptyset}
+
( \sdap M) \bce^{\sN{N=4,\, s=2 }}_{2 2}
\nonumber\\
=&
(- \sdap M) |\Ytfrom 1 1 2\rangle
=
(+ \sdap M) |\Ytfrom 2 2 1\rangle
\nonumber\\
=&
\frac{1}{-2 \sdap M}
\sum_{a\, b} b^{\sN{N=4, s=2, a}}\,
\left( D^{\sN{ N=4, s=2, 1}}\, D^{\sN{ N=4, s=1 }}\right)_{a b}
e^{\sN {N=4, s=0, b} }
\nonumber\\
&+
\frac{1}{-2 \sdap M}
\sum_{a\, b} b^{\sN{N=4, s=2, a}}\,
\left( D^{\sN{ N=4, s=2, 1}}\, U^{\sN{ N=4, s=1 }}\right)_{a b}
e^{\sN {N=4, s=2, b} }_{i i}
.
\end{align}

All the other variations can then be expressed using these states,
f.x.
\begin{align}
\delta^{j} b^{\sN{N=4~ s=2\rightarrow 1}}_{i }
=&
\delta_{i j}\, (- \sdap M) b^{\sN{N=4~ s=2 \rightarrow 0}}_{\emptyset}
+
( \sdap M) b^{\sN{N=4~ s=2 }}_{i j}
\nonumber\\
=&
\delta_{i j} 
(- \sdap M) \left( |\Ytfrom 1 1 2> - |\Ytfrom i i 2\rangle \right)
+
(1 - \delta_{i j} )\, 
(\sdap M) |\Yt i j > 
.
\end{align}

In particular these states have the same norm of the previously
considered, i.e.
\begin{equation}
\langle\Ytfrom i i 2| \Ytfrom i i 2\rangle
=
\langle\Ytfrom 1 1 2| \Ytfrom 1 1 2\rangle
=
348
.
\end{equation}


\subsection{A typical high level $N$ state}
There are $4$  $s=2$ indices $SO(25)$ $\ydiagram{2}$ level $N=8$ states
with $i_1 \ne i_2$ (so these are true spin 2 tensor states).
They are

\myscaleM{ 
\parbox{\linewidth}{ 
  \begin{align}
\mkern-200mu    
\bce^\sN{N=8,\, s=2->2}_{i_1 i_2}
=
 &
\left( 
\begin{matrix} 
+45046685248\,{1}^{i_{1}}\,{1}^{i_{2}}\,\left(5 , 1\right)
&
-116302234048\,{1}^{i_{1}}\,{1}^{i_{2}}\,\left(4 , 2\right)
&
+85370416576\,{1}^{i_{1}}\,{1}^{i_{2}}\,\left(3 , 3\right)
&
-12373813824\,\left(1 , 1\right)\,{1}^{i_{1}}\,{1}^{i_{2}}\,\left(3 , 1\right)
\\
+6956634688\,{1}^{i_{1}}\,{1}^{i_{2}}\,\left(5 , 1\right)
&
-6154773088\,{1}^{i_{1}}\,{1}^{i_{2}}\,\left(4 , 2\right)
&
+2616926056\,{1}^{i_{1}}\,{1}^{i_{2}}\,\left(3 , 3\right)
&
-1137379944\,\left(1 , 1\right)\,{1}^{i_{1}}\,{1}^{i_{2}}\,\left(3 , 1\right)
\\
+100558816\,{1}^{i_{1}}\,{1}^{i_{2}}\,\left(5 , 1\right)
&
-38085279616\,{1}^{i_{1}}\,{1}^{i_{2}}\,\left(4 , 2\right)
&
+31889928192\,{1}^{i_{1}}\,{1}^{i_{2}}\,\left(3 , 3\right)
&
-9232582608\,\left(1 , 1\right)\,{1}^{i_{1}}\,{1}^{i_{2}}\,\left(3 , 1\right)
\\
+18032524384\,{1}^{i_{1}}\,{1}^{i_{2}}\,\left(5 , 1\right)
&
-46541020384\,{1}^{i_{1}}\,{1}^{i_{2}}\,\left(4 , 2\right)
&
+34045910608\,{1}^{i_{1}}\,{1}^{i_{2}}\,\left(3 , 3\right)
&
-4890952992\,\left(1 , 1\right)\,{1}^{i_{1}}\,{1}^{i_{2}}\,\left(3 , 1\right)
\end{matrix}
\right . 
\nonumber
\end{align}
} 
} 


\myscaleM{ 
\parbox{\linewidth}{ 
\begin{align}
&
\begin{matrix} 
0
&
+6409364192\,{1}^{i_{1}}\,{1}^{i_{2}}\,\left(2 , 1\right)^2
&
-115708932\,\left(1 , 1\right)^3\,{1}^{i_{1}}\,{1}^{i_{2}}
&
-5233831680\,{1}^{i_{1}}\,{2}^{i_{2}}\,\left(4 , 1\right)
\\
0
&
-176828848\,{1}^{i_{1}}\,{1}^{i_{2}}\,\left(2 , 1\right)^2
&
+15278358\,\left(1 , 1\right)^3\,{1}^{i_{1}}\,{1}^{i_{2}}
&
-6330836880\,{1}^{i_{1}}\,{2}^{i_{2}}\,\left(4 , 1\right)
\\
+7513517400\,\left(1 , 1\right)\,{1}^{i_{1}}\,{1}^{i_{2}}\,\left(2 , 2\right)
&
+2257854464\,{1}^{i_{1}}\,{1}^{i_{2}}\,\left(2 , 1\right)^2
&
+519575656\,\left(1 , 1\right)^3\,{1}^{i_{1}}\,{1}^{i_{2}}
&
+2734770240\,{1}^{i_{1}}\,{2}^{i_{2}}\,\left(4 , 1\right)
\\
0
&
+3363868736\,{1}^{i_{1}}\,{1}^{i_{2}}\,\left(2 , 1\right)^2
&
+169953144\,\left(1 , 1\right)^3\,{1}^{i_{1}}\,{1}^{i_{2}}
&
-1180051440\,{1}^{i_{1}}\,{2}^{i_{2}}\,\left(4 , 1\right)
\end{matrix}
\nonumber
\end{align}
} 
} 


\myscaleM{ 
\parbox{\linewidth}{ 
\begin{align}
&
\begin{matrix} 
-3473944320\,{1}^{i_{1}}\,{2}^{i_{2}}\,\left(3 , 2\right)
&
+685630176\,\left(1 , 1\right)\,{1}^{i_{1}}\,\left(2 , 1\right)\,{2}^{i_{2}}
&
-5233831680\,{1}^{i_{2}}\,{2}^{i_{1}}\,\left(4 , 1\right)
&
-3473944320\,{1}^{i_{2}}\,{2}^{i_{1}}\,\left(3 , 2\right)
\\
+3311438280\,{1}^{i_{1}}\,{2}^{i_{2}}\,\left(3 , 2\right)
&
+792981756\,\left(1 , 1\right)\,{1}^{i_{1}}\,\left(2 , 1\right)\,{2}^{i_{2}}
&
-6330836880\,{1}^{i_{2}}\,{2}^{i_{1}}\,\left(4 , 1\right)
&
+3311438280\,{1}^{i_{2}}\,{2}^{i_{1}}\,\left(3 , 2\right)
\\
+1815197760\,{1}^{i_{1}}\,{2}^{i_{2}}\,\left(3 , 2\right)
&
-1473163008\,\left(1 , 1\right)\,{1}^{i_{1}}\,\left(2 , 1\right)\,{2}^{i_{2}}
&
+2734770240\,{1}^{i_{2}}\,{2}^{i_{1}}\,\left(4 , 1\right)
&
+1815197760\,{1}^{i_{2}}\,{2}^{i_{1}}\,\left(3 , 2\right)
\\
-783256560\,{1}^{i_{1}}\,{2}^{i_{2}}\,\left(3 , 2\right)
&
+942293808\,\left(1 , 1\right)\,{1}^{i_{1}}\,\left(2 , 1\right)\,{2}^{i_{2}}
&
-1180051440\,{1}^{i_{2}}\,{2}^{i_{1}}\,\left(4 , 1\right)
&
-783256560\,{1}^{i_{2}}\,{2}^{i_{1}}\,\left(3 , 2\right)
\end{matrix}
\nonumber
\end{align}
} 
} 


\myscaleM{ 
\parbox{\linewidth}{ 
\begin{align}
&
\begin{matrix} 
+685630176\,\left(1 , 1\right)\,{1}^{i_{2}}\,\left(2 , 1\right)\,{2}^{i_{1}}
&
-99511053472\,{1}^{i_{1}}\,\left(3 , 1\right)\,{3}^{i_{2}}
&
+61355799120\,{1}^{i_{1}}\,\left(2 , 2\right)\,{3}^{i_{2}}
&
+7806831960\,\left(1 , 1\right)^2\,{1}^{i_{1}}\,{3}^{i_{2}}
\\
+792981756\,\left(1 , 1\right)\,{1}^{i_{2}}\,\left(2 , 1\right)\,{2}^{i_{1}}
&
-5772838032\,{1}^{i_{1}}\,\left(3 , 1\right)\,{3}^{i_{2}}
&
+2988972120\,{1}^{i_{1}}\,\left(2 , 2\right)\,{3}^{i_{2}}
&
+354792960\,\left(1 , 1\right)^2\,{1}^{i_{1}}\,{3}^{i_{2}}
\\
-1473163008\,\left(1 , 1\right)\,{1}^{i_{2}}\,\left(2 , 1\right)\,{2}^{i_{1}}
&
+8367533376\,{1}^{i_{1}}\,\left(3 , 1\right)\,{3}^{i_{2}}
&
-9882502560\,{1}^{i_{1}}\,\left(2 , 2\right)\,{3}^{i_{2}}
&
-153262080\,\left(1 , 1\right)^2\,{1}^{i_{1}}\,{3}^{i_{2}}
\\
+942293808\,\left(1 , 1\right)\,{1}^{i_{2}}\,\left(2 , 1\right)\,{2}^{i_{1}}
&
-38907399776\,{1}^{i_{1}}\,\left(3 , 1\right)\,{3}^{i_{2}}
&
+24952444560\,{1}^{i_{1}}\,\left(2 , 2\right)\,{3}^{i_{2}}
&
+66132480\,\left(1 , 1\right)^2\,{1}^{i_{1}}\,{3}^{i_{2}}
\end{matrix}
\nonumber
\end{align}
} 
} 


\myscaleM{ 
\parbox{\linewidth}{ 
\begin{align}
&
\begin{matrix} 
-99511053472\,{1}^{i_{2}}\,\left(3 , 1\right)\,{3}^{i_{1}}
&
+61355799120\,{1}^{i_{2}}\,\left(2 , 2\right)\,{3}^{i_{1}}
&
+7806831960\,\left(1 , 1\right)^2\,{1}^{i_{2}}\,{3}^{i_{1}}
&
+150274348752\,{2}^{i_{1}}\,{2}^{i_{2}}\,\left(3 , 1\right)
\\
-5772838032\,{1}^{i_{2}}\,\left(3 , 1\right)\,{3}^{i_{1}}
&
+2988972120\,{1}^{i_{2}}\,\left(2 , 2\right)\,{3}^{i_{1}}
&
+354792960\,\left(1 , 1\right)^2\,{1}^{i_{2}}\,{3}^{i_{1}}
&
+16085639412\,{2}^{i_{1}}\,{2}^{i_{2}}\,\left(3 , 1\right)
\\
+8367533376\,{1}^{i_{2}}\,\left(3 , 1\right)\,{3}^{i_{1}}
&
-9882502560\,{1}^{i_{2}}\,\left(2 , 2\right)\,{3}^{i_{1}}
&
-153262080\,\left(1 , 1\right)^2\,{1}^{i_{2}}\,{3}^{i_{1}}
&
-32120724816\,{2}^{i_{1}}\,{2}^{i_{2}}\,\left(3 , 1\right)
\\
-38907399776\,{1}^{i_{2}}\,\left(3 , 1\right)\,{3}^{i_{1}}
&
+24952444560\,{1}^{i_{2}}\,\left(2 , 2\right)\,{3}^{i_{1}}
&
+66132480\,\left(1 , 1\right)^2\,{1}^{i_{2}}\,{3}^{i_{1}}
&
+58850361816\,{2}^{i_{1}}\,{2}^{i_{2}}\,\left(3 , 1\right)
\end{matrix}
\nonumber
\end{align}
} 
} 


\myscaleM{ 
\parbox{\linewidth}{ 
\begin{align}
&
\begin{matrix} 
-87445609944\,\left(2 , 2\right)\,{2}^{i_{1}}\,{2}^{i_{2}}
&
-12571597368\,\left(1 , 1\right)^2\,{2}^{i_{1}}\,{2}^{i_{2}}
&
+11953413472\,{1}^{i_{1}}\,\left(2 , 1\right)\,{4}^{i_{2}}
&
+11953413472\,{1}^{i_{2}}\,\left(2 , 1\right)\,{4}^{i_{1}}
\\
-10240343064\,\left(2 , 2\right)\,{2}^{i_{1}}\,{2}^{i_{2}}
&
-1081090683\,\left(1 , 1\right)^2\,{2}^{i_{1}}\,{2}^{i_{2}}
&
+1035492232\,{1}^{i_{1}}\,\left(2 , 1\right)\,{4}^{i_{2}}
&
+1035492232\,{1}^{i_{2}}\,\left(2 , 1\right)\,{4}^{i_{1}}
\\
+18825004152\,\left(2 , 2\right)\,{2}^{i_{1}}\,{2}^{i_{2}}
&
+2893492344\,\left(1 , 1\right)^2\,{2}^{i_{1}}\,{2}^{i_{2}}
&
+4531582624\,{1}^{i_{1}}\,\left(2 , 1\right)\,{4}^{i_{2}}
&
+4531582624\,{1}^{i_{2}}\,\left(2 , 1\right)\,{4}^{i_{1}}
\\
-35606501052\,\left(2 , 2\right)\,{2}^{i_{1}}\,{2}^{i_{2}}
&
-296059644\,\left(1 , 1\right)^2\,{2}^{i_{1}}\,{2}^{i_{2}}
&
-4919416424\,{1}^{i_{1}}\,\left(2 , 1\right)\,{4}^{i_{2}}
&
-4919416424\,{1}^{i_{2}}\,\left(2 , 1\right)\,{4}^{i_{1}}
\end{matrix}
\nonumber
\end{align}
} 
} 


\myscaleM{ 
\parbox{\linewidth}{ 
\begin{align}
&
\begin{matrix} 
-8099419328\,\left(2 , 1\right)\,{2}^{i_{1}}\,{3}^{i_{2}}
&
-8099419328\,\left(2 , 1\right)\,{2}^{i_{2}}\,{3}^{i_{1}}
&
+16509498048\,\left(1 , 1\right)\,{1}^{i_{1}}\,{5}^{i_{2}}
&
+16509498048\,\left(1 , 1\right)\,{1}^{i_{2}}\,{5}^{i_{1}}
\\
+552215832\,\left(2 , 1\right)\,{2}^{i_{1}}\,{3}^{i_{2}}
&
+552215832\,\left(2 , 1\right)\,{2}^{i_{2}}\,{3}^{i_{1}}
&
+1609754688\,\left(1 , 1\right)\,{1}^{i_{1}}\,{5}^{i_{2}}
&
+1609754688\,\left(1 , 1\right)\,{1}^{i_{2}}\,{5}^{i_{1}}
\\
+1032786624\,\left(2 , 1\right)\,{2}^{i_{1}}\,{3}^{i_{2}}
&
+1032786624\,\left(2 , 1\right)\,{2}^{i_{2}}\,{3}^{i_{1}}
&
-4350695184\,\left(1 , 1\right)\,{1}^{i_{1}}\,{5}^{i_{2}}
&
-4350695184\,\left(1 , 1\right)\,{1}^{i_{2}}\,{5}^{i_{1}}
\\
+4354329776\,\left(2 , 1\right)\,{2}^{i_{1}}\,{3}^{i_{2}}
&
+4354329776\,\left(2 , 1\right)\,{2}^{i_{2}}\,{3}^{i_{1}}
&
+12675913584\,\left(1 , 1\right)\,{1}^{i_{1}}\,{5}^{i_{2}}
&
+12675913584\,\left(1 , 1\right)\,{1}^{i_{2}}\,{5}^{i_{1}}
\end{matrix}
\nonumber
\end{align}
} 
} 


\myscaleM{ 
\parbox{\linewidth}{ 
\begin{align}
&
\begin{matrix} 
-8899090200\,\left(1 , 1\right)\,{2}^{i_{1}}\,{4}^{i_{2}}
&
-8899090200\,\left(1 , 1\right)\,{2}^{i_{2}}\,{4}^{i_{1}}
&
-6971443712\,\left(1 , 1\right)\,{3}^{i_{1}}\,{3}^{i_{2}}
&
-140858127872\,{4}^{i_{1}}\,{4}^{i_{2}}
\\
-2447956350\,\left(1 , 1\right)\,{2}^{i_{1}}\,{4}^{i_{2}}
&
-2447956350\,\left(1 , 1\right)\,{2}^{i_{2}}\,{4}^{i_{1}}
&
+842309928\,\left(1 , 1\right)\,{3}^{i_{1}}\,{3}^{i_{2}}
&
-4971342632\,{4}^{i_{1}}\,{4}^{i_{2}}
\\
+469024200\,\left(1 , 1\right)\,{2}^{i_{1}}\,{4}^{i_{2}}
&
+469024200\,\left(1 , 1\right)\,{2}^{i_{2}}\,{4}^{i_{1}}
&
+5878816896\,\left(1 , 1\right)\,{3}^{i_{1}}\,{3}^{i_{2}}
&
+38667063376\,{4}^{i_{1}}\,{4}^{i_{2}}
\\
-29106609450\,\left(1 , 1\right)\,{2}^{i_{1}}\,{4}^{i_{2}}
&
-29106609450\,\left(1 , 1\right)\,{2}^{i_{2}}\,{4}^{i_{1}}
&
+36244429904\,\left(1 , 1\right)\,{3}^{i_{1}}\,{3}^{i_{2}}
&
+68057946724\,{4}^{i_{1}}\,{4}^{i_{2}}
\end{matrix}
\nonumber
\end{align}
} 
} 


\myscaleM{ 
\parbox{\linewidth}{ 
\begin{align}
&
\begin{matrix} 
+76450201952\,{3}^{i_{2}}\,{5}^{i_{1}}
&
+76450201952\,{3}^{i_{1}}\,{5}^{i_{2}}
&
+582046080\,{2}^{i_{2}}\,{6}^{i_{1}}
&
+582046080\,{2}^{i_{1}}\,{6}^{i_{2}}
\\
+2302898112\,{3}^{i_{2}}\,{5}^{i_{1}}
&
+2302898112\,{3}^{i_{1}}\,{5}^{i_{2}}
&
+704042280\,{2}^{i_{2}}\,{6}^{i_{1}}
&
+704042280\,{2}^{i_{1}}\,{6}^{i_{2}}
\\
-20510913216\,{3}^{i_{2}}\,{5}^{i_{1}}
&
-20510913216\,{3}^{i_{1}}\,{5}^{i_{2}}
&
-304129440\,{2}^{i_{2}}\,{6}^{i_{1}}
&
-304129440\,{2}^{i_{1}}\,{6}^{i_{2}}
\\
-41261960384\,{3}^{i_{2}}\,{5}^{i_{1}}
&
-41261960384\,{3}^{i_{1}}\,{5}^{i_{2}}
&
+7644749040\,{2}^{i_{2}}\,{6}^{i_{1}}
&
+7644749040\,{2}^{i_{1}}\,{6}^{i_{2}}
\end{matrix}
\nonumber
\end{align}
} 
} 


\myscaleM{ 
\parbox{\linewidth}{ 
\begin{align}
&
\left. 
\begin{matrix} 
-7458513760\,{1}^{i_{2}}\,{7}^{i_{1}}
&
-7458513760\,{1}^{i_{1}}\,{7}^{i_{2}}
\\
-598874560\,{1}^{i_{2}}\,{7}^{i_{1}}
&
-598874560\,{1}^{i_{1}}\,{7}^{i_{2}}
\\
+842905280\,{1}^{i_{2}}\,{7}^{i_{1}}
&
+842905280\,{1}^{i_{1}}\,{7}^{i_{2}}
\\
-1322543680\,{1}^{i_{2}}\,{7}^{i_{1}}
&
-1322543680\,{1}^{i_{1}}\,{7}^{i_{2}}
\end{matrix}
\right) 
.
\end{align}
} 
} 

Notice that these states are independent but not orthogonal.

This example shows a typical feature of the low $s$ high level $N$ states:
the presence of enormous numbers which cannot be eliminated by any
obvious state recombinations.
This happens only for $s\le \oh N$ where in generic dimension
the number of constraints exceed the number of independent variables.
In critical dimension there are however solutions which are obtained
for example using the echelon approach.
This requires making a number of row combinations of the order of
independent variables which grow exponentially, thus transforming
small numbers of the order of the independent variables into numbers
with thousands of figures at level $N\sim 20$.

This is the cause or at least one of the causes of the presence of
chaos in certain classes of amplitudes as discussed below.

\subsection{The $\ydiagram{2,1}$ irrep}

Again we can start looking at the $SO(D-1)$ polarization tensors.
They satisfy
\begin{align}
\epsilon_{\Yto I J K} =& -\epsilon_{\Yto K J I},
\nonumber\\
\sum_I \epsilon_{\Yto I I K} =& 0
, 
\end{align}
along with
\begin{equation}
\epsilon_{\Yto I J K} - \epsilon_{\Yto J I K} = \epsilon_{\Yto I K J}
,
\end{equation}
whose consistency can be checked by setting $J=K$.
As a first step we use the previous relations from the
$SO(D-2)$ point of view in a way to reveal the independent components
($i, j,k$ are all different)
\begin{align}
\epsilon_{\Yto i j k}&,
&
&
\nonumber\\
\epsilon_{\Yto 1 j i} =& - \epsilon_{\Yto i j 1},
&
\epsilon_{\Yto i 1 j} =&  \epsilon_{\Yto i j 1} - \epsilon_{\Yto j i 1},
\nonumber\\
\sum_i \epsilon_{\Yto i i 1}&=0,
&
\sum_i \epsilon_{\Yto i i j} + \epsilon_{\Yto 1 1 j}&=0.
\label{eq:eps_Y21_from_soD-2}
\end{align}
The equations of the second line shows that only $\epsilon_{\Yto 1 j
i}$ is independent.
Again there is no canonical way of solving the equations in the last
line.
A possible solution which we discuss below is (no sum over $i$)
\begin{align}
\epsilon_{\Yto i i 1}&=-\epsilon_{\Yto 2 2 1},~~~~(i\ne 2)
&
\epsilon_{\Yto i i j}&= - \epsilon_{\Yto 1 1 j}
.
\end{align}

Let us start building the previous $SO(D-1)$ states from the $SO(D-2)$ states 
with the most straightforward approach as done for $\Yt {\,} {\,}$.
We consider the states
\begin{equation}
|{\Yto i j k}_{SO(D-1)} \rangle
\equiv
|{\Yto i j k}_{SO(D-2)} \gg
\equiv
|{\Yto i j k}_{\GL(D-2)} \gg
,~~~~
i \ne j \ne k \ne i 
,
\end{equation}
where the condition $i \ne j \ne k $ allows to forget about the trace
condition and consider the $\GL(D-2)$ states as $SO(D-2)$ states
as $SO(D-1)$ states.
As before in the following 
$|*\rangle$ are the $SO(D-1)$ states and
$|*\gg$ are the $SO(D-2)$ states.

On these states we apply a sequence of $\delta$s as
\begin{align}
\begin{matrix}
|{\Yto i j k}_{SO(D-1)} \rangle
& 
\equiv
&
|{\Yto i j k}_{SO(D-2)} \gg
\\\\
\downarrow \delta^{i} / ({-\sdap M}) & &
\\\\
|{\Yto 1 j k}_{SO(D-1)} \rangle
& 
\equiv
&
|{\Yt j k}_{[1]\, SO(D-2)} \gg + |{\Yoo j k}_{[1]\, SO(D-2)} \gg
\\\\
\downarrow \delta^{j} / ({-\sdap M}) & &
\\\\
-\sqrt{3}\, |{\Yto j j k}_{ (1) SO(D-1)} \rangle
=
|{\Yto 1 1 k}_{\GL(D-1)} \rangle - |{\Yto j j k}_{\GL(D-1)} \rangle
& 
\equiv
&
|{\Yo k}_{[1][1] \GL(D-2)} \gg
- 
|{\Yto j j k}_{ \GL(D-2)} \gg
\end{matrix}
,
\label{eq:Y21_SOD-1_lc}
\end{align}
where the normalization factor $\sqrt{3}$ is discussed below.

To obtain the full irrep we still need to consider the states of the
first equation in the last line of eq. \eqref{eq:eps_Y21_from_soD-2}.
There are no canonical states.
One possible choice, discussed above reads for $j\ne 2$
\begin{align}
\sqrt{3}\, |{\Yto j j 1}_{ (2) SO(D-1)} \rangle
=
|{\Yto j j 1}_{\GL(D-1)} \rangle - |{\Yto 2 2 1}_{\GL(D-1)} \rangle
& 
\equiv
&
|{\Yt j j}_{[1]\, SO(D-2)} \gg
-\,
|{\Yt 2 2}_{[1]\, SO(D-2)} \gg
.
\end{align}

The normalization factors are easily obtained using the simplest
possible representation of $\ydiagram{2, 1}$, i.e.
the one obtained by applying the Young symmetrizer $\Yto I J K$ to
$a^{\dagger I}\, a^{\dagger J}\, b^{\dagger K}\, | 0\rangle$ with
with $[a^I,\, a^{\dagger J}] = n\, \delta^{I J}$
and 
 $[b^I,\, b^{\dagger J}] = m\, \delta^{I J}$.
We get that
\begin{align}
Y_{\Yto I J K} a^{\dagger I}\, a^{\dagger J}\, b^{\dagger K}\, | 0\rangle
= | {\Yto I J K} \rangle
~~\Rightarrow~~
\langle Y_{\Yto I J K} | Y_{\Yto I J K} \rangle = 8 n^2 m,
\end{align}
while
$
\parallel |{\Yto j j 1}_{\GL(D-1)} \rangle - |{\Yto 2 2 1}_{\GL(D-1)} \rangle
\parallel^2
=
2*6 n^2 m
.
$

The explicit expressions for the previous states is
\begin{align}
|{\Yto i j k}_{SO(D-1)} \rangle
=&
| \Yto i j k \gg
\nonumber\\
=&
\sum_a b^{\sN{ N,\, s=3,\, a}}\, e^{\sN{ N,\, s=3,\, a}}_{i j k}
,
\nonumber\\
|{\Yto 1 j k}_{SO(D-1)} \rangle
=& 
| \Yt j k \gg + | \Yoo j k \gg
\nonumber\\
=&
\frac{-1}{\sdap M}
\sum_{a,b} b^{\sN{ N,\, s=3,\, a}}\, D^{\sN{ N,\, s=3,\, 1}}_{a b}
e^{\sN{ N,\, s=2,\, b}}_{j k}
,
\nonumber\\
|{\Yto j j k}_{ (1) SO(D-1)} \rangle
=&
\frac{1}{\sqrt{3}}\left(
|{\Yto j j k}_{ \GL(D-2)} \gg 
- 
|{\Yo k}_{[1][1] \GL(D-2)} \gg
\right)
\nonumber\\
=&
\frac{1}{\sqrt{3}}
\left(\frac{-1}{\sdap M}\right)^2
\Big[
\sum_{a,b} b^{\sN{ N,\, s=3,\, a}}\,
(D^{\sN{ N,\, s=3,\, 1}} D^{\sN{ N,\, s=2,\, 1}})_{a b}\,
e^{\sN{ N,\, s=1,\, b}}_{k}
\nonumber\\
&+
\sum_{a,b} b^{\sN{ N,\, s=3,\, a}}\,
(D^{\sN{ N,\, s=3,\, 1}} U^{\sN{ N,\, s=2}})_{a b}\,
e^{\sN{ N,\, s=3,\, b}}_{j k j}
\Big]
,
\nonumber\\
|{\Yto j j 1}_{ (2) SO(D-1)} \rangle
=&
\frac{1}{\sqrt{3}}\left(
|{\Yt j j}_{[1]\, SO(D-2)} \gg
-\,
|{\Yt 2 2}_{[1]\, SO(D-2)} \gg
\right)
\nonumber\\
=&
\frac{1}{\sqrt{3}}
\frac{-1}{\sdap M}
\sum_{a,b} b^{\sN{ N,\, s=3,\, a}}\, D^{\sN{ N,\, s=3,\, 1}}_{a b}
( - e^{\sN{ N,\, s=2,\, b}}_{i i} + e^{\sN{ N,\, s=2,\, b}}_{2 2} )
,
\end{align}
where $b^{\sN {N,\, s=3,\, a}}$ are the projected coefficients
using the Young symmetrizier $Y_{\ydiagram{2,1}}$
similarly  as in eq. \eqref{eq:Y2_b_has_symmetrizer}.


\subsection{The  $\ydiagram{3}$ irrep}
As before we can start looking at the $SO(D-1)$ polarization tensors.
The totally  symmetric polarizations $\epsilon_{\Ytr I J K} $ satisfy
\begin{align}
\sum_I \epsilon_{\Ytr I I K} =& 0
.
\end{align}
As a first step we use the previous relations from the
$SO(D-2)$ point of view in a way to reveal the independent components
($i, j,k$ are all different)
\begin{align}
\epsilon_{\Ytr i j k}&,
&
&
\nonumber\\
\sum_i \epsilon_{\Ytr i i 1}&=0,
&
\sum_i \epsilon_{\Ytr i i j} + \epsilon_{\Ytr 1 1 j}&=0.
\label{eq:eps_Y3_from_soD-2}
\end{align}
Again there is no canonical way of solving the equations in the last
line.
A possible solution which we discuss below is (no sum over $i$)
\begin{align}
\epsilon_{\Ytr i i 1}&=-\epsilon_{\Ytr 1 1 1},
&
\epsilon_{\Ytr i i j}&= - \epsilon_{\Ytr 1 1 j}
.
\end{align}

Let us start building the previous $SO(D-1)$ states from the $SO(D-2)$ states 
with the most straightforward approach as done for $\Yt {\,} {\,}$.
We consider the states
\begin{equation}
|{\Ytr i j k}_{SO(D-1)} \rangle
\equiv
|{\Ytr i j k}_{SO(D-2)} \gg
\equiv
|{\Ytr i j k}_{\GL(D-2)} \gg
,~~~~
i \ne j \ne k \ne i 
,
\end{equation}
where as before the condition $i \ne j \ne k $ allows to forget about the trace
condition and consider the $\GL(D-2)$ states as $SO(D-2)$ states
as $SO(D-1)$ states.
As usual in the following 
$|*\rangle$ are the $SO(D-1)$ states and
$|*\gg$ are the $SO(D-2)$ states.

On these states we apply a sequence of $\delta$s as
\begin{align}
\begin{matrix}
|{\Ytr i j k}_{SO(D-1)} \rangle
& 
\equiv
&
|{\Ytr i j k}_{SO(D-2)} \gg
\\\\
\downarrow \delta^{i} / ({-\sdap M}) & &
\\\\
|{\Ytr 1 j k}_{SO(D-1)} \rangle
& 
\equiv
&
|{\Yt j k}_{[1]\, SO(D-2)} \gg 
\\\\
\downarrow \delta^{j} / ({-\sdap M}) & &
\\\\
-2\, |{\Ytr j j k}_{ (1) SO(D-1)} \rangle
=
|{\Ytr 1 1 k}_{\GL(D-1)} \rangle - |{\Ytr j j k}_{\GL(D-1)} \rangle
& 
\equiv
&
|{\Yo k}_{[1][1] \GL(D-2)} \gg
- 
|{\Ytr j j k}_{ \GL(D-2)} \gg
\\\\
\downarrow \delta^{k} / ({-\sdap M}) & &
\\\\
-2\sqrt{\frac{2}{3}}\,
( |{\Ytr 1 k k }_{ (1) SO(D-1)} \rangle
- |{\Ytr 1 j j}_{ (1) SO(D-1)} \rangle )
&&
\\
\mkern-50mu    
=
2 |{\Ytr 1 1 k}_{\GL(D-1)} \rangle - |{\Ytr 1 1 1}_{\GL(D-1)} \rangle
+ |{\Ytr j j 1}_{\GL(D-1)} \rangle
&&
\\
& 
\equiv
&
|{\Yo k}_{[1][1] \GL(D-2)} \gg
- 
|{\Ytr j j k}_{ \GL(D-2)} \gg
\end{matrix}
,
\label{eq:Y3_SOD-1_lc}
\end{align}
where we now discuss the normalization factors $2$ and
$\sqrt{\frac{3}{8}}$.
As for the $\Yt I J$ case 
we can take whichever explicit representation to compute the
normalizations.
Explicitly for $I\ne J\ne K$
$ |\Ytr I J K \rangle  = a^{\dagger I}\, a^{\dagger J}\, a^{\dagger K}\, |0\rangle$
with $[a^I,\, a^{\dagger J}] = n\, \delta^{I J}$ 
and $I\ne J$.
The normalization is then
$\langle \Ytr I J K | \Ytr I J K\rangle = n^3$
which is valid for the states
$ |\Ytr i j k \rangle$ and $ |\Ytr 1 j k \rangle$.
Then the states
$ |{\Ytr j j k}_{(1)} \rangle  =
\oh \left( {a^{j \dagger}}^2\, a^{\dagger k}
- {a^{1 \dagger}}^2\, a^{\dagger k}\, \right) \, |0\rangle
$
and
$ |{\Ytr 1 k k } \rangle  =
\sqrt{\frac{3}{8}}
\left( {a^{k \dagger}}^2\, a^{\dagger 1}
- \frac{1}{3} {a^{1 \dagger}}^3 \right) \, |0\rangle
$
have the same normalization.
The reason of the factor $\frac{1}{3}$ is that $GL$ states must have
the same normalization and

The explicit expressions for the previous states is
\begin{align}
|{\Ytr i j k}_{SO(D-1)} \rangle
=&
| \Ytr i j k \gg
\nonumber\\
=&
\sum_a b^{\sN{ N,\, s=3,\, a}}\, e^{\sN{ N,\, s=3,\, a}}_{i j k}
,
\nonumber\\
|{\Ytr 1 j k}_{SO(D-1)} \rangle
=& 
| \Yt j k \gg 
\nonumber\\
=&
\frac{-1}{\sdap M}
\sum_{a,b} b^{\sN{ N,\, s=3,\, a}}\, D^{\sN{ N,\, s=3,\, 1}}_{a b}
e^{\sN{ N,\, s=2,\, b}}_{j k}
,
\nonumber\\
|{\Ytr j j k}_{ (1) SO(D-1)} \rangle
=&
\frac{1}{2}\left(
|{\Ytr j j k}_{ \GL(D-2)} \gg 
- 
|{\Yo k}_{[1][1] \GL(D-2)} \gg
\right)
\nonumber\\
=&
\frac{1}{2}
\left(\frac{-1}{\sdap M}\right)^2
\Bigg[
\sum_{a,b} b^{\sN{ N,\, s=3,\, a}}\,
(D^{\sN{ N,\, s=3,\, 1}} D^{\sN{ N,\, s=2,\, 1}})_{a b}\,
e^{\sN{ N,\, s=1,\, b}}_{k}
\nonumber\\
&+
\sum_{a,b} b^{\sN{ N,\, s=3,\, a}}\,
(D^{\sN{ N,\, s=3,\, 1}} U^{\sN{ N,\, s=2}})_{a b}\,
e^{\sN{ N,\, s=3,\, b}}_{j k j}
\Bigg]
,
\nonumber\\
{\rm Mix}_{k,\, j}
=&
2 |{\Ytr k k 1}_{ (1) SO(D-1)} \rangle
+ |{\Ytr j j 1}_{ (1) SO(D-1)} \rangle
\nonumber\\
=&
- \sqrt{\frac{3}{8}}
\left( \frac{-1}{\sdap M} \right)^3
\Bigg[
\sum_{a,b} b^{\sN{ N,\, s=3,\, a}}\,
( D^{\sN{ N,\, s=3,\, 1}} D^{\sN{ N,\, s=2,\, 1}} D^{\sN{ N,\, s=1,\,1}} )_{a b}
e^{\sN{ N,\, s=0,\, b}}
\nonumber\\
&
+
\sum_{a,b} b^{\sN{ N,\, s=3,\, a}}\,
( D^{\sN{ N,\, s=3,\, 1}} D^{\sN{ N,\, s=2,\, 1}} U^{\sN{ N,\, s=1}} )_{a b}
e^{\sN{ N,\, s=2,\, b}}_{k k}
\nonumber\\
&
+
\sum_{a,b} b^{\sN{ N,\, s=3,\, a}}\,
( D^{\sN{ N,\, s=3,\, 1}} U^{\sN{ N,\, s=2}} D^{\sN{ N,\, s=3,\, 2}} )_{a b}
e^{\sN{ N,\, s=2,\, b}}_{j j}
\nonumber\\
&
+
\sum_{a,b} b^{\sN{ N,\, s=3,\, a}}\,
( D^{\sN{ N,\, s=3,\, 1}} U^{\sN{ N,\, s=2}} A^{\sN{ N,\, s=3,\, 1 3}} )_{a b}
e^{\sN{ N,\, s=2,\, b}}_{k k}
\Bigg]
.
\end{align}
As before $b^{\sN {N,\, s=3,\, a}}$ are the projected coefficients
using the Young symmetrizier $Y_{\ydiagram{3}}$
similarly as done in eq. \eqref{eq:Y2_b_has_symmetrizer}.

We can then compute the $|{\Ytr k k 1}_{ (1) SO(D-1)} \rangle$ state
by making the combination
\begin{equation}
|{\Ytr k k 1}_{ (1) SO(D-1)} \rangle
=
\frac{3}{2}
\left(
{\rm Mix}_{k,\, j}
-
\oh {\rm Mix}_{j,\, k}
\right)
,
\end{equation}
which also can be used to check the consistency of the procedure since
the final state depends on $k$ only while the initial on both $k$ and
$j$.

\subsection{The totally antisymmetric irreps}

These irreps are the simplest to deal with.
To build the full irrep only one step is needed.
Explicitly we have
On these states we apply a sequence of $\delta$s as
\begin{align}
\begin{matrix}
|{\Yooo {i_1} {\vdots} {i_n} }_{SO(D-1)} \rangle
& 
\equiv
&
|{\Yooo {i_1} {\vdots} {i_n} }_{SO(D-2)} \gg
\\\\
\downarrow \delta^{i_1} / ({-\sdap M}) & &
\\\\
|{\Yooo {1} {\vdots} {i_n} }_{SO(D-1)} \rangle
& 
\equiv
&
|{\Yoo  {\vdots} {i_n}}_{[1]\, SO(D-2)} \gg 
\label{eq:Y111_SOD-1_lc}
\end{matrix}
\end{align}
The explicit expressions for the previous states is
\begin{align}
|{\Yooo {i_1} {\vdots} {i_n} }_{SO(D-1)} \rangle
=&
| \Yooo {i_1} {\vdots} {i_n} \gg
\nonumber\\
=&
\sum_a b^{\sN{ N,\, s=n,\, a}}\, e^{\sN{ N,\, s=n,\, a}}_{i_1 \dots i_n}
,
\nonumber\\
|{\Yooo {1} {\vdots} {i_n}}_{SO(D-1)} \rangle
=& 
| \Yoo {\vdots} {i_n} \gg 
\nonumber\\
=&
\frac{-1}{\sdap M}
\sum_{a,b} b^{\sN{ N,\, s=n,\, a}}\, D^{\sN{ N,\, s=n-1,\, 1}}_{a b}
e^{\sN{ N,\, s=n-1,\, b}}_{{i_2} {\dots} {i_n}}
,
\end{align}
where $b^{\sN {N,\, s=n,\, a}}$ are the projected coefficients
using a Young symmetrizier like $Y_{\ydiagram{1,1,1}}$
similarly  as in eq. \eqref{eq:Y2_b_has_symmetrizer}.


\subsection{An example of how to build $S_s$ irreps}
We want to describe how the symmetric group irreps are built in the
approach taken in this paper.
Let us take as example the $SO(D-1)$ irrep $\ydiagram{3,1}$ at level
$N=6$.
We start from the basis of possible $s=4$ indices tensors at level $N=6$
\begin{align}
  T_{N=6,\, s=4}
  =
  \Big\{&
    \left(1 , 1\right)\,{1}^{i_{1}}\,{1}^{i_{2}}\,{1}^{i_{3}}\,{1}^{i_{4}} ,\,
         {1}^{i_{1}}\,{1}^{i_{2}}\,{1}^{i_{3}}\,{3}^{i_{4}} ,\,
         {1}^{i_{1}}\,{1}^{i_{2}}\,{1}^{i_{4}}\,{3}^{i_{3}} ,
         \nonumber\\
         &
         {1}^{i_{1}}\,{1}^{i_{3}}\,{1}^{i_{4}}\,{3}^{i_{2}} ,\,
         {1}^{i_{2}}\,{1}^{i_{3}}\,{1}^{i_{4}}\,{3}^{i_{1}} ,\,
         {1}^{i_{1}}\,{1}^{i_{2}}\,{2}^{i_{3}}\,{2}^{i_{4}} ,
         \nonumber\\
         &
         {1}^{i_{1}}\,{1}^{i_{3}}\,{2}^{i_{2}}\,{2}^{i_{4}} ,\,
         {1}^{i_{2}}\,{1}^{i_{3}}\,{2}^{i_{1}}\,{2}^{i_{4}} ,\,
         {1}^{i_{1}}\,{1}^{i_{4}}\,{2}^{i_{2}}\,{2}^{i_{3}} ,
         \nonumber\\
         &
         {1}^{i_{2}}\,{1}^{i_{4}}\,{2}^{i_{1}}\,{2}^{i_{3}} ,\,
         {1}^{i_{3}}\,{1}^{i_{4}}\,{2}^{i_{1}}\,{2}^{i_{2}} \Big\}
         ,
         \label{eq:N6S4BigBasis}
\end{align}
which is a sum of different $SO(D-2)$  irreps.
They may become $SO(D-2)$ irreps after Young symmetrizers are used.

We then look for almost true states with  $s=4$ indexes,
i.e. states in $V_{N=6,\,s=4}= span\,   T_{N=6,\, s=4}$
whose number of indices does not increase under the boost $M^{i -}$.
They are a mixture of $SO(D-1)$ irreps since no Young symmetrizer has
been yet applied.
When all indexes are different they are mixtures of irreps with number
of indexes equal
or less than $s=4$ indexes.
When some indexes are equal they are mixture
of irreps with number
of indexes that may be greater than $s=4$ indexes.

The basis of these states is
\begin{align}
  \left(
  \begin{array}{ccccccccccc}
        3&2&2&-16&-16&0&0&0&0&0&27\cr
        3&2&-16&2&-16&0&0&0&0&27&0\cr
        3&2&-16&-16&2&0&0&0&27&0&0\cr 
        3&-16&2&2&-16&0&0&27&0&0&0\cr
        3&-16&2&-16&2&0&27&0&0&0&0\cr
        3&-16&-16&2&2&27&0&0&0&0&0\cr
        \end{array}
  \right)
  ,
\end{align}
where each line is a state and the coefficients refer to the basis
$T_{N=6,\, s=4}$ given in eq. \ref{eq:N6S4BigBasis}.

\ytableausetup{boxsize=1em}
Now we project any previous state, i.e. any line using
$Y_{ \begin{ytableau} i_1 & i_2 & i_3 \\ i_4 \end{ytableau} }$
and we obtain only one independent state
\begin{align}
  \left(
  0 , 9 , 0 , 0 , -9 , -{{27}\over{4}} , -{{27}\over{4}} , 0 , 0 \
  , {{27}\over{4}} , {{27}\over{4}} \right)
  .
\end{align}
Applying repeatedly the swaps $i_1 \leftrightarrow i_k$ ($k=2,3,4$)
we build  the vector space where the $S_4$ irrep is represented
\begin{align}
  \left(
\begin{array}{ccccccccccc}
0&9&0&0&-9&-{{27}\over{4}}&-{{27}\over{4}}&0&0&{{27}\over{4}}&{{27}\over{4}}\\
0&9&-9&0&0&0&-{{27}\over{4}}&-{{27}\over{4}}&{{27}\over{4}}&{{27}\over{4}}&0\\
0&-9&0&9&0&{{27}\over{4}}&0&{{27}\over{4}}&-{{27}\over{4}}&0&-{{27}\over{4}}\\
   \end{array}
\right)
  ,
\label{eq:Y31 N6 s4 without tensors}
\end{align}
which has the dimension $\frac{4!}{4\cdot 2}=3$ as computed by hook rule.
A simpler set of states is the one with integer entries with
relatively prime numbers which read

\begin{align}
b^{[ N=6,\, s=4->4]}=
 &
\left( 
\begin{matrix} 
0
&
3
&
3
&
0
&
0
&
-3
&
-3
&
-4
&
0
&
0
&
4
\\
0
&
0
&
3
&
3
&
-3
&
-3
&
0
&
0
&
0
&
-4
&
4
\\
0
&
-3
&
0
&
-3
&
3
&
0
&
3
&
0
&
4
&
0
&
-4
\end{matrix}
\right)
,
\end{align}
%

or with the tensor structures shown explicitly

\myscaleM{ 
\parbox{\linewidth}{ 
\begin{align}
\bce^\sN{N=6,\, s=4->4}=
 &
\left( 
\begin{matrix} 
0
&
+3\,{1}^{i_{3}}\,{1}^{i_{4}}\,{2}^{i_{1}}\,{2}^{i_{2}}
&
+3\,{1}^{i_{2}}\,{1}^{i_{4}}\,{2}^{i_{1}}\,{2}^{i_{3}}
&
0
\\
0
&
0
&
+3\,{1}^{i_{2}}\,{1}^{i_{4}}\,{2}^{i_{1}}\,{2}^{i_{3}}
&
+3\,{1}^{i_{1}}\,{1}^{i_{4}}\,{2}^{i_{2}}\,{2}^{i_{3}}
\\
0
&
-3\,{1}^{i_{3}}\,{1}^{i_{4}}\,{2}^{i_{1}}\,{2}^{i_{2}}
&
0
&
-3\,{1}^{i_{1}}\,{1}^{i_{4}}\,{2}^{i_{2}}\,{2}^{i_{3}}
\end{matrix}
\right . 
\nonumber
\label{eq:Y31 N6 s4 with tensors}
\end{align}
} 
} 


\myscaleM{ 
\parbox{\linewidth}{ 
\begin{align}
&
\begin{matrix} 
0
&
-3\,{1}^{i_{1}}\,{1}^{i_{3}}\,{2}^{i_{2}}\,{2}^{i_{4}}
&
-3\,{1}^{i_{1}}\,{1}^{i_{2}}\,{2}^{i_{3}}\,{2}^{i_{4}}
&
-4\,{1}^{i_{2}}\,{1}^{i_{3}}\,{1}^{i_{4}}\,{3}^{i_{1}}
\\
-3\,{1}^{i_{2}}\,{1}^{i_{3}}\,{2}^{i_{1}}\,{2}^{i_{4}}
&
-3\,{1}^{i_{1}}\,{1}^{i_{3}}\,{2}^{i_{2}}\,{2}^{i_{4}}
&
0
&
0
\\
+3\,{1}^{i_{2}}\,{1}^{i_{3}}\,{2}^{i_{1}}\,{2}^{i_{4}}
&
0
&
+3\,{1}^{i_{1}}\,{1}^{i_{2}}\,{2}^{i_{3}}\,{2}^{i_{4}}
&
0
\end{matrix}
\nonumber
\end{align}
} 
} 


\myscaleM{ 
\parbox{\linewidth}{ 
\begin{align}
&
\left. 
\begin{matrix} 
0
&
0
&
+4\,{1}^{i_{1}}\,{1}^{i_{2}}\,{1}^{i_{3}}\,{3}^{i_{4}}
\\
0
&
-4\,{1}^{i_{1}}\,{1}^{i_{2}}\,{1}^{i_{4}}\,{3}^{i_{3}}
&
+4\,{1}^{i_{1}}\,{1}^{i_{2}}\,{1}^{i_{3}}\,{3}^{i_{4}}
\\
+4\,{1}^{i_{1}}\,{1}^{i_{3}}\,{1}^{i_{4}}\,{3}^{i_{2}}
&
0
&
-4\,{1}^{i_{1}}\,{1}^{i_{2}}\,{1}^{i_{3}}\,{3}^{i_{4}}
\end{matrix}
\right)
\nonumber
,  \end{align}
} 
} 
\begin{equation}~\end{equation}

At the same time we can explicitly compute the associated $S_4$ irrep.
In particular we need only the action of the swaps $(1, k)$
($k=2,3,4$) since all
the other actions can be computed using them.
Their explicit matrix representation on the previous states is
\begin{align} 
R[(1, 2)]= 
& 
\ifx\endpmatrix\undefined\pmatrix\else{\begin{pmatrix}\fi 0&0&-1\cr 0&1&0\cr -1&0&0\cr \ifx\endpmatrix\undefined\else\end{pmatrix}}\fi 
, 
\end{align}

\begin{align} 
R[(1, 3)]= 
& 
\ifx\endpmatrix\undefined\pmatrix{\else{\begin{pmatrix}\fi 0&1&0\cr 1&0&0\cr 0&0&1\cr \ifx\endpmatrix\undefined}\else\end{pmatrix}}\fi 
, 
\end{align} 

\begin{align} 
R[(1, 4)]= 
& 
\ifx\endpmatrix\undefined\pmatrix{\else{\begin{pmatrix}\fi -1&0&0\cr -1&1&0\cr 1&0&1\cr \ifx\endpmatrix\undefined}\else\end{pmatrix}}\fi 
. 
\end{align} 

Starting from these states we can build their descendants,
i.e. their images at level $N=6$ but with $s=3,2,1,0$ which are
obtained by repeatedly applying the boost $M^{i -}$ and keeping only
the states with one index less that the states we started from.
These states are the building blocks of the $SO(D-1)$ irrep but they
are not states in the $SO(D-1)$ irrep.
They must be combined to get the states of the $SO(D-1)$ irrep as done
in eq. \eqref{eq:Y2_SOD-1_lc}
or in eq. \eqref{eq:irrep_Y2_explicit}
where the descendant at $s=0$ is combined
with the state at level $s=2$.

In the case at hand the $\GL(D-1)$ irrep
$\begin{ytableau} I_1 & I_2 & I_3 \\ I_4 \end{ytableau}$
($I=1, \dots D-1$) splits
into
$4$ indices tensor
$\begin{ytableau} i_1 & i_2 & i_3 \\ i_4 \end{ytableau}$
,
$3$ indices tensors
$\begin{ytableau} i_1 & i_2 & i_3 \\ 1 \end{ytableau}
\oplus
\begin{ytableau} i_1 & i_2 & 1 \\ i_3 \end{ytableau}$,
$2$ indices tensors
$\begin{ytableau} i_1 & i_2 & 1 \\ 1 \end{ytableau}
\oplus
\begin{ytableau} i_1 & 1 & 1 \\ i_2 \end{ytableau}$
and
$1$ index tensor
$\begin{ytableau} i_1 & 1 & 1 \\ 1 \end{ytableau}$.

The state transforming as
$\begin{ytableau} i_1 & i_2 & i_3 \\ i_4 \end{ytableau}$
is the one we started our construction
\eqref{eq:Y31 N6 s4 without tensors}, 
the others are the descendants obtained by the action of $M^{i -}$
boost.

They are explicitly given in the next equations.
In the following equations each row corresponds to the image of the
state described by the corresponding row in eq.
\ref{eq:Y31 N6 s4 with tensors}

 \myscale{ 
\parbox{\linewidth}{ 
\begin{align*}
b^\sN{N=6~ s=4->3}=
 &
\left( 
\begin{matrix} 
8
&
-3
&
-3
&
0
&
12
&
-2
&
-2
&
12
&
-6
&
12
&
-6
\\
0
&
-3
&
0
&
3
&
0
&
4
&
-14
&
18
&
-18
&
14
&
-4
\\
0
&
0
&
3
&
-3
&
0
&
14
&
-4
&
-14
&
4
&
-18
&
18
\end{matrix}
\right . 
\end{align*}
} 
} 


\myscale{ 
\parbox{\linewidth}{ 
\begin{align*}
&
\left. 
\begin{matrix} 
-16
&
-4
&
-4
\\
-12
&
0
&
12
\\
12
&
-12
&
0
\end{matrix}
\right) 
,  \end{align*}
} 
} 
\begin{equation}\end{equation}

 \myscale{ 
\parbox{\linewidth}{ 
\begin{align*}
b^\sN{N=6~ s=4->2}=
 &
\left( 
\begin{matrix} 
0
&
0
&
0
&
0
&
0
&
0
&
0
&
0
&
0
&
0
&
0
\\
12
&
12
&
-3
&
-8
&
40
&
-24
&
-8
&
6
&
24
&
-8
&
136
\\
-12
&
-12
&
3
&
-40
&
8
&
8
&
24
&
-6
&
-24
&
-136
&
8
\end{matrix}
\right . 
\end{align*}
} 
} 


\myscale{ 
\parbox{\linewidth}{ 
\begin{align*}
&
\left. 
\begin{matrix} 
0
&
0
\\
-100
&
-100
\\
100
&
100
\end{matrix}
\right) 
,  \end{align*}
} 
} 

\begin{align}
b^\sN{N=6~ s=4->1}=
 &
\left( 
\begin{matrix} 
0
&
0
&
0
&
0
&
0
&
0
&
0
&
0
&
0
\\
0
&
0
&
0
&
0
&
0
&
0
&
0
&
0
&
0
\\
-16
&
-6
&
-5
&
30
&
15
&
0
&
10
&
-5
&
-36
\end{matrix}
\right) 
,  \end{align}

\begin{align}
b^\sN{N=6~ s=4->0}=
 &
\left( 
\begin{matrix} 
0
&
0
&
0
&
0
&
0
&
0
&
0
\\
0
&
0
&
0
&
0
&
0
&
0
&
0
\\
0
&
0
&
0
&
0
&
0
&
0
&
0
\end{matrix}
\right) 
.  \end{align}

 or with the tensor structures shown explicitly

 \myscaleM{ 
\parbox{\linewidth}{ 
\begin{align}
\bce^\sN{N=6~ s=4->3}_{i_1 i_2 i_3}=
 &
\left( 
\begin{matrix} 
+8\,{1}^{i_{1}}\,{1}^{i_{2}}\,{1}^{i_{3}}\,\left(2 , 1\right)
&
-3\,\left(1 , 1\right)\,{1}^{i_{1}}\,{1}^{i_{2}}\,{2}^{i_{3}}
&
-3\,\left(1 , 1\right)\,{1}^{i_{1}}\,{1}^{i_{3}}\,{2}^{i_{2}}
&
0
\\
0
&
-3\,\left(1 , 1\right)\,{1}^{i_{1}}\,{1}^{i_{2}}\,{2}^{i_{3}}
&
0
&
+3\,\left(1 , 1\right)\,{1}^{i_{2}}\,{1}^{i_{3}}\,{2}^{i_{1}}
\\
0
&
0
&
+3\,\left(1 , 1\right)\,{1}^{i_{1}}\,{1}^{i_{3}}\,{2}^{i_{2}}
&
-3\,\left(1 , 1\right)\,{1}^{i_{2}}\,{1}^{i_{3}}\,{2}^{i_{1}}
\end{matrix}
\right . 
\nonumber
\end{align}
} 
} 


\myscaleM{ 
\parbox{\linewidth}{ 
\begin{align}
&
\begin{matrix} 
+12\,{2}^{i_{1}}\,{2}^{i_{2}}\,{2}^{i_{3}}
&
-2\,{1}^{i_{3}}\,{2}^{i_{2}}\,{3}^{i_{1}}
&
-2\,{1}^{i_{2}}\,{2}^{i_{3}}\,{3}^{i_{1}}
&
+12\,{1}^{i_{3}}\,{2}^{i_{1}}\,{3}^{i_{2}}
\\
0
&
+4\,{1}^{i_{3}}\,{2}^{i_{2}}\,{3}^{i_{1}}
&
-14\,{1}^{i_{2}}\,{2}^{i_{3}}\,{3}^{i_{1}}
&
+18\,{1}^{i_{3}}\,{2}^{i_{1}}\,{3}^{i_{2}}
\\
0
&
+14\,{1}^{i_{3}}\,{2}^{i_{2}}\,{3}^{i_{1}}
&
-4\,{1}^{i_{2}}\,{2}^{i_{3}}\,{3}^{i_{1}}
&
-14\,{1}^{i_{3}}\,{2}^{i_{1}}\,{3}^{i_{2}}
\end{matrix}
\nonumber
\end{align}
} 
} 


\myscaleM{ 
\parbox{\linewidth}{ 
\begin{align}
&
\begin{matrix} 
-6\,{1}^{i_{1}}\,{2}^{i_{3}}\,{3}^{i_{2}}
&
+12\,{1}^{i_{2}}\,{2}^{i_{1}}\,{3}^{i_{3}}
&
-6\,{1}^{i_{1}}\,{2}^{i_{2}}\,{3}^{i_{3}}
&
-16\,{1}^{i_{2}}\,{1}^{i_{3}}\,{4}^{i_{1}}
\\
-18\,{1}^{i_{1}}\,{2}^{i_{3}}\,{3}^{i_{2}}
&
+14\,{1}^{i_{2}}\,{2}^{i_{1}}\,{3}^{i_{3}}
&
-4\,{1}^{i_{1}}\,{2}^{i_{2}}\,{3}^{i_{3}}
&
-12\,{1}^{i_{2}}\,{1}^{i_{3}}\,{4}^{i_{1}}
\\
+4\,{1}^{i_{1}}\,{2}^{i_{3}}\,{3}^{i_{2}}
&
-18\,{1}^{i_{2}}\,{2}^{i_{1}}\,{3}^{i_{3}}
&
+18\,{1}^{i_{1}}\,{2}^{i_{2}}\,{3}^{i_{3}}
&
+12\,{1}^{i_{2}}\,{1}^{i_{3}}\,{4}^{i_{1}}
\end{matrix}
\nonumber
\end{align}
} 
} 


\myscaleM{ 
\parbox{\linewidth}{ 
\begin{align}
&
\left. 
\begin{matrix} 
-4\,{1}^{i_{1}}\,{1}^{i_{3}}\,{4}^{i_{2}}
&
-4\,{1}^{i_{1}}\,{1}^{i_{2}}\,{4}^{i_{3}}
\\
0
&
+12\,{1}^{i_{1}}\,{1}^{i_{2}}\,{4}^{i_{3}}
\\
-12\,{1}^{i_{1}}\,{1}^{i_{3}}\,{4}^{i_{2}}
&
0
\end{matrix}
\right) 
,  \end{align}
} 
} 

 \myscaleM{ 
\parbox{\linewidth}{ 
\begin{align}
b^\sN{N=6~ s=4->2}_{i_1 i_2}=
 &
\left( 
\begin{matrix} 
0
&
0
&
0
&
0
\\
+12\,{1}^{i_{1}}\,{1}^{i_{2}}\,\left(3 , 1\right)
&
+12\,{1}^{i_{1}}\,{1}^{i_{2}}\,\left(2 , 2\right)
&
-3\,\left(1 , 1\right)^2\,{1}^{i_{1}}\,{1}^{i_{2}}
&
-8\,{1}^{i_{1}}\,\left(2 , 1\right)\,{2}^{i_{2}}
\\
-12\,{1}^{i_{1}}\,{1}^{i_{2}}\,\left(3 , 1\right)
&
-12\,{1}^{i_{1}}\,{1}^{i_{2}}\,\left(2 , 2\right)
&
+3\,\left(1 , 1\right)^2\,{1}^{i_{1}}\,{1}^{i_{2}}
&
-40\,{1}^{i_{1}}\,\left(2 , 1\right)\,{2}^{i_{2}}
\end{matrix}
\right . 
\nonumber
\end{align}
} 
} 


\myscaleM{ 
\parbox{\linewidth}{ 
\begin{align}
&
\begin{matrix} 
0
&
0
&
0
&
0
\\
+40\,{1}^{i_{2}}\,\left(2 , 1\right)\,{2}^{i_{1}}
&
-24\,\left(1 , 1\right)\,{1}^{i_{1}}\,{3}^{i_{2}}
&
-8\,\left(1 , 1\right)\,{1}^{i_{2}}\,{3}^{i_{1}}
&
+6\,\left(1 , 1\right)\,{2}^{i_{1}}\,{2}^{i_{2}}
\\
+8\,{1}^{i_{2}}\,\left(2 , 1\right)\,{2}^{i_{1}}
&
+8\,\left(1 , 1\right)\,{1}^{i_{1}}\,{3}^{i_{2}}
&
+24\,\left(1 , 1\right)\,{1}^{i_{2}}\,{3}^{i_{1}}
&
-6\,\left(1 , 1\right)\,{2}^{i_{1}}\,{2}^{i_{2}}
\end{matrix}
\nonumber
\end{align}
} 
} 


\myscaleM{ 
\parbox{\linewidth}{ 
\begin{align}
&
\begin{matrix} 
0
&
0
&
0
&
0
\\
+24\,{3}^{i_{1}}\,{3}^{i_{2}}
&
-8\,{2}^{i_{2}}\,{4}^{i_{1}}
&
+136\,{2}^{i_{1}}\,{4}^{i_{2}}
&
-100\,{1}^{i_{2}}\,{5}^{i_{1}}
\\
-24\,{3}^{i_{1}}\,{3}^{i_{2}}
&
-136\,{2}^{i_{2}}\,{4}^{i_{1}}
&
+8\,{2}^{i_{1}}\,{4}^{i_{2}}
&
+100\,{1}^{i_{2}}\,{5}^{i_{1}}
\end{matrix}
\nonumber
\end{align}
} 
} 


\myscaleM{ 
\parbox{\linewidth}{ 
\begin{align}
&
\left. 
\begin{matrix} 
0
\\
-100\,{1}^{i_{1}}\,{5}^{i_{2}}
\\
+100\,{1}^{i_{1}}\,{5}^{i_{2}}
\end{matrix}
\right) 
,  \end{align}
} 
} 

 \myscaleM{ 
\parbox{\linewidth}{ 
\begin{align}
b^\sN{N=6~ s=4->1}_{i_1}=
 &
\left( 
\begin{matrix} 
0
&
0
&
0
&
0
\\
0
&
0
&
0
&
0
\\
-16\,{1}^{i_{1}}\,\left(4 , 1\right)
&
-6\,{1}^{i_{1}}\,\left(3 , 2\right)
&
-5\,\left(1 , 1\right)\,{1}^{i_{1}}\,\left(2 , 1\right)
&
+30\,{2}^{i_{1}}\,\left(3 , 1\right)
\end{matrix}
\right . 
\nonumber
\end{align}
} 
} 


\myscaleM{ 
\parbox{\linewidth}{ 
\begin{align}
&
\begin{matrix} 
0
&
0
&
0
&
0
\\
0
&
0
&
0
&
0
\\
+15\,\left(2 , 2\right)\,{2}^{i_{1}}
&
0
&
+10\,\left(2 , 1\right)\,{3}^{i_{1}}
&
-5\,\left(1 , 1\right)\,{4}^{i_{1}}
\end{matrix}
\nonumber
\end{align}
} 
} 


\myscaleM{ 
\parbox{\linewidth}{ 
\begin{align}
&
\left. 
\begin{matrix} 
0
\\
0
\\
-36\,{6}^{i_{1}}
\end{matrix}
\right) 
,  \end{align}
} 
} 

 \myscaleM{ 
\parbox{\linewidth}{ 
\begin{align}
b^\sN{N=6~ s=4->0}_\emptyset=
 &
\left( 
\begin{matrix} 
0
&
0
&
0
&
0
\\
0
&
0
&
0
&
0
\\
0
&
0
&
0
&
0
\end{matrix}
\right . 
\nonumber
\end{align}
} 
} 


\myscaleM{ 
\parbox{\linewidth}{ 
\begin{align}
&
\left. 
\begin{matrix} 
0
&
0
&
0
\\
0
&
0
&
0
\\
0
&
0
&
0
\end{matrix}
\right) 
.  \end{align}
} 
} 



 



 



 



Notice that it may well happen that the same irrep appears multiple
times.
In this case we have not tried to get the best combinations but simply reported
the result of the algorithm.


\subsection{Special cases: the Regge and subleading Regge trajectory}

The leading and subleading Regge trajectories
can be treated explicitly without using any CAS.
Actually it is by far better to do so when the number of indices $s$
is big since the vector spaces increase their dimensions.

The basis are readily found to be
\begin{align}
  T_{N,\, s=N}
  =&
  \{
  1^{i_1}\,  1^{i_2}\dots  1^{i_N}
  \}
  ,
  \nonumber\\
  T_{N,\, s=N-1}
  =&
  \{
  2^{i_1}\,  1^{i_2}\dots  1^{i_{N-1}},\,
  1^{i_1}\,  2^{i_2}\dots  1^{i_{N-1}},\,
  \dots
  1^{i_1}\,  1^{i_2}\dots  1^{i_{N-1}}
  \}
  .
\end{align}

The leading Regge trajectory at level $N$ is easily done since
there is only  one element of the basis.
In the rest frame it is not possible to increase the number of
indices and therefore it is a true $s=N$ tensor.
In the same way the $S_N$ irrep is trivial and given by
\begin{align}
  R[(1, k)]= ( 1 ) 
  ,
\end{align}
since all possible swaps map the previous base element in itself.

It is also immediate to see that the descendant with $s=N-1$ is
proportional to
\begin{align}
  \delta^{i \downarrow}
    1^{i_1}\, 1^{i_2}\dots  1^{i_N}
    =
    &
    2\,\delta^{i_1 i_2}\,
    2^{i}\,  1^{i_3}\dots  1^{i_{N}}
    +
    2\,\delta^{i_1 i_3}\,
    2^{i}\,  1^{i_3}\dots  1^{i_{N-1}}
    \dots
    \nonumber\\
    &
    -
    2\,\delta^{i i_1}\,
    (
    2^{i_2}\,  1^{i_3}\dots  1^{i_{N}} +
    \dots
    1^{i_2}\,  1^{i_3}\dots  2^{i_{N}}
    )
    - \dots
    \nonumber\\
    =&
    -
    2(N-1)\,\delta^{i i_1}\,
    2^{(i_2}\,  1^{i_3}\dots  1^{i_{N})}
    \dots
    .
\end{align}

Since all structures are equivalent we can consider the state with
indices $i_1\dots i_{N-1}$.
The complement in the vector space $T_{N,\, s=N-1}$ of this vector is the
set of states which are true $s=N-1$ tensors. 
They are given by the
following $(N-1) \times N $ matrix \wrt the $T_{N,\, s=N-1}$ basis
or by the explicit states
\begin{align}
  b^{[N,\, s=N-1]}
  =
  \begin{pmatrix}
  1 & 0  & \dots  0 & -1 \\
  0 & 1  & \dots  0 & -1 \\
  \vdots & \vdots & \vdots &\vdots \\
  0 & 0  & \dots  1 & -1
  \end{pmatrix}
\Rightarrow
  \bce^{[N,\, s=N-1]}_{i_1\dots i_{N-1}}
=
\begin{pmatrix}
    2^{i_1}\,  1^{i_2}\dots  1^{i_{N-1}} -  1^{i_1}\,  1^{i_2}\dots  2^{i_{N-1}} 
\\
    1^{i_1}\,  2^{i_2}\dots  1^{i_{N-1}} -  1^{i_1}\,  1^{i_2}\dots  2^{i_{N-1}} 
\\
\vdots
\\
    1^{i_1}\,  1^{i_2}\dots  2^{i_{N-1}} -  1^{i_1}\,  1^{i_2}\dots  2^{i_{N-1}} 
  \end{pmatrix}
  .
\end{align}
It is then easy to compute the $S_{N-1}$ irrep with result
\begin{align}
R[(1, 2)]= 
&
\begin{pmatrix} 
  0 & 1 & \dots 0 & 0 \\
  1 & 0 & \dots 0 & 0 \\
  0 & 0 & 1 \dots 0 & 0 \\
  0 & 0 &\dots 0 & 1 
 \end{pmatrix} 
, 
\nonumber\\
R[(1, N-1)]= 
&
\begin{pmatrix} 
  -1 & 0 & \dots 0 & 0 \\
  -1 & 1 & \dots 0 & 0 \\
  -1 & 0 & 1 \dots 0 & 0 \\
  0 & 0 &\dots -1 & 1 
 \end{pmatrix} 
.
\end{align}

All descendants must then be computed case by case.

\section{Chaos in three point amplitudes with two tachyons from
  lower spin}
\label{sec:chaos_from_big_numbers}

Chaos in string amplitudes was originally observed in three point
amplitudes with two tachyons.  However these on shell amplitudes are
completely determined by kinematics \cite{Bianchi:2023uby,Bianchi:2024fsi}.
Let us start with a mixture of massive particles $M_s$ of equal mass and
described by a transverse polarization tensor
$\epsilon_{\mu_1\dots \mu_s}$.
We do not require $\epsilon$ to be an irrep but only transverse and
this is why we wrote mixture.
The ``massive particle'' has momentum $k_{\sN 1}$ and two tachyons have
momenta $k_{\sN 2}$ and $k_{\sN 3}$.

Let us then exam the invariants.
All $k_{\sN r} \cdot k_{\sN t}$ with $r,t =1,2,3$ are fixed by
kinematics and on shell relations to be function of the masses.
We are left with only one invariant which is not fixed by kinematics
\begin{align}
  \epsilon_{\mu_1\dots \mu_s}\,
  k_{\sN  2}^{\mu_1}\, \dots k_{\sN  2}^{\mu_s}
  .
\end{align}
This happens because $\epsilon$ is transverse and we can always
replace $k_{\sN 3} = - k_{\sN 1} - k_{\sN 2}$.
This means that we can only see the coupling of totally symmetric
polarizations
\begin{equation}
  \epsilon_{\mu_1\dots \mu_s}
  ~\Rightarrow~
  \epsilon_{(\mu_1\dots \mu_s)}
. 
\end{equation}  

To proceed let us go the ``massive particle rest frame'' then we can
clearly see the mixture by decomposing the polarization tensor in irreps
\begin{align}
\epsilon_{I_1\dots I_s}
=
\epsilon_{I_1\dots I_s}^{( { \YsymmOV s  } )}
+
c_2\,
\delta_{( I_1 I_2}\, \epsilon_{I_3\dots I_s)}^{( { \YsymmOV {s-2}  } )}
+
c_4\,
\delta_{( I_1 I_2}\, \delta_{ I_3 I_4}\, \epsilon_{I_5\dots I_s)}^{(
{ \YsymmOV {s-4}  } )}
+\dots
\end{align}
with $1\le I \le D-1$
and all $c_{2 k}$ are fixed by group theory.

We can now choose a restricted kinematics as
\begin{equation}
k^0_{\sN 2}= E,~~
k^1_{\sN 2}= 0,~~
k^2_{\sN 2}= p_{out}\,\cos \theta,~~
k^0_{\sN 2}= p_{out}\,\sin \theta
,
\end{equation}
and
a restricted class of polarization where only
$\epsilon_{I_1=2 \dots I_s=2}\ne 0$.

With these restrictions the amplitude is then given by
\begin{align}
A_{M_s \rightarrow T T}
\sim
p_{out}^s
  \sum_{k} c_{2 k}\,
\epsilon_{2\dots 2}^{( { \YsymmOV {s-2 k}  } )}\,
\cos^{s -2 k} \theta    
.
\end{align}

Everything is fixed by kinematics or group theory but
$\epsilon_{2\dots 2}^{( { \YsymmOV {s-2 k}  } )}$.
If all mixtures, i.e. DDF states, at level $N$ and ``spin'' $s$ had
roughly the same
$\epsilon_{2\dots 2}^{( { \YsymmOV {s-2 k}  } )}$
then the amplitudes would be roughly the same.

The explicit construction of the states reveals that this is not the
case.
The origin of the chaotic behavior of these amplitudes is therefore
not in the string itself but rather in the chaotic mixture of irreps
in the DDF states.

Actually the previous approach suggests a way of extracting some
normalizations of the different irreps.

Start from the DDF state with $s$ indexes at level $N=\sum_{l=1}^s n_l$
\begin{align}
\uA^2_{-n_1}\dots \uA^2_{-n_s} |\uk_T\rangle_
{ {\YsymmOV {s} }_{\,\, GL(24)}}
,
\end{align}
which transform as a $\YsymmOV {s}$ of $GL(24)$ since we do not impose
any trace constraint.
This state is actually a mixture of $SO(25)$ states
\begin{align}
\uA^2_{-n_1}\dots \uA^2_{-n_s} |\uk_T\rangle_{ {\YsymmOV {s} }_{\,\, GL(24)} }
=&
\sum_{S=0}^s
\sum_{M =0}^{S}
\sum_{L =0}^{N-s}
c_{S\, L\, M}
\,
\left|
\YmixOVUN {S}{L}{M}{}_{SO(25)}
\right\rangle_{ {\YsymmOV {s} }_{\,\, GL(24)}}
,
\end{align}
where the $| *\rangle$ are the properly normalized states as discussed
in the examples above.
We can then choose a restricted kinematics as
\begin{equation}
k^0_{\sN 2}= E,~~
k^1_{\sN 2}= p_{out}\,\cos \theta,~~
k^2_{\sN 2}= p_{out}\,\sin \theta\,\cos \phi,~~
k^3_{\sN 2}= p_{out}\,\sin \theta\,\sin \phi,~~
,
\end{equation}
and get the amplitude
\begin{align}
A_{M_s \rightarrow T T}
\sim
p_{out}^s
\sum_{S=0}^s
\sum_{M =0}^{S}
\sum_{L =0}^{N-s}
c_{S\, L\, M}\,
\epsilon_{ \YmixOVUN {S}{L}{M}{} }\,
\cos^{L+M}\theta\,
\sin^S\theta\,
\cos^S\phi
,
\end{align}
from which it is possible to extract the $c_{S\,L\, M=0}$ coefficients
since the $\epsilon$s are actually $1$ for properly normalized states.

This approach can be extended easily also to the case of the
``pistol'' irreps by using as outgoing particles one photon and one tachyon.

\section{Conclusions}

In this paper we have made a brute force attack on the bosonic string
spectrum and, more importantly, to the explicit \lc expressions of states
of the irreps.

Among the main results there are the table \ref{tab:1} of all irreps and
multiplicities up to level $10$,
eq. \ref{eq:scalars_and_vectors_up_to_19} of the multiplicities of
scalars and vectors up to level $19$
and eq. \ref{eq:scalars_up_to_22} of the multiplicities of
scalars up to level $22$.

We have also reported in this paper the full
results for the level $N=3$ and $N=4$ in appendixes \ref{app:level3}
and \ref{app:level4}.
All the other levels are in separated TeX files since they are very big.

In appendix \ref{app:scalars_up_to_10} we have given the explicit
form of the scalars up to level $10$.

From these explicit results we have noticed  the
presence of enormous numbers (which seem to grow more than
exponentially with the level) in the $\GL(D-1)$ states with a small $s$
irrep.

Finally in section \ref{sec:chaos_from_big_numbers} we have argued
that this is the cause of chaos in some three point massive string
amplitudes.
It is not clear whether this is the unique cause since it could be
that some ``chaotic'' coefficients enter the four point amplitudes
which cannot be traced back to this origin.

It would then be interesting to extended these results to the
superstrings and off shell using the Brower states.

Another point worth exploring is whether there are other causes of
chaos in four point amplitudes.

\section*{Acknowledgments}
We would like to thank Dripto Biswas and Raffaele Marotta for
discussions.
This research is partially supported by the MUR PRIN contract
2020KR4KN2 “String Theory as a bridge between Gauge Theories and
Quantum Gravity” and by the INFN project ST\&FI “String Theory \&
Fundamental Interactions”.

\printbibliography[heading=bibintoc]

\appendix
\section{Constraints on increasing and decreasing operators from Lorentz algebra}
\label{app:Lorentz_algebra}

We want to discuss the constraints from Lorentz algebra on the matrix
representation of the decreasing $\delta^{i \downarrow}$, $\delta_A^{i \downarrow}$
and increasing $\delta^{i \uparrow}$ operators.

Given a level $N$ and $s$ indexes we have basis elements
$e^{\sN {N, s, a}}_{i_1\dots i_s} \in T_{N, s}$.

In the following we keep $N$ fixed and therefore we write simply
$e^{\sN {s, a}}_{i_1\dots i_s}$.
This is also true for the matrices, e.g.
$U^{\sN {N, s}}\rightarrow U^{\sN s}$.

Using these basis elements we can define a reducible representation of
the symmetric group $S_s$ as
\begin{equation}
e^{\sN {s, a} }_{\sigma( i_1\dots i_s)}
\equiv
e^{\sN {s, a} }_{ i_{\sigma(1)} \dots i_{\sigma(s)} }
=
(M^{\sN s}_\sigma)_{a b}\,
e^{\sN {s, b} }_{i_1\dots i_s}
.
\end{equation}  

The action of an increasing $\delta^{l \uparrow}$ operator is
defined as
\begin{align}
\delta^{l \uparrow}\, 
e^{\sN {s, a} }_{i_1\dots i_s}
=
U^{\sN s}_{a b}\,
e^{\sN {s+1, b} }_{i_1\dots i_s\, l}
.
\end{align}
The action of  decreasing $\delta^{m \downarrow}$ operator is more
complex and defined as
\begin{align}
\delta^{m \downarrow}\, 
e^{\sN {s, a} }_{i_1\dots i_s}
=&
\delta_{m, i_1}
D^{\sN {s,1}}_{a b}\,
e^{\sN {s-1, b} }_{i_2\dots i_s}
+
\delta_{m, i_2}
D^{\sN {s,2}}_{a b}\,
e^{\sN {s-1, b} }_{i_1\, i_3\dots i_s}
+
\dots
\nonumber\\
&
+
\delta_{m, i_s}
D^{\sN {s,s}}_{a b}\,
e^{\sN {s-1, b} }_{i_1\, i_2\dots i_{s-1}}
\nonumber\\
=&
\sum_{p=1}^{s}
\delta_{m, i_p}
D^{\sN {s, p}}_{a b}\,
e^{\sN {s-1, b} }_{i_1\dots i_{p-1}\,i_{p+1} \dots i_{s}}
.
\end{align}
The action of  decreasing $\delta_A^{m \downarrow}$ operator is even more
complex and defined as
\begin{align}
\delta_A^{m \downarrow}\, 
e^{\sN {s, a} }_{i_1\dots i_s}
=&
\delta_{i_1, i_2}
A^{\sN {s,1 2}}_{a b}\,
e^{\sN {s-1, b} }_{m\,i_3\dots i_s}
+
\delta_{i_1, i_3}
A^{\sN {s,1 3}}_{a b}\,
e^{\sN {s-1, b} }_{m\, i_2\dots i_s}
+
\dots
\nonumber\\
&
+
\delta_{i_p, i_q}
A^{\sN {s, p q}}_{a b}\,
e^{\sN {s-1, b} }_{m i_1\dots i_{p-1}\,i_{p+1} \dots i_{q-1}\,i_{q+1}\dots i_{s}}
+
\dots
\nonumber\\
=&
\sum_{p=1}^{s-1} \sum_{q=p+1}^s\,
\delta_{i_p, i_q}
A^{\sN {s, p q}}_{a b}\,
e^{\sN {s-1, b} }_{m i_1\dots i_{p-1}\,i_{p+1} \dots i_{q-1}\,i_{q+1}\dots i_{s}}
.
\end{align}

Not all $D$s and $A$s matrices are independent.
Actually only
$D^{\sN {s,1}}_{a b}$ and
$A^{\sN {s,1 2}}_{a b}$ are independent and the ones reported in the
supplementary material.

In facts let us consider the cycle $\sigma=(1 2 \dots p)$ which acts on the
indexes as $i_1 \rightarrow i_2 \rightarrow\dots i_p \rightarrow 1$,
we have
\begin{align}
\delta^{m \downarrow}\, 
e^{\sN {s, a} }_{\sigma(i_1\dots i_s)}
=&
\delta^{m \downarrow}\, 
e^{\sN {s, a} }_{i_p i_1\dots i_{p-1} i_{p+1}\dots i_s}
=
\delta_{m, i_p}
D^{\sN {s,1}}_{a b}\,
e^{\sN {s-1, b} }_{i_1\dots i_{p-1} i_{p+1}\dots i_s}
+
\dots
\nonumber\\
=
(M^{\sN s}_\sigma)_{a b}\,
\delta^{m \downarrow}\, 
e^{\sN {s, b} }_{i_1\dots i_s}
=&
\dots
+
\delta_{m, i_p}
(M^{\sN s}_\sigma)_{a c}\,
D^{\sN {s, p}}_{c b}\,
e^{\sN {s-1, b} }_{i_1\dots i_{p-1}\,i_{p+1} \dots i_{s}}
+
\dots
,
\end{align}
so we get
\begin{align}
D^{\sN {s,1}}
=
M^{\sN s}_{(1\dots p)}\,
D^{\sN {s, p}}
~~\Rightarrow~~
D^{\sN {s,p}}
=
M^{\sN s}_{(p\dots 1)}\,
D^{\sN {s, 1}}
.
\end{align}

For the case of the $A$ we need to consider the permutation $\sigma_{p q}$
$i_1\rightarrow i_p$, $i_2\rightarrow i_q$,
$i_3 \dots i_{p+1} \rightarrow i_1 i_2 \dots i_{p-1}$
and
$i_{p+2} \dots i_q \rightarrow i_{p+1} \dots i_{q-1}$
then
\begin{align}
\delta_A^{m \downarrow}\, 
e^{\sN {s, a} }_{\sigma_{p q}(i_1\dots i_s)}
=&
\delta_A^{m \downarrow}\, 
e^{\sN {s, a} }_{i_p i_q i_1\dots i_{p-1}\,i_{p+1} \dots i_{q-1}\,i_{q+1}\dots i_{s}}
=
\delta_{i_p, i_q}
A^{\sN {s, 1 2}}_{a b}\,
e^{\sN {s-1, b} }_{m i_1\dots i_{p-1}\,i_{p+1} \dots i_{q-1}\,i_{q+1}\dots i_{s}}
+
\dots
\nonumber\\
(M^{\sN s}_{\sigma_{p q}})_{a b}\,
\delta_A^{m \downarrow}\, 
e^{\sN {s, b} }_{i_1\dots i_s}
=&
\dots
+
\delta_{i_p, i_q}\,
(M^{\sN s}_{\sigma_{p q}})_{a c}\,
A^{\sN {s, p q}}_{c b}\,
e^{\sN {s-1, b} }_{m i_1\dots i_{p-1}\,i_{p+1} \dots i_{q-1}\,i_{q+1}\dots i_{s}}
+
\dots
,
\end{align}
so we get
\begin{align}
A^{\sN {s,1 2}}
=
M^{\sN s}_{\sigma_{p q}}\,
A^{\sN {s, p q}}
~~\Rightarrow~~
A^{\sN {s,p q}}
=
M^{\sN s}_{\sigma_{p q}^{-1}}\,
A^{\sN {s, 1 2}}
,
\end{align}
where $\sigma_{p q}^{-1}$ acts as
$i_1 \dots i_{p-1} \rightarrow i_3 \dots i_{p+1}$,
$i_p \rightarrow i_1$,
$i_{p+1} \dots i_{q-1} \rightarrow i_{p+2} \dots i_q$
and
$i_2 \rightarrow i_2$.

We can now compute the constraints from Lorentz algebra.
We recall that the covariant expression for the Lorentz generators is
\begin{equation}
  M^{\mu\nu}
  =
  x_0^\mu p_0^\nu - x_0^\nu p_0^\nu
  +
  i \sum_{n\ne 0} \frac{ \alpha^\mu_n \alpha^\nu_{-n} }{n}
  ,
\end{equation}
so that the \lc expression for the \nzm part of the generators of interest is
\begin{equation}
M^{i +}|_{l c \,\&\, \nzm}=0,
~~~~
M^{i -}|_{l c \,\&\, \nzm}=
  i \sum_{n\ne 0} \frac{ \alpha^i_{n (l c)} \hat \alpha^-_{-n (l c)} }{n},
~~\Rightarrow~~
M^{i 1}|_{l c \,\&\, \nzm}=
\frac{-1}{\sqrt{2}}
M^{i -}|_{l c \,\&\, \nzm}
,
\end{equation}  
with
$
  {\hat \alpha}^-_n
  =
  \frac{1}{2 \alpha^+_{0 (l c)} }
  \sum_m \alpha_{n-m (l c)}^i   \alpha_{m (l c)}^i
$.
If we use the commutation
\begin{equation}
[M^{ m 1}, M^{ l 1}] = i M^{m l}
,
\end{equation}
the definition $\delta^m = i \alpha^+_0 M^{m -}_{l c \,\&\, \nzm}$
and the fact that we are in rest frame so $\alpha^+_0 = \sap M$ (where
$M$ is the mass of the state) we
can write
\begin{align}
[\delta^m,\, \delta^l]
=&
[\sap M\, i M^{m-}_{l c \,\&\, \nzm},\, \sap M\, i M^{l-}_{l c \,\&\, \nzm}]
=
(-2 \ap\, M^2) [M^{m 1},\, M^{l 1}]
=
(-2 \ap\, M^2)\, i M^{m l}_{l c \,\&\, \nzm}
.
\label{eq:delta_delta_normalization}
\end{align}

In particular the action of $i M^{l m }_{l c \,\&\, \nzm}$ on a basis
element is simply 
\begin{equation}
i M^{m l }_{l c \,\&\, \nzm} e^{\sN {s, a} }_{i_1\dots i_s}
=
-
\sum_{p=1}^s
\delta_{m, i_p}
e^{\sN {s, a} }_{i_1\dots i_{p-1} l i_{p+1} \dots i_s}
+
\sum_{p=1}^s
\delta_{l, i_p}
e^{\sN {s, a} }_{i_1\dots i_{p-1} m i_{p+1} \dots i_s}
.
\label{eq:Mml_action_on_basis_element}
\end{equation}  

When computing $[\delta^m, \, \delta^l]$ we get a contribution which
increases the number of indexes by two, one which keeps the number of
indexes constant and one which decreases the number of indexes by two.

Because of Lorentz algebra and
eq. \eqref{eq:Mml_action_on_basis_element}
the two contributions which changes the
number of indexes must vanish while the other which keeps constant the
number of indexes is related to  swap $l \leftrightarrow m$ and
therefore to the matrix $M^{\sN s}_{l m}$.

Let us start from the contribution which increases the number of
indexes,
\begin{align}
[\delta^{m \uparrow}, \, \delta^ {l \uparrow}]\,
e^{\sN {s, a} }_{i_1\dots i_s}
=&
(U^{\sN s} U^{\sN {s+1}})_{a b}\, e^{\sN {s+2, b} }_{i_1\dots i_s l m}
-
(U^{\sN s} U^{\sN {s+1}})_{a b}\, e^{\sN {s+2, b} }_{i_1\dots i_s m l}
\nonumber\\
=&
\left( U^{\sN s} U^{\sN {s+1}}\, ( 1 - M^{\sN {s+2}}_{(s+1,s+2)} )\right)_{a b}\,
e^{\sN {s+2, b} }_{i_1\dots i_s l m}
=
0
,
\end{align}
where in the line we have used the matrix associated with the swap
$(s+1,\, s+2)$
\begin{equation}
e^{\sN {s+2, a} }_{i_1\dots i_s m l}
=
( M^{\sN {s+2}}_{(s+1,s+2)} )_{a b}\, 
e^{\sN {s+2, b} }_{i_1\dots i_s l m}
.
\end{equation}
It follows the matricial constraint
\begin{equation}
U^{\sN s} U^{\sN {s+1}}\, ( 1 - M^{\sN {s+2}}_{(s+1,s+2)} )
=
0
.
\label{eq:constraint_UU}
\end{equation}

We can now  consider the contribution which keeps the number of
indexes,
\begin{align}
[\delta^{m \uparrow}, \, \delta^ {l \downarrow}]\,
e^{\sN {s, a} }_{i_1\dots i_s}
+&
[\delta^{m \downarrow}, \, \delta^ {l \uparrow}]\,
e^{\sN {s, a} }_{i_1\dots i_s}
=
\nonumber\\
=&
\sum_{p=1}^s\,
\delta_{l\, i_p}\,
(D^{\sN {s, p}} U^{\sN {s-1}})_{a b}\,
e^{\sN {s, b} }_{i_1\dots i_{p-1}\, i_{p+1} \dots i_s m}
\nonumber\\
&+
\sum_{p=1}^{s-1}\,\sum_{q=p+1}^{s}\,
\delta_{i_p\, i_q}\,
(A^{\sN {s, p\, q}} U^{\sN {s-1}})_{a b}\,
e^{\sN {s, b} }_{l i_1\dots i_{p-1}\, i_{p+1} \dots
i_{q-1}\,i_{q+1} \dots i_s m}
\nonumber\\
&+
\sum_{p=1}^s\,
\delta_{m\, i_p}\,
(U^{\sN {s}} D^{\sN {s+1, p}})_{a b}\,
e^{\sN {s, b} }_{i_1\dots i_{p-1}\, i_{p+1} \dots i_s l}
+
\delta_{m\, l}\,
(U^{\sN {s}} D^{\sN {s+1, s+1}})_{a b}\,
e^{\sN {s, b} }_{i_1\dots i_s}
\nonumber\\
&+
\sum_{p=1}^{s-1}\,\sum_{q=p+1}^{s}\,
\delta_{i_p\, i_q}\,
(U^{\sN {s}} A^{\sN {s+1, p q}})_{a b}\,
e^{\sN {s, b} }_{m i_1\dots i_{p-1}\, i_{p+1} \dots
i_{q-1}\,i_{q+1} \dots i_s l}
\nonumber\\
&+
\sum_{p=1}^{s}\
\delta_{l,\, i_p}\,
(U^{\sN {s}} A^{\sN {s+1, p\, s+1}})_{a b}\,
e^{\sN {s, b} }_{m i_1\dots i_{p-1}\, i_{p+1} \dots i_s }
\nonumber\\
&-
(m \leftrightarrow l)
.
\end{align}
The terms proportional to $\delta_{i_p\, i_q}$ must cancel since they
are not in eq. \eqref{eq:Mml_action_on_basis_element}
and this implies
\begin{align}
\left(
A^{\sN {s, p\, q}}\, U^{\sN {s-1}} 
+
U^{\sN {s}}\, A^{\sN {s+1, p q}}\, M^{\sN s}_{(1 s)}
\right)
\left( 1 - M^{\sN s}_{(1 s)} \right)
=
0,
\label{eq:constraint_AU_UA}
\end{align}
where
$1 - M^{\sN s}_{(1 s)}$ implements the antisymmetry in $m l$.

Now if we look to the contribution proportional to $\delta_{l i_p}$
and compare with \eqref{eq:Mml_action_on_basis_element}
and \eqref{eq:delta_delta_normalization}
we get
\begin{align}
&
(D^{\sN {s, p}} U^{\sN {s-1}})_{a b}\,
e^{\sN {s, b} }_{i_1\dots i_{p-1}\, i_{p+1} \dots i_s m}
-
(U^{\sN {s}} D^{\sN {s+1, p}})_{a b}\,
e^{\sN {s, b} }_{i_1\dots i_{p-1}\, i_{p+1} \dots i_s m}
\nonumber\\
&+
(U^{\sN {s}} A^{\sN {s+1, p\, s+1}})_{a b}\,
e^{\sN {s, b} }_{m i_1\dots i_{p-1}\, i_{p+1} \dots i_s }
\nonumber\\
=&
(- 2 \ap\, M^2)\,
e^{\sN {s, a} }_{i_1\dots i_{p-1} m i_{p+1} \dots i_s}
,
\end{align}
which implies the constraint which can be written in matricial form as
\begin{align}
D^{\sN {s, p}} U^{\sN {s-1}}
-
U^{\sN {s}} D^{\sN {s+1, p}}
+
U^{\sN {s}} A^{\sN {s+1, p\, s+1}}\, M^{\sN s}_{(1\dots s)}
=
(- 2 \ap\, M^2)\,
M^{\sN s}_{(p\dots s)}
,
\label{eq:constraint_DU_UD}
\end{align}
where
$M^{\sN s}_{(1\dots s)}$ implements the change
from $m i_1\dots i_{p-1}\, i_{p+1} \dots i_s$
to $i_1\dots i_{p-1}\, i_{p+1} \dots  i_s m$
and
$M^{\sN s}_{(p\dots s)}$ implements the change
from $m i_{p+1} \dots i_s$
to $i_{p+1} \dots i_s m$.

\section{Dimensions of some $SO(25)$ and $\GL(*)$ irreps}
\label{app:irrep_dims}

We start with a Young diagram $Y_\lambda$
with $\mu_1\ge \mu_2  \dots \ge \mu_n$ rows, i.e.
\ytableausetup{mathmode, boxframe=normal, boxsize=2em}
\begin{align}
    \begin{ytableau}
1 & 2 & 3 & \none[\dots] & \scriptstyle \mu_1 - 1 & \scriptstyle \mu_1 \\
1 & 2 & 3 & \none[\dots]& \scriptstyle \mu_2 \\
\none[\vdots] & \none[\vdots] & \none[\vdots] \\
1 & \none[\dots]& \scriptstyle \mu_n &\none &\none \\
\end{ytableau}
  ,
\end{align}
which has $s=\sum_{k=1}^n \mu_k$ boxes.

We use the following general formula for computing the dimensions of
an irrep of $SO(2 n+1) $
(the limit to $n$ labels is due to the existence of the Hodge duality)
\begin{equation}
  dim_{SO(2 n +1 )}(Y_\lambda)
  =
  \frac{ \prod_{1\le i<j \le n}  (R_i+ R_j) }{ \prod_{1\le i<j \le n}  (r_i+ r_j)}
  \frac{ \prod_{i=1}^n  R_i }{ \prod_{i=1}^n  r_i}
  ,
\end{equation}
where we have defined the vectors
\begin{equation}
  \begin{array} {c | c | c}
    r & \mu & R
    \\
    \hline
    n+\oh & \mu_1 & n+\oh+\mu_1
    \\
    n-\oh & \mu_2 & n-\oh+\mu_2
    \\
    \vdots  &   \vdots  &   \vdots 
    \\
    \oh & \mu_n & \oh+\mu_n
  \end{array}
  .
\end{equation}

In the same way we can use hook formula for computing the dimension of
the previous Young diagram for $S_s$ irreps
\begin{align}
  dim_{S_s}(Y_\lambda)
  = \frac{s!}{\prod h_\lambda (i,j)}
  ,
\end{align}
where the product is over all cells $(i,j)$ of the Young diagram.
The hook $\displaystyle H_{\lambda }(i,j)$
is the set of cells $\displaystyle (a,b)$
such that $a = i$ and ${\displaystyle b\geq j}$
or ${\displaystyle a\geq i}$ and ${\displaystyle b=j}$.
The hook length $h_\lambda ( i , j )$ is the number of cells in
${\displaystyle H_{\lambda }(i,j)}$

\ytableausetup{mathmode,  boxframe=normal, boxsize=0.5em}
The result for the $SO(25)$ and
$S_s$ dimensions is the following for $s=1, 2, 3$
\begin{align}
  \bullet & ~(1,\,1)\, (N\ge 0)
  &&
  &&
\nonumber\\
\ydiagram{1} & ~(25,\,1)\, (N\ge 1)
,
&&
&&
\nonumber\\
  \ydiagram{2} & ~(324,\,1)\, (N\ge 2)
  &\ydiagram{1,1} & ~(300,\,1)(N\ge 3)
  ,
  &&
\nonumber\\
  \ydiagram{3} & ~(2900,\,1)\, (N\ge 3)
  &\ydiagram{2,1} & ~(5175,\,2)\, (N\ge 4)
  &\ydiagram{1,1,1} & ~(2300,\,1)\, (N\ge 6)
  ,
  \nonumber\\
\end{align}
and for $s=4$
\begin{align}
  \ydiagram{4} & ~(20150,\,1)\, (N\ge 4)
  &\ydiagram{3,1} & ~(52026,\,3)\, (N\ge 5)
  &\ydiagram{2,2} & ~(32175,\,2)\, (N\ge 6)
  \nonumber\\
  \ydiagram{2,1,1} & ~(44550,\,3)\, (N\ge 7)
  &\ydiagram{1,1,1,1} & ~(12650,\,1)\, (N\ge 10)
  ,
  &&
  \nonumber\\
\end{align}
and for $s=5$
\begin{align}
  \ydiagram{5} & ~(115830,\,1)\, (N\ge 5)
  &\ydiagram{4,1} & ~(385020,\,4)\, (N\ge 6)
  &\ydiagram{3,2} & ~(430650,\,5)\,(N\ge 7)
  \nonumber\\
  \ydiagram{3,1,1} & ~(476905,\,6)\, (N\ge 8)
  &\ydiagram{2,2,1} & ~(368550,\,5)\, (N\ge 9)
  &\ydiagram{2,1,1,1} & ~(260820,\,4)\, (N\ge 11)
  \nonumber\\
  \ydiagram{1,1,1,1,1} & ~(53130,\,1)\,(N\ge 15)
,
  &&
  &&
  \nonumber\\
\end{align}
and for $s=6$
\begin{align}
  \ydiagram{6} & ~(573300,\,1)\, (N\ge 6)
  &\ydiagram{5,1} & ~(2302300,\,5)\, (N\ge 7)
  &\ydiagram{4,2} & ~(3580500,\,9)\,(N\ge 8)
  \nonumber\\
  \ydiagram{3,3} & ~(1848924,\,5)\, (N\ge 9)
  &\ydiagram{4,1,1} & ~(3670524,\,10)\, (N\ge 9)
  &\ydiagram{3,2,1} & ~(5252625\,16)\, (N\ge 10)
  \nonumber\\
  \ydiagram{2,2,2} & ~(1462500,\,5)\, (N\ge 12)
  &\ydiagram{2,2,1,1} & ~(2421900,\,9) \,(N\ge 13)
  &\ydiagram{2,1,1,1,1} & ~(1138500,\,5)\, (N\ge 16)
  \nonumber\\
  \ydiagram{1,1,1,1,1} & ~(177100,\,1)\, (N\ge 21)
  .
  &&
  &&
\end{align}
and for $s=7$
\begin{align}
  \ydiagram{7} & ~(2510820,\,1)\, (N\ge 7)
  &\ydiagram{6,1} & ~(11705850,\,6)\, (N\ge 8)
  &\ydiagram{5,2} & ~(22808500,\,7)\,(N\ge 9)
  \nonumber\\
  \ydiagram{4,3} & ~(20470230,\,14)\, (N\ge 9)
  &\ydiagram{5,1,1} & ~(22542300,\,15)\, (N\ge 9)
  .
  &&
\end{align}
and for $s=8$
\begin{align}
  \ydiagram{8} & ~(9924525,\,1)\, (N\ge 8)
  &\ydiagram{7,1} & ~(52272675,\,6)\, (N\ge 9)
  &\ydiagram{6,2} & ~(120656250,\,20)\,(N\ge 9)
  .
\end{align}
and for $s=9$
\begin{align}
  \ydiagram{9} & ~(35937525,\,1)\, (N\ge 9)
  &\ydiagram{8,1} & ~(209664780,\,6)\, (N\ge 10)
  &&
  .
\end{align}
and for $s=10$
\begin{align}
  \ydiagram{10} & ~(120609840,\,1)\, (N\ge 10)
&&
  &&
  .
\end{align}

\section{Level 3}
\label{app:level3}

\ytableausetup{boxsize=1em}

In the following we give
either all states for some chosen $SO(*)$ irreps
or the explicit top level states in $GL(*)$ irreps.
In both case states are in the rest frame.
This means that for $GL(*)$ traces must still be subtracted
when some indexes are equal
and the states can be boosted as discussed in the main text.
\subsection{Basis}

\myscaleT{ 
\parbox{\linewidth}{ 
\begin{align}
T_{3, 0}= 
 \{ 
&
\left(2 , 1\right)
\} 
 \end{align} 

} 

} 

\myscaleT{ 
\parbox{\linewidth}{ 
\begin{align}
T_{3, 1}= 
 \{ 
&
\left(1 , 1\right)\,({1}^{i_{1}})
,\,
({3}^{i_{1}})
\} 
 \end{align} 

} 

} 

\myscaleT{ 
\parbox{\linewidth}{ 
\begin{align}
T_{3, 2}= 
 \{ 
&
({1}^{i_{2}})\,({2}^{i_{1}})
,\,
({1}^{i_{1}})\,({2}^{i_{2}})
\} 
 \end{align} 

} 

} 

\myscaleT{ 
\parbox{\linewidth}{ 
\begin{align}
T_{3, 3}= 
 \{ 
&
({1}^{i_{1}})\,({1}^{i_{2}})\,({1}^{i_{3}})
\} 
 \end{align} 

} 

} 

\subsection{$SO(25)$ tensors with 0 indexes}

No irreps with spin 0 are present.

\subsection{$SO(25)$ tensors with 1 indexes}

No irreps with spin 1 are present.

\subsection{$SO(25)$ tensors with 2 indexes}

We give the expansion of the $SO(25)$ tensors 
on the basis $T_{3,\, s}$ with $ 0 \le s \le 2$
given above.

\subsubsection{Irrep $ \ydiagram{1,1} $}

The expression for the given irrep for the coefficients 
 on the basis elements reads as follows.


\myscaleM{ 
\parbox{\linewidth}{ 
\begin{align}
b^{\sN {N=3,\, s=2->2} }=
 &
\left( 
\begin{matrix}
1
&
-1
\end{matrix}
\right) 
.  \end{align}
} 
} 



The irrep matrices  associated with the swaps $1\leftrightarrow k$ read
  as follows.

\myscaleR{ 
\parbox{\linewidth}{ 
\begin{align} 
R[(1, 2)]= 
& 
\begin{pmatrix} -1\cr \end{pmatrix} 
. 
\end{align} 
 
} 

} 



The expression including explicitly the basis elements for symmetric tensor number 1 reads as follows.

\myscaleM{ 
\parbox{\linewidth}{ 
\begin{align}
|{\Yoo {i_1} {i_2}} _{(n=1 )} \rangle=
&
\left( 
\begin{matrix}
-({1}^{i_{2}})\,({2}^{i_{1}})
&
+({1}^{i_{1}})\,({2}^{i_{2}})
\end{matrix}
\right) 
,  \end{align}
} 
} 

and

\myscaleM{ 
\parbox{\linewidth}{ 
\begin{align}
|{\Yoo 1 {i_1} } _{(n=1 )} \rangle=
&
\left( 
\begin{matrix}
-{{\left(1 , 1\right)\,({1}^{i_{1}})}\over{4}}
&
+{{({3}^{i_{1}})}\over{2}}
\end{matrix}
\right) 
,  \end{align}
} 
} 

with squared norm

\begin{equation}%
\parallel\, |\Yoo I J _{(n=1 )} \rangle \, \parallel^2
          =
4
.
\end{equation}

\subsection{$SO(25)$ tensors with 3 indexes}

We give the expansion of the $SO(25)$ tensors 
on the basis $T_{3,\, s}$ with $ 0 \le s \le 3$
given above.

\subsubsection{Irrep $ \ydiagram{3} $}

The expression for the given irrep for the coefficients 
 on the basis elements reads as follows.


\myscaleM{ 
\parbox{\linewidth}{ 
\begin{align}
b^{\sN {N=3,\, s=3->3} }=
 &
\left( 
\begin{matrix}
1
\end{matrix}
\right) 
.  \end{align}
} 
} 



The irrep matrices  associated with the swaps $1\leftrightarrow k$ read
  as follows.

\myscaleR{ 
\parbox{\linewidth}{ 
\begin{align} 
R[(1, 2)]= 
& 
\begin{pmatrix} 1\cr \end{pmatrix} 
, 
\end{align} 
 
} 

} 


\myscaleR{ 
\parbox{\linewidth}{ 
\begin{align} 
R[(1, 3)]= 
& 
\begin{pmatrix} 1\cr \end{pmatrix} 
. 
\end{align} 
 
} 

} 



The expression including explicitly the basis elements for symmetric tensor number 1 reads as follows.

\myscaleM{ 
\parbox{\linewidth}{ 
\begin{align}
|{\Ytr {i_1} {i_2} {i_3}} _{(n=1 )} \rangle=
&
\left( 
\begin{matrix}
+({1}^{i_{1}})\,({1}^{i_{2}})\,({1}^{i_{3}})
\end{matrix}
\right) 
,  \end{align}
} 
} 

and

\myscaleM{ 
\parbox{\linewidth}{ 
\begin{align}
|{\Ytr 1 {i_1} {i_2}} _{(n=1 )} \rangle=
&
\left( 
\begin{matrix}
+{{({1}^{i_{2}})\,({2}^{i_{1}})}\over{2}}
&
+{{({1}^{i_{1}})\,({2}^{i_{2}})}\over{2}}
\end{matrix}
\right) 
,  \end{align}
} 
} 

and

\myscaleM{ 
\parbox{\linewidth}{ 
\begin{align}
|{\Ytr {i_1} {i_1} {i_2}} _{(1)\,(n=1 )} \rangle=
&
\left( 
\begin{matrix}
+{{\left(1 , 1\right)\,({1}^{i_{2}})}\over{16}}
&
-{{({1}^{i_{1}})^2\,({1}^{i_{2}})}\over{2}}
&
+{{3\,({3}^{i_{2}})}\over{8}}
\end{matrix}
\right) 
,  \end{align}
} 
} 

and

\myscaleM{ 
\parbox{\linewidth}{ 
\begin{align}
|{\Ytr {1} {i_1} {i_1}} _{(1)\,(n=1 )} \rangle=
&
\left( 
\begin{matrix}
-{{\left(2 , 1\right)}\over{2^{{{5}\over{2}}}\,\sqrt{3}}}
&
+{{\sqrt{3}\,({1}^{i_{1}})\,({2}^{i_{1}})}\over{2^{{{3}\over{2}}}}}
\end{matrix}
\right) 
,  \end{align}
} 
} 

with squared norm

\begin{equation}%
\parallel\, |\Ytr I J K_{(n=1 )} \rangle \, \parallel^2
          =
1
.
\end{equation}


\section{Level 4}
\label{app:level4}
In the following we give the explicit expansions for the states in $\GL(*)$ irreps in the rest frame.
This means that traces must still be subtracted and the states can be boosted as discussed in the main text.

In the following we give
either all states for some chosen $SO(*)$ irreps
or the explicit top level states in $GL(*)$ irreps.
In both case states are in the rest frame.
This means that for $GL(*)$ traces must still be subtracted
when some indexes are equal
and the states can be boosted as discussed in the main text.
\subsection{Basis}

\myscaleT{ 
\parbox{\linewidth}{ 
\begin{align}
T_{4, 0}= 
 \{ 
&
\left(1 , 1\right)^2
,\,
\left(2 , 2\right)
,\,
\left(3 , 1\right)
\} 
 \end{align} 

} 

} 

\myscaleT{ 
\parbox{\linewidth}{ 
\begin{align}
T_{4, 1}= 
 \{ 
&
({1}^{i_{1}})\,\left(2 , 1\right)
,\,
\left(1 , 1\right)\,({2}^{i_{1}})
,\,
({4}^{i_{1}})
\} 
 \end{align} 

} 

} 

\myscaleT{ 
\parbox{\linewidth}{ 
\begin{align}
T_{4, 2}= 
 \{ 
&
\left(1 , 1\right)\,({1}^{i_{1}})\,({1}^{i_{2}})
,\,
({2}^{i_{1}})\,({2}^{i_{2}})
,\,
({1}^{i_{2}})\,({3}^{i_{1}})
,\,
({1}^{i_{1}})\,({3}^{i_{2}})
\} 
 \end{align} 

} 

} 

\myscaleT{ 
\parbox{\linewidth}{ 
\begin{align}
T_{4, 3}= 
 \{ 
&
({1}^{i_{2}})\,({1}^{i_{3}})\,({2}^{i_{1}})
,\,
({1}^{i_{1}})\,({1}^{i_{3}})\,({2}^{i_{2}})
,\,
({1}^{i_{1}})\,({1}^{i_{2}})\,({2}^{i_{3}})
\} 
 \end{align} 

} 

} 

\myscaleT{ 
\parbox{\linewidth}{ 
\begin{align}
T_{4, 4}= 
 \{ 
&
({1}^{i_{1}})\,({1}^{i_{2}})\,({1}^{i_{3}})\,({1}^{i_{4}})
\} 
 \end{align} 

} 

} 

\subsection{$SO(25)$ tensors with 0 indexes}

We give the expansion of the $SO(25)$ tensors 
on the basis $T_{4,\, s}$ with $ 0 \le s \le 0$
given above.

\subsubsection{Irrep $ \ydiagram{0} $}

The expression for the given irrep for the coefficients 
 on the basis elements reads as follows.


\myscaleM{ 
\parbox{\linewidth}{ 
\begin{align}
b^{\sN {N=4,\, s=0->0} }=
 &
\left( 
\begin{matrix}
-1
&
-7
&
10
\end{matrix}
\right) 
,  \end{align}
} 
} 



The expression including explicitly the basis elements for scalar number 1 reads as follows.

\myscaleM{ 
\parbox{\linewidth}{ 
\begin{align}
|\bullet_{(n=1 )} \rangle=
&
\left( 
\begin{matrix}
-\left(1 , 1\right)^2
&
-7\,\left(2 , 2\right)
&
+10\,\left(3 , 1\right)
\end{matrix}
\right) 
,  \end{align}
} 
} 

with squared norm

\begin{equation}%
\parallel\, |\bullet_{(n=1 )} \rangle \, \parallel^2
          =
21600
.
\end{equation}

\subsection{$SO(25)$ tensors with 1 indexes}

No irreps with spin 1 are present.

\subsection{$SO(25)$ tensors with 2 indexes}

We give the expansion of the $SO(25)$ tensors 
on the basis $T_{4,\, s}$ with $ 0 \le s \le 2$
given above.

\subsubsection{Irrep $ \ydiagram{2} $}

The expression for the given irrep for the coefficients 
 on the basis elements reads as follows.


\myscaleM{ 
\parbox{\linewidth}{ 
\begin{align}
b^{\sN {N=4,\, s=2->2} }=
 &
\left( 
\begin{matrix}
-1
&
-7
&
4
&
4
\end{matrix}
\right) 
.  \end{align}
} 
} 



The irrep matrices  associated with the swaps $1\leftrightarrow k$ read
  as follows.

\myscaleR{ 
\parbox{\linewidth}{ 
\begin{align} 
R[(1, 2)]= 
& 
\begin{pmatrix} 1\cr \end{pmatrix} 
. 
\end{align} 
 
} 

} 



The expression including explicitly the basis elements for symmetric tensor number 1 reads as follows.

\myscaleM{ 
\parbox{\linewidth}{ 
\begin{align}
|{\Yt {i_1} {i_2}} _{(n=1 )} \rangle=
&
\left( 
\begin{matrix}
-\left(1 , 1\right)\,({1}^{i_{1}})\,({1}^{i_{2}})
&
-7\,({2}^{i_{1}})\,({2}^{i_{2}})
&
+4\,({1}^{i_{2}})\,({3}^{i_{1}})
&
+4\,({1}^{i_{1}})\,({3}^{i_{2}})
\end{matrix}
\right) 
,  \end{align}
} 
} 

and

\myscaleM{ 
\parbox{\linewidth}{ 
\begin{align}
|{\Yt 1 {i_1} } _{(n=1 )} \rangle=
&
\left( 
\begin{matrix}
+{{2\,({1}^{i_{1}})\,\left(2 , 1\right)}\over{\sqrt{6}}}
&
-{{9\,\left(1 , 1\right)\,({2}^{i_{1}})}\over{2\,\sqrt{6}}}
&
+{{2\,({4}^{i_{1}})}\over{\sqrt{6}}}
\end{matrix}
\right) 
,  \end{align}
} 
} 

and

\myscaleM{ 
\parbox{\linewidth}{ 
\begin{align}
|{\Yt {i_1} {i_1} } _{(1)\,(n=1 )} \rangle=
&
\left( 
\begin{matrix}
-{{3\,\left(1 , 1\right)^2}\over{16}}
&
+{{\left(1 , 1\right)\,({1}^{i_{1}})^2}\over{2}}
&
+{{\left(2 , 2\right)}\over{4}}
&
+{{7\,({2}^{i_{1}})^2}\over{2}}
&
-{{\left(3 , 1\right)}\over{4}}
&
-4\,({1}^{i_{1}})\,({3}^{i_{1}})
\end{matrix}
\right) 
,  \end{align}
} 
} 

with squared norm

\begin{equation}%
\parallel\, |\Yt I J _{(n=1 )} \rangle \, \parallel^2
          =
348
.
\end{equation}

\subsection{$SO(25)$ tensors with 3 indexes}

We give the expansion of the $SO(25)$ tensors 
on the basis $T_{4,\, s}$ with $ 0 \le s \le 3$
given above.

\subsubsection{Irrep $ \ydiagram{2,1} $}

The expression for the given irrep for the coefficients 
 on the basis elements reads as follows.


\myscaleM{ 
\parbox{\linewidth}{ 
\begin{align}
b^{\sN {N=4,\, s=3->3} }=
 &
\left( 
\begin{matrix}
1
&
0
&
-1
\\
0
&
1
&
-1
\end{matrix}
\right) 
.  \end{align}
} 
} 



The irrep matrices  associated with the swaps $1\leftrightarrow k$ read
  as follows.

\myscaleR{ 
\parbox{\linewidth}{ 
\begin{align} 
R[(1, 2)]= 
& 
\begin{pmatrix} 0&1\cr 1&0\cr \end{pmatrix} 
, 
\end{align} 
 
} 

} 


\myscaleR{ 
\parbox{\linewidth}{ 
\begin{align} 
R[(1, 3)]= 
& 
\begin{pmatrix} -1&0\cr -1&1\cr \end{pmatrix} 
. 
\end{align} 
 
} 

} 



The expression including explicitly the basis elements for $\ydiagram{2,1}$ tensor number 1 reads as follows.

\myscaleM{ 
\parbox{\linewidth}{ 
\begin{align}
|{\Yto {i_1} {i_2} {i_3}} _{(n=1 )} \rangle=
&
\left( 
\begin{matrix}
-({1}^{i_{2}})\,({1}^{i_{3}})\,({2}^{i_{1}})
&
+({1}^{i_{1}})\,({1}^{i_{2}})\,({2}^{i_{3}})
\end{matrix}
\right) 
,  \end{align}
} 
} 

and

\myscaleM{ 
\parbox{\linewidth}{ 
\begin{align}
|{\Yto 1 {i_1} {i_2}} _{(n=1 )} \rangle=
&
\left( 
\begin{matrix}
+{{\left(1 , 1\right)\,({1}^{i_{1}})\,({1}^{i_{2}})}\over{2\,\sqrt{6}}}
&
-{{({2}^{i_{1}})\,({2}^{i_{2}})}\over{\sqrt{6}}}
&
+{{({1}^{i_{2}})\,({3}^{i_{1}})}\over{\sqrt{6}}}
&
-{{({1}^{i_{1}})\,({3}^{i_{2}})}\over{\sqrt{6}}}
\end{matrix}
\right) 
,  \end{align}
} 
} 

and

\myscaleM{ 
\parbox{\linewidth}{ 
\begin{align}
|{\Yto {i_1} {i_1} {i_2}} _{(1)\,(n=1 )} \rangle=
&
\left( 
\begin{matrix}
-{{({1}^{i_{2}})\,\left(2 , 1\right)}\over{3^{{{3}\over{2}}}}}
&
+{{({1}^{i_{1}})\,({1}^{i_{2}})\,({2}^{i_{1}})}\over{\sqrt{3}}}
&
-{{({1}^{i_{1}})^2\,({2}^{i_{2}})}\over{\sqrt{3}}}
&
+{{2\,({4}^{i_{2}})}\over{3^{{{3}\over{2}}}}}
\end{matrix}
\right) 
,  \end{align}
} 
} 

and

\myscaleM{ 
\parbox{\linewidth}{ 
\begin{align}
|{\Yto {i_1} {i_1} {1}} _{(2)\,(n=1 )} \rangle=
&
\left( 
\begin{matrix}
-{{\left(1 , 1\right)\,({1}^{2})^2}\over{2\,\sqrt{3}\,\sqrt{6}}}
&
+{{\left(1 , 1\right)\,({1}^{i_{1}})^2}\over{2\,\sqrt{3}\,\sqrt{6}}}
&
+{{({2}^{2})^2}\over{\sqrt{3}\,\sqrt{6}}}
&
-{{({2}^{i_{1}})^2}\over{\sqrt{3}\,\sqrt{6}}}
\end{matrix}
\right) 
,  \end{align}
} 
} 

with squared norm

\begin{equation}%
\parallel\, |\Yto I J K _{(n=1 )} \rangle \, \parallel^2
          =
4
.
\end{equation}

\subsection{$SO(25)$ tensors with 4 indexes}

We give the expansion of the $SO(25)$ tensors 
on the basis $T_{4,\, s}$ with $ 0 \le s \le 4$
given above.

\subsubsection{Irrep $ \ydiagram{4} $}

The expression for the given irrep for the coefficients 
 on the basis elements reads as follows.


\myscaleM{ 
\parbox{\linewidth}{ 
\begin{align}
b^{\sN {N=4,\, s=4->4} }=
 &
\left( 
\begin{matrix}
1
\end{matrix}
\right) 
.  \end{align}
} 
} 



The irrep matrices  associated with the swaps $1\leftrightarrow k$ read
  as follows.

\myscaleR{ 
\parbox{\linewidth}{ 
\begin{align} 
R[(1, 2)]= 
& 
\begin{pmatrix} 1\cr \end{pmatrix} 
, 
\end{align} 
 
} 

} 


\myscaleR{ 
\parbox{\linewidth}{ 
\begin{align} 
R[(1, 3)]= 
& 
\begin{pmatrix} 1\cr \end{pmatrix} 
, 
\end{align} 
 
} 

} 


\myscaleR{ 
\parbox{\linewidth}{ 
\begin{align} 
R[(1, 4)]= 
& 
\begin{pmatrix} 1\cr \end{pmatrix} 
. 
\end{align} 
 
} 

} 


Since the irrep has not being fully built we give the only sensible
  descendant. 


The expression including explicitly the basis elements for 4 indexes reads as follows.

\myscaleM{ 
\parbox{\linewidth}{ 
\begin{align}
(b \cdot e)^{\sN {N=4,\, s=4->4} }=
 &
\left( 
\begin{matrix}
+({1}^{i_{1}})\,({1}^{i_{2}})\,({1}^{i_{3}})\,({1}^{i_{4}})
\end{matrix}
\right) 
,  \end{align}
} 
} 



The expression including explicitly the basis elements for 3 indexes reads as follows.

\myscaleM{ 
\parbox{\linewidth}{ 
\begin{align}
(b \cdot e)^{\sN {N=4,\, s=4->3} }=
 &
\left( 
\begin{matrix}
+({1}^{i_{2}})\,({1}^{i_{3}})\,({2}^{i_{1}})
&
+({1}^{i_{1}})\,({1}^{i_{3}})\,({2}^{i_{2}})
&
+({1}^{i_{1}})\,({1}^{i_{2}})\,({2}^{i_{3}})
\end{matrix}
\right) 
.  \end{align}
} 
} 



\section{Explicit form of scalars up to level $N=10$}
\label{app:scalars_up_to_10}

We give the explicit expressions for the scalars up to level $10$ and
the expressions where the coefficients are factorized over primes.
These show quite  big prime numbers which increase rapidly with the level.

\newgeometry{a4paper,landscape,left=0.1in,right=0.5in,top=1in,bottom=0.5in,%
nohead
}

\begin{align*}
|\bullet_{(N=4,\, n=1)} \rangle
=
+10&\, \left(3 , 1\right) -7\, \left(2 , 2\right) -1\, \left(1 , 1\right)^2 
\nonumber\\
=+2*5&\, \left(3 , 1\right) -7\, \left(2 , 2\right) -1\, \left(1 , 1\right)^2 
\nonumber\\
\parallel |\bullet_{(N=4,\, n=1)} \rangle \parallel ^2
=&
21600
.
\end{align*}

\begin{align*}
|\bullet_{(N=6,\, n=1)} \rangle
=
-84&\, \left(1 , 1\right)\,\left(3 , 1\right) +54\, \left(1 , 1\right)\,\left(2 , 2\right) +24\, \left(2 , 1\right)^2 +5\, \left(1 , 1\right)^3 
\nonumber\\
+24&\, \left(5 , 1\right) -336\, \left(4 , 2\right) +280\, \left(3 , 3\right) 
\nonumber\\
=-2^{2}*3*7&\, \left(1 , 1\right)\,\left(3 , 1\right) +2*3^{3}\, \left(1 , 1\right)\,\left(2 , 2\right) +2^{3}*3\, \left(2 , 1\right)^2 +5\, \left(1 , 1\right)^3 
\nonumber\\
+2^{3}*3&\, \left(5 , 1\right) -2^{4}*3*7\, \left(4 , 2\right) +2^{3}*5*7\, \left(3 , 3\right) 
\nonumber\\
\parallel |\bullet_{(N=6,\, n=1)} \rangle \parallel ^2
=&
133632000
.
\end{align*}

\begin{align*}
\bullet_{(N=8,\, n=1)} \rangle
=
+36960&\, \left(2 , 2\right)^2 +10560\, \left(1 , 1\right)^2\,\left(2 , 2\right) +480\, \left(1 , 1\right)^4 
\nonumber\\
+86400&\, \left(3 , 1\right)^2 -105600\, \left(2 , 2\right)\,\left(3 , 1\right) -9600\, \left(1 , 1\right)^2\,\left(3 , 1\right) 
\nonumber\\
-38400&\, \left(1 , 1\right)\,\left(5 , 1\right) +147840\, \left(4 , 4\right) +15360\, \left(1 , 1\right)\,\left(3 , 3\right) 
\nonumber\\
+9600&\, \left(7 , 1\right) -163200\, \left(5 , 3\right) 
\nonumber\\
=+2^{5}*3*5*7*11&\, \left(2 , 2\right)^2 +2^{6}*3*5*11\, \left(1 , 1\right)^2\,\left(2 , 2\right) +2^{5}*3*5\, \left(1 , 1\right)^4 
\nonumber\\
+2^{7}*3^{3}*5^{2}&\, \left(3 , 1\right)^2 -2^{7}*3*5^{2}*11\, \left(2 , 2\right)\,\left(3 , 1\right) -2^{7}*3*5^{2}\, \left(1 , 1\right)^2\,\left(3 , 1\right) 
\nonumber\\
-2^{9}*3*5^{2}&\, \left(1 , 1\right)\,\left(5 , 1\right) +2^{7}*3*5*7*11\, \left(4 , 4\right) +2^{10}*3*5\, \left(1 , 1\right)\,\left(3 , 3\right) 
\nonumber\\
+2^{7}*3*5^{2}&\, \left(7 , 1\right) -2^{7}*3*5^{2}*17\, \left(5 , 3\right) 
\nonumber\\
\parallel |\bullet_{(N=8,\, n=1)} \rangle \parallel ^2
=&
2511129600
.
\end{align*}
\begin{align*}
\bullet_{(N=8,\, n=2,\, NO)} \rangle
=
+1924&\, \left(1 , 1\right)^2\,\left(2 , 2\right) +2720\, \left(1 , 1\right)\,\left(2 , 1\right)^2 +157\, \left(1 , 1\right)^4 
\nonumber\\
+8960&\, \left(2 , 2\right)\,\left(3 , 1\right) -4160\, \left(1 , 1\right)^2\,\left(3 , 1\right) -1636\, \left(2 , 2\right)^2 
\nonumber\\
+27904&\, \left(1 , 1\right)\,\left(3 , 3\right) +5120\, \left(2 , 1\right)\,\left(3 , 2\right) -2560\, \left(3 , 1\right)^2 
\nonumber\\
+83856&\, \left(4 , 4\right) -36640\, \left(1 , 1\right)\,\left(4 , 2\right) -12160\, \left(2 , 1\right)\,\left(4 , 1\right) 
\nonumber\\
+9600&\, \left(6 , 2\right) -97920\, \left(5 , 3\right) +8960\, \left(1 , 1\right)\,\left(5 , 1\right) 
\nonumber\\
=+2^{2}*13*37&\, \left(1 , 1\right)^2\,\left(2 , 2\right) +2^{5}*5*17\, \left(1 , 1\right)\,\left(2 , 1\right)^2 +157\, \left(1 , 1\right)^4 
\nonumber\\
+2^{8}*5*7&\, \left(2 , 2\right)\,\left(3 , 1\right) -2^{6}*5*13\, \left(1 , 1\right)^2\,\left(3 , 1\right) -2^{2}*409\, \left(2 , 2\right)^2 
\nonumber\\
+2^{8}*109&\, \left(1 , 1\right)\,\left(3 , 3\right) +2^{10}*5\, \left(2 , 1\right)\,\left(3 , 2\right) -2^{9}*5\, \left(3 , 1\right)^2 
\nonumber\\
+2^{4}*3*1747&\, \left(4 , 4\right) -2^{5}*5*229\, \left(1 , 1\right)\,\left(4 , 2\right) -2^{7}*5*19\, \left(2 , 1\right)\,\left(4 , 1\right) 
\nonumber\\
+2^{7}*3*5^{2}&\, \left(6 , 2\right) -2^{7}*3^{2}*5*17\, \left(5 , 3\right) +2^{8}*5*7\, \left(1 , 1\right)\,\left(5 , 1\right).
\nonumber\\
\end{align*}
Notice however that the previous two scalars are not othogonal (NO).
Using Gram-Schmidt procedure the second can be made orthogonal as
\begin{align}
|\bullet_{(n=2 )} \rangle=
&
\left.
\begin{matrix}
-3825\,\left(1 , 1\right)^4
&
-94112\,\left(1 , 1\right)\,\left(2 , 1\right)^2
&
-31212\,\left(1 , 1\right)^2\,\left(2 , 2\right)
&
+180360\,\left(2 , 2\right)^2
&
+111792\,\left(1 , 1\right)^2\,\left(3 , 1\right)
\end{matrix}
\right . 
\nonumber \\ 
&
\begin{matrix}
-663600\,\left(2 , 2\right)\,\left(3 , 1\right)
&
+377872\,\left(3 , 1\right)^2
&
-177152\,\left(2 , 1\right)\,\left(3 , 2\right)
&
-914048\,\left(1 , 1\right)\,\left(3 , 3\right)
&
+420736\,\left(2 , 1\right)\,\left(4 , 1\right)
\end{matrix}
\nonumber \\ 
&
\begin{matrix}
+1267744\,\left(1 , 1\right)\,\left(4 , 2\right)
&
-2406400\,\left(4 , 4\right)
&
-438592\,\left(1 , 1\right)\,\left(5 , 1\right)
&
+2841584\,\left(5 , 3\right)
&
-332160\,\left(6 , 2\right)
\end{matrix}
\nonumber \\ 
&
\left. 
\begin{matrix}
+32144\,\left(7 , 1\right)
\end{matrix}
\right..
\nonumber\\
\parallel\, |\bullet_{(n=2 )} \rangle \, \parallel^2
=&
63715835525529600
.
\end{align}

\begin{align*}
\bullet_{(N=10,\, n=1)} \rangle
=
-7001360&\, \left(1 , 1\right)^3\,\left(2 , 2\right) -5890880\, \left(1 , 1\right)^2\,\left(2 , 1\right)^2 -317384\, \left(1 , 1\right)^5 
\nonumber\\
+10143680&\, \left(1 , 1\right)^3\,\left(3 , 1\right) -9604800\, \left(1 , 1\right)\,\left(2 , 2\right)^2 +68240640\, \left(2 , 1\right)^2\,\left(2 , 2\right) 
\nonumber\\
-110560640&\, \left(1 , 1\right)\,\left(3 , 1\right)^2 +155847360\, \left(1 , 1\right)\,\left(2 , 2\right)\,\left(3 , 1\right) -160044800\, \left(2 , 1\right)^2\,\left(3 , 1\right) 
\nonumber\\
+22608640&\, \left(1 , 1\right)^2\,\left(3 , 3\right) -599759360\, \left(3 , 2\right)^2 -176368640\, \left(1 , 1\right)\,\left(2 , 1\right)\,\left(3 , 2\right) 
\nonumber\\
+251264000&\, \left(1 , 1\right)\,\left(2 , 1\right)\,\left(4 , 1\right) -319298560\, \left(3 , 1\right)\,\left(3 , 3\right) +453156480\, \left(2 , 2\right)\,\left(3 , 3\right) 
\nonumber\\
-6512000&\, \left(1 , 1\right)^2\,\left(4 , 2\right) -1250283520\, \left(4 , 1\right)^2 +1676595200\, \left(3 , 2\right)\,\left(4 , 1\right) 
\nonumber\\
-431083520&\, \left(2 , 1\right)\,\left(4 , 3\right) -105996800\, \left(3 , 1\right)\,\left(4 , 2\right) +132272640\, \left(2 , 2\right)\,\left(4 , 2\right) 
\nonumber\\
-1130586240&\, \left(2 , 2\right)\,\left(5 , 1\right) -35996800\, \left(1 , 1\right)^2\,\left(5 , 1\right) -6471680\, \left(1 , 1\right)\,\left(4 , 4\right) 
\nonumber\\
+152270720&\, \left(1 , 1\right)\,\left(5 , 3\right) +236597760\, \left(2 , 1\right)\,\left(5 , 2\right) +864734720\, \left(3 , 1\right)\,\left(5 , 1\right) 
\nonumber\\
-136550400&\, \left(1 , 1\right)\,\left(6 , 2\right) +588779520\, \left(2 , 1\right)\,\left(6 , 1\right) -506908416\, \left(5 , 5\right) 
\nonumber\\
-6693120&\, \left(9 , 1\right) -104961920\, \left(1 , 1\right)\,\left(7 , 1\right) +534097920\, \left(6 , 4\right) 
\nonumber\\
=-2^{4}*5*87517&\, \left(1 , 1\right)^3\,\left(2 , 2\right) -2^{6}*5*41*449\, \left(1 , 1\right)^2\,\left(2 , 1\right)^2 -2^{3}*97*409\, \left(1 , 1\right)^5 
\nonumber\\
+2^{6}*5*31699&\, \left(1 , 1\right)^3\,\left(3 , 1\right) -2^{6}*3^{2}*5^{2}*23*29\, \left(1 , 1\right)\,\left(2 , 2\right)^2 +2^{8}*3*5*13*1367\, \left(2 , 1\right)^2\,\left(2 , 2\right) 
\nonumber\\
-2^{7}*5*172751&\, \left(1 , 1\right)\,\left(3 , 1\right)^2 +2^{6}*3*5*67*2423\, \left(1 , 1\right)\,\left(2 , 2\right)\,\left(3 , 1\right) -2^{8}*5^{2}*17*1471\, \left(2 , 1\right)^2\,\left(3 , 1\right) 
\nonumber\\
+2^{8}*5*17*1039&\, \left(1 , 1\right)^2\,\left(3 , 3\right) -2^{9}*5*234281\, \left(3 , 2\right)^2 -2^{10}*5*7^{2}*19*37\, \left(1 , 1\right)\,\left(2 , 1\right)\,\left(3 , 2\right) 
\nonumber\\
+2^{10}*5^{3}*13*151&\, \left(1 , 1\right)\,\left(2 , 1\right)\,\left(4 , 1\right) -2^{10}*5*7*59*151\, \left(3 , 1\right)\,\left(3 , 3\right) +2^{7}*3^{2}*5*7*11239\, \left(2 , 2\right)\,\left(3 , 3\right) 
\nonumber\\
-2^{7}*5^{3}*11*37&\, \left(1 , 1\right)^2\,\left(4 , 2\right) -2^{12}*5*41*1489\, \left(4 , 1\right)^2 +2^{12}*5^{2}*7*2339\, \left(3 , 2\right)\,\left(4 , 1\right) 
\nonumber\\
-2^{12}*5*7*31*97&\, \left(2 , 1\right)\,\left(4 , 3\right) -2^{9}*5^{2}*7^{2}*13^{2}\, \left(3 , 1\right)\,\left(4 , 2\right) +2^{9}*3^{2}*5*5741\, \left(2 , 2\right)\,\left(4 , 2\right) 
\nonumber\\
-2^{7}*3*5*7*84121&\, \left(2 , 2\right)\,\left(5 , 1\right) -2^{7}*5^{2}*7*1607\, \left(1 , 1\right)^2\,\left(5 , 1\right) -2^{14}*5*79\, \left(1 , 1\right)\,\left(4 , 4\right) 
\nonumber\\
+2^{7}*5*7*41*829&\, \left(1 , 1\right)\,\left(5 , 3\right) +2^{9}*3^{4}*5*7*163\, \left(2 , 1\right)\,\left(5 , 2\right) +2^{9}*5*151*2237\, \left(3 , 1\right)\,\left(5 , 1\right) 
\nonumber\\
-2^{11}*3*5^{2}*7*127&\, \left(1 , 1\right)\,\left(6 , 2\right) +2^{12}*3*5*7*37^{2}\, \left(2 , 1\right)\,\left(6 , 1\right) -2^{8}*3*7*94291\, \left(5 , 5\right) 
\nonumber\\
-2^{8}*3^{2}*5*7*83&\, \left(9 , 1\right) -2^{7}*5*7^{2}*3347\, \left(1 , 1\right)\,\left(7 , 1\right) +2^{12}*3*5*8693\, \left(6 , 4\right) 
\nonumber\\
\end{align*}

\begin{align*}
\bullet_{(N=10,\, n=2,\, NO)} \rangle
=
+19677330&\, \left(1 , 1\right)^3\,\left(2 , 2\right) -47163960\, \left(1 , 1\right)^2\,\left(2 , 1\right)^2 -210843\, \left(1 , 1\right)^5 
\nonumber\\
+22957560&\, \left(1 , 1\right)^3\,\left(3 , 1\right) +361295910\, \left(1 , 1\right)\,\left(2 , 2\right)^2 +560211360\, \left(2 , 1\right)^2\,\left(2 , 2\right) 
\nonumber\\
-1162515840&\, \left(1 , 1\right)\,\left(2 , 1\right)\,\left(3 , 2\right) -64935960\, \left(1 , 1\right)\,\left(3 , 1\right)^2 -1309078560\, \left(2 , 1\right)^2\,\left(3 , 1\right) 
\nonumber\\
+5874354480&\, \left(2 , 2\right)\,\left(3 , 3\right) +296376720\, \left(1 , 1\right)^2\,\left(3 , 3\right) -4296644160\, \left(3 , 2\right)^2 
\nonumber\\
+11735649600&\, \left(3 , 2\right)\,\left(4 , 1\right) +1977019200\, \left(1 , 1\right)\,\left(2 , 1\right)\,\left(4 , 1\right) -5660451840\, \left(3 , 1\right)\,\left(3 , 3\right) 
\nonumber\\
-1848428040&\, \left(2 , 2\right)\,\left(4 , 2\right) -37692000\, \left(1 , 1\right)^2\,\left(4 , 2\right) -8170393680\, \left(4 , 1\right)^2 
\nonumber\\
+356204340&\, \left(1 , 1\right)\,\left(4 , 4\right) -3849526080\, \left(2 , 1\right)\,\left(4 , 3\right) +3083070000\, \left(3 , 1\right)\,\left(4 , 2\right) 
\nonumber\\
+6242937120&\, \left(3 , 1\right)\,\left(5 , 1\right) -8371400640\, \left(2 , 2\right)\,\left(5 , 1\right) -631954800\, \left(1 , 1\right)^2\,\left(5 , 1\right) 
\nonumber\\
+1978054608&\, \left(5 , 5\right) +1725927720\, \left(1 , 1\right)\,\left(5 , 3\right) +2933664960\, \left(2 , 1\right)\,\left(5 , 2\right) 
\nonumber\\
-2201854560&\, \left(6 , 4\right) -2474722800\, \left(1 , 1\right)\,\left(6 , 2\right) +3351233280\, \left(2 , 1\right)\,\left(6 , 1\right) 
\nonumber\\
+127541520&\, \left(9 , 1\right) +155847360\, \left(8 , 2\right) -628636680\, \left(1 , 1\right)\,\left(7 , 1\right) 
\nonumber\\
=+2*3^{4}*5*17*1429&\, \left(1 , 1\right)^3\,\left(2 , 2\right) -2^{3}*3^{2}*5*131011\, \left(1 , 1\right)^2\,\left(2 , 1\right)^2 -3^{4}*19*137\, \left(1 , 1\right)^5 
\nonumber\\
+2^{3}*3^{3}*5*29*733&\, \left(1 , 1\right)^3\,\left(3 , 1\right) +2*3^{3}*5*181*7393\, \left(1 , 1\right)\,\left(2 , 2\right)^2 +2^{5}*3*5*491*2377\, \left(2 , 1\right)^2\,\left(2 , 2\right) 
\nonumber\\
-2^{7}*3*5*605477&\, \left(1 , 1\right)\,\left(2 , 1\right)\,\left(3 , 2\right) -2^{3}*3*5*541133\, \left(1 , 1\right)\,\left(3 , 1\right)^2 -2^{5}*3*5*29*157*599\, \left(2 , 1\right)^2\,\left(3 , 1\right) 
\nonumber\\
+2^{4}*3*5*613*39929&\, \left(2 , 2\right)\,\left(3 , 3\right) +2^{4}*3*5*71*17393\, \left(1 , 1\right)^2\,\left(3 , 3\right) -2^{6}*3*5*4475671\, \left(3 , 2\right)^2 
\nonumber\\
+2^{6}*3*5^{2}*2444927&\, \left(3 , 2\right)\,\left(4 , 1\right) +2^{6}*3^{2}*5^{2}*13*59*179\, \left(1 , 1\right)\,\left(2 , 1\right)\,\left(4 , 1\right) -2^{10}*3*5*401*919\, \left(3 , 1\right)\,\left(3 , 3\right) 
\nonumber\\
-2^{3}*3*5*15403567&\, \left(2 , 2\right)\,\left(4 , 2\right) -2^{5}*3^{3}*5^{3}*349\, \left(1 , 1\right)^2\,\left(4 , 2\right) -2^{4}*3^{2}*5*19*61*9791\, \left(4 , 1\right)^2 
\nonumber\\
+2^{2}*3^{2}*5*1978913&\, \left(1 , 1\right)\,\left(4 , 4\right) -2^{6}*3^{3}*5*41*10867\, \left(2 , 1\right)\,\left(4 , 3\right) +2^{4}*3*5^{4}*102769\, \left(3 , 1\right)\,\left(4 , 2\right) 
\nonumber\\
+2^{5}*3^{2}*5*7^{2}*103*859&\, \left(3 , 1\right)\,\left(5 , 1\right) -2^{6}*3*5*2953^{2}\, \left(2 , 2\right)\,\left(5 , 1\right) -2^{4}*3^{2}*5^{2}*175543\, \left(1 , 1\right)^2\,\left(5 , 1\right) 
\nonumber\\
+2^{4}*3*1931*21341&\, \left(5 , 5\right) +2^{3}*3*5*11*17*76913\, \left(1 , 1\right)\,\left(5 , 3\right) +2^{6}*3*5*3055901\, \left(2 , 1\right)\,\left(5 , 2\right) 
\nonumber\\
-2^{5}*3*5*43*107*997&\, \left(6 , 4\right) -2^{4}*3^{3}*5^{2}*11*37*563\, \left(1 , 1\right)\,\left(6 , 2\right) +2^{8}*3*5*773*1129\, \left(2 , 1\right)\,\left(6 , 1\right) 
\nonumber\\
+2^{4}*3^{3}*5*137*431&\, \left(9 , 1\right) +2^{6}*3*5*67*2423\, \left(8 , 2\right) -2^{3}*3^{3}*5*7^{3}*1697\, \left(1 , 1\right)\,\left(7 , 1\right) 
\nonumber\\
\end{align*}

\begin{align*}
\bullet_{(N=10,\, n=3,\, NO)} \rangle
=
+1835730&\, \left(1 , 1\right)^3\,\left(2 , 2\right) +33390240\, \left(1 , 1\right)^2\,\left(2 , 1\right)^2 +634392\, \left(1 , 1\right)^5 
\nonumber\\
-28381140&\, \left(1 , 1\right)^3\,\left(3 , 1\right) -131463540\, \left(1 , 1\right)\,\left(2 , 2\right)^2 -183165840\, \left(2 , 1\right)^2\,\left(2 , 2\right) 
\nonumber\\
+489240960&\, \left(1 , 1\right)\,\left(2 , 1\right)\,\left(3 , 2\right) +77508240\, \left(1 , 1\right)\,\left(3 , 1\right)^2 +499892640\, \left(2 , 1\right)^2\,\left(3 , 1\right) 
\nonumber\\
-2298663120&\, \left(2 , 2\right)\,\left(3 , 3\right) +40012320\, \left(1 , 1\right)^2\,\left(3 , 3\right) +1749995520\, \left(3 , 2\right)^2 
\nonumber\\
-4571773440&\, \left(3 , 2\right)\,\left(4 , 1\right) -889324800\, \left(1 , 1\right)\,\left(2 , 1\right)\,\left(4 , 1\right) +2111964960\, \left(3 , 1\right)\,\left(3 , 3\right) 
\nonumber\\
+693257760&\, \left(2 , 2\right)\,\left(4 , 2\right) -185952000\, \left(1 , 1\right)^2\,\left(4 , 2\right) +3215222400\, \left(4 , 1\right)^2 
\nonumber\\
+838163040&\, \left(1 , 1\right)\,\left(4 , 4\right) +1907243520\, \left(2 , 1\right)\,\left(4 , 3\right) -1060488000\, \left(3 , 1\right)\,\left(4 , 2\right) 
\nonumber\\
-2530764480&\, \left(3 , 1\right)\,\left(5 , 1\right) +3258178560\, \left(2 , 2\right)\,\left(5 , 1\right) +289576200\, \left(1 , 1\right)^2\,\left(5 , 1\right) 
\nonumber\\
+1811059488&\, \left(5 , 5\right) -1919744880\, \left(1 , 1\right)\,\left(5 , 3\right) -1681467840\, \left(2 , 1\right)\,\left(5 , 2\right) 
\nonumber\\
-2022084480&\, \left(6 , 4\right) +1215583200\, \left(1 , 1\right)\,\left(6 , 2\right) -1237192320\, \left(2 , 1\right)\,\left(6 , 1\right) 
\nonumber\\
-51550560&\, \left(9 , 1\right) +155847360\, \left(7 , 3\right) +250591920\, \left(1 , 1\right)\,\left(7 , 1\right) 
\nonumber\\
=+2*3^{3}*5*13*523&\, \left(1 , 1\right)^3\,\left(2 , 2\right) +2^{5}*3*5*13*5351\, \left(1 , 1\right)^2\,\left(2 , 1\right)^2 +2^{3}*3^{4}*11*89\, \left(1 , 1\right)^5 
\nonumber\\
-2^{2}*3^{2}*5*29*5437&\, \left(1 , 1\right)^3\,\left(3 , 1\right) -2^{2}*3^{3}*5*13*61*307\, \left(1 , 1\right)\,\left(2 , 2\right)^2 -2^{4}*3^{3}*5*11*13*593\, \left(2 , 1\right)^2\,\left(2 , 2\right) 
\nonumber\\
+2^{7}*3*5*13*17*1153&\, \left(1 , 1\right)\,\left(2 , 1\right)\,\left(3 , 2\right) +2^{4}*3*5*322951\, \left(1 , 1\right)\,\left(3 , 1\right)^2 +2^{5}*3*5*13*80111\, \left(2 , 1\right)^2\,\left(3 , 1\right) 
\nonumber\\
-2^{4}*3*5*13*701*1051&\, \left(2 , 2\right)\,\left(3 , 3\right) +2^{5}*3*5*31*2689\, \left(1 , 1\right)^2\,\left(3 , 3\right) +2^{12}*3*5*7*13*313\, \left(3 , 2\right)^2 
\nonumber\\
-2^{9}*3*5*13*29*1579&\, \left(3 , 2\right)\,\left(4 , 1\right) -2^{8}*3*5^{2}*7*13*509\, \left(1 , 1\right)\,\left(2 , 1\right)\,\left(4 , 1\right) +2^{5}*3*5*7*628561\, \left(3 , 1\right)\,\left(3 , 3\right) 
\nonumber\\
+2^{5}*3^{2}*5*13*29*1277&\, \left(2 , 2\right)\,\left(4 , 2\right) -2^{8}*3*5^{3}*13*149\, \left(1 , 1\right)^2\,\left(4 , 2\right) +2^{7}*3*5^{2}*13*25763\, \left(4 , 1\right)^2 
\nonumber\\
+2^{5}*3*5*11*13*12211&\, \left(1 , 1\right)\,\left(4 , 4\right) +2^{9}*3*5*7*13*2729\, \left(2 , 1\right)\,\left(4 , 3\right) -2^{6}*3^{2}*5^{3}*11*13*103\, \left(3 , 1\right)\,\left(4 , 2\right) 
\nonumber\\
-2^{6}*3*5*37*71249&\, \left(3 , 1\right)\,\left(5 , 1\right) +2^{10}*3^{3}*5*7^{2}*13*37\, \left(2 , 2\right)\,\left(5 , 1\right) +2^{3}*3*5^{2}*482627\, \left(1 , 1\right)^2\,\left(5 , 1\right) 
\nonumber\\
+2^{5}*3^{2}*7*929*967&\, \left(5 , 5\right) -2^{4}*3*5*7998937\, \left(1 , 1\right)\,\left(5 , 3\right) -2^{6}*3^{2}*5*13*97*463\, \left(2 , 1\right)\,\left(5 , 2\right) 
\nonumber\\
-2^{7}*3*5*13*81013&\, \left(6 , 4\right) +2^{5}*3^{5}*5^{2}*13^{2}*37\, \left(1 , 1\right)\,\left(6 , 2\right) -2^{7}*3*5*7*13*73*97\, \left(2 , 1\right)\,\left(6 , 1\right) 
\nonumber\\
-2^{5}*3^{3}*5*11933&\, \left(9 , 1\right) +2^{6}*3*5*67*2423\, \left(7 , 3\right) +2^{4}*3*5*1044133\, \left(1 , 1\right)\,\left(7 , 1\right) 
\nonumber\\
\end{align*}
Notice however that the previous three scalars are not othogonal (NO).
Using Gram-Schmidt procedure the second and third ones can be made orthogonal as

\restoregeometry

\end{document}